\documentclass{aa}  

\usepackage{graphicx}
\usepackage{txfonts}
\usepackage{longtable}
\usepackage{rotating}
\usepackage{lscape}
\usepackage{epsfig}
\usepackage{times}
\usepackage{natbib}
\usepackage{setspace}
\usepackage{multirow}
\def\kms{$\rm km\;s^{-1}$}

\def\eg{e.g.,}

\def\apjl{ApJL}
\def\apjs{ApJS}

\def\aap{A\&A}

\def\tef{$T_{\rm eff}\,$}
\def\logg{$\log{g}\,$}

\begin{document}

\title{Infrared Telescope Facility (IRTF) spectral library II:}
\subtitle{New indices in Y, J, H, and L atmospheric windows}

\author{
L. Morelli\inst{1}
\and
V. D. Ivanov\inst{2}
\and
A. Pizzella\inst{3,4}
\and
D. Gasparri\inst{1}
\and
L. Coccato \inst{2}
\and
E. M. Corsini\inst{3,4}
\and
E. Dalla Bont\`a\inst{3,4}
\and
P. Fran\c{c}ois\inst{5}
\and
M. Cesetti\inst{6}
}
\institute{
Instituto de Astronom\'ia y Ciencias Planetarias, Universidad de Atacama, 
Copiap\'o, Chile
\email{lorenzo.morelli@uda.cl}
\and
European Southern Observatory, Karl-Schwarzschild-Str. 2, 85748 Garching 
bei M\"unchen, Germany
\and
Dipartimento di Fisica e Astronomia ``G. Galilei'', Universit\`a di Padova, vicolo 
dell'Osservatorio 3, 35122 Padova, Italy
\and
INAF--Osservatorio Astronomico di Padova, vicolo dell'Osservatorio 5, 
35122 Padova, Italy
\and
GEPI, Observatoire de Paris, PSL Research University, CNRS, 
Univ. Paris Diderot, Sorbonne Paris Cit\'e, 61 Avenue de l’Observatoire, 
75014 Paris, France
\and
Independent Scholar, Viale degli Aironi 4, 30021 Caorle (VE), Italy \\
}
\date{}

 
\abstract
{Stellar population studies in the infrared (IR) wavelength range have two 
main advantages with respect to the optical regime: they probe different populations, 
because most of the light in the IR comes from redder and generally 
older stars, and they allow 
us to see through dust because IR light is less affected by 
extinction. Unfortunately, IR modeling work was 
halted by the lack of adequate stellar libraries, but this has changed 
in the recent years.}
{Our project investigates the sensitivity of various spectral features 
in the 1--5\,$\mu$m wavelength range to the physical properties of stars
(\tef, [Fe/H], \logg) and aims to objectively define 
spectral indices that can characterize the age and metallicity
of unresolved stellar populations.}
{We implemented a method that uses derivatives of the indices as
functions of \tef, [Fe/H] or \logg across the entire
available wavelength range to reveal the most sensitive indices to
these parameters and the ranges in which these indices work.}
{Here, we complement the previous work in the I and K bands,
    reporting a new system of 14, 12, 22, and 12 indices for
    Y, J, H, and L atmospheric windows, respectively, and describe
    their behavior.  We list the equivalent widths of these indices
    for the Infrared Telescope Facility (IRTF) spectral library stars.}
{Our analysis indicates that features sensitive to the effective
    temperature are present and measurable in all the investigated
    atmospheric windows at the spectral resolution and in the
  metallicity range of the IRTF library for a signal-to-noise ratio
  greater than 20-30.  The surface gravity is more challenging and only
  indices in the H and J windows are best suited for this. The
  metallicity range of the stars with available spectra is too narrow
  to search for suitable diagnostics. For the spectra of unresolved
  galaxies, the defined indices are valuable tools in tracing the
  properties of the stars in the IR-dominant stellar populations.}

\keywords{Infrared: stars -- Line: identification -- Stars: abundances
-- Stars: supergiants -- Stars: late-type -- Stars: fundamental
parameters}

\maketitle
%


\section{Introduction}

The extension of stellar evolutionary models towards the observationally
challenging infrared (IR) regime is driven by the fact that  optical
and  IR light originate from different types of stars, allowing us to
obtain independent constraints on young and old stellar populations.
Another reason to venture into the IR region is to penetrate the dust in
galaxies with heavy extinction \citep[\eg][]{1998ApJ...505..639E,Iva00}.
Furthermore, new sets of galaxy spectra with better quality than 
before become available in the IR, posing further challenges to 
modeling \citep[\eg][]{2019A&A...621A..60F}.

The interpretation of these data requires both understanding of the
complex physical processes in galaxies and a significant observational
effort to build libraries of IR stellar spectra. In recent times,
technological improvements have allowed significant advances in both
these areas. The Infrared Telescope Facility (IRTF) stellar library
\citep{2005ApJ...623.1115C,2009ApJS..185..289R} offered 226 stellar
and brown dwarf spectra with a resolving power of $R \sim 2000$ over
$\lambda$=1-5\,$\mu$m. \citet{2012ApJ...760...71C},
  \citet{2015MNRAS.449.2853R,2016A&A...589A..73R}, and
  \citet{2016MNRAS.463.3409V} incorporated the IRTF library into their
  models.
To improve the short metallicity range of the IRTF
   library, \citet{2017ApJS..230...23V} presented a new IRTF library
   with an extended metallicity range ($-1.7 \le [Fe/H] \le 0.6$ dex). 
This library allows us to expand the stellar population models to wide
metallicity and wavelength ranges \citep{2018ApJ...854..139C}.

 The X-SHOOTER Spectral Library
 \citep[XSL,][]{che11,2019A&A...627A.138A} and its later releases
   \citep{chen2014,2020A&A...634A.133G} also extends into the IR
 ($\lambda$=0.3-2.5\,$\mu$m), although with a somewhat higher resolving
 power of $R \sim 10,000$.
In \citet[][hereafter Paper I]{Ces13}, we investigated the sensitivity
of the features in the I and K atmospheric windows to the stellar
physical parameters using the method of ``sensitivity functions'' --
derivatives of the changes in the spectra with respect to the
effective temperature \tef, surface gravity \logg, and to some extent
(because of the limited metallicity spread in the IRTF library) to the
metallicity [Fe/H].  Indeed, the hydrogen, Na, Ca, and CO features
turned out to be the best \tef indicators, and we highlighted that
  some Mg and Fe features can serve the same purpose. We confirmed
that the well-known Ca lines and CO bands were the best \logg
indicators and we demonstrated the surface gravity sensitivity of the
FeClTi feature at 0.908--0.910\,$\mu$m. The most promising abundance
indicators were of course the Ca, Fe, and Mg lines, but the limited [Fe/H]
range (from $-$0.6 to 0.4 dex) prevented us from deriving firm
conclusions.

In this paper, we complete the exploration of the IRTF libraries, defining 
indices optimized for stellar population analysis in the Y, J, H, and L
atmospheric windows. Our work constitutes a preparation for the wealth
of data expected from the new generation of instruments like the Multi-Object Optical and Near-IR
Spectrograph 
\citep[MOONS;][]{2011Msngr.145...11C} at the Very Large Telescope, the Infrared
Multi-Object Spectrograph  \citep[IRMS;][]{2010SPIE.7735E..5PM} at the Thirty Meter Telescope, and
the ELT Adaptive optics for GaLaxy
Evolution  \citep[EAGLE;][]{2010SPIE.7735E..2DC} at the Extreme Large Telescope (ELT) that will make it
possible to measure spectral features of $\alpha$- and iron-group elements
of a large number of galaxies, and even of individual red super giants to tens of
megaparsecs \citep{Eva11}. Here, we provide the basis for these near-future
facilities. Furthermore, we also investigate the mid-IR L window
features that will be accessible with the James Webb Space Telescope.

  Section\,\ref{sec:IRTF} briefly describes the properties of the IRTF
  spectral library. The main spectral features of the Y, J, K, and L
  atmospheric windows, sensitivity map analysis, and their use as
  spectral diagnostics are discussed in
  Sects.\,\ref{sec:main_features}, \ref{sec:sensitivity}, and
  \ref{sec:diagnostics}, respectively.  Index definition and
  measurements and their sensitivity to the galaxy broadening velocity
  are described in Sects.\,\ref{sec:ind_mesure} and
  \ref{sec:sigma_broadening} respectively.  Section\,\ref{conclusions}
  describes and discusses our results.  Due to the large
    number of results we included in the text only those for the Y band,
    and we moved all the figures for the J, H, and L bands to Sect.
    \ref{app:app}. The plots of the model spectra and of the
    sensitivity maps are shown in Appendices \ref{app:plots} and
    \ref{app:sens} respectively.  The plots of the broadening effects
    are shown in Appendix \ref{app:SigmaV} and the behavior of the
    spectral indices in Appendix \ref{app:EWs}. Finally the values of
    the measured indices in the different bands are reported in
    Appendix \ref{app:table}.

\section{IRTF spectral library and analysis method}\label{sec:IRTF}

The IRTF spectral library \citep{2005ApJ...623.1115C,2009ApJS..185..289R} 
contains flux calibrated spectra of about 200 objects, with a typical 
uncertainty of about 5\,\%.
The analysis of the spectra and methodology adopted in this work are
the same as in Paper I. In this section, we give a short summary,
while we refer to Paper I for a detailed description of the
performed steps.

The spectra were re-binned and normalized to unity at regions free of
absorption features, in order to reduces the scatter in our final
analysis. We then assembled the available spectral types (Spt), 
\tef, \logg, [Fe/H], $M_V$ , and parallaxes for the sample 
stars from the literature (Paper I, Table\,3), typically derived 
from high-resolution optical spectroscopy.

Following the convention adopted in Paper I, we quantified the spectral 
types  with two parametrizations. One is based on the 
corresponding effective temperature and the other on a quantitative 
analog SpT of the spectral classes, as follows: 
SpT=0 for F0 stars, 1 for G0, 2 for K0, and so on, with decimals for the 
sub-types ({\rm e.g.}, SpT=0.5 for F5; 1.8 for G8; 3.05 for M0.5).

Our method is a generalization of the sensitivity indices
applied first by \citet[][see their Tables\,2 and 3]{Wor94}, which are simple
derivatives of the index strength with respect to the physical
parameter of interest (e.g., \tef, [Fe/H], \logg): a
positive/negative sensitivity index implies that the feature becomes stronger/weaker
with an increasing parameter value, while zero means no dependence.

We improved this method, applying it directly to the spectra in the
wavelength space. To provide continuity across the entire parametric
space we first generated semi-continuous sets of interpolated
  spectra (called model spectra in Paper I), which were
constructed by least-square fitting of the normalized intensity with a
second-order polynomial as a function of spectral type (SpT), surface
gravity, or temperature for each wavelength bin independently, and
then combining in the wavelength space the polynomial fits for all
wavelength bins. Finally, the  sensitivity map was obtained by calculating the
derivatives of the interpolated spectra with respect to
spectral type, surface gravity, or temperature.

Following the sensitivity map, we identify spectral regions
sensitive to a given physical parameter (they appear as sharp features
on the respective sensitivity map) and we defined the indices
accordingly to encompass these regions. Adopting the same approach
  as Paper I, we decided to define narrow width indices instead of
  intermediate width indices. Broader indices do boost the signal-to-noise ratio (S/N), but they were mostly investigated with lower
  resolution spectrographs
  \citep[\eg][]{1970ApJ...159..973W,Lancon2007} of the past
  generation, so we decided to concentrate on the narrow width indices taking
  advantage of the higher resolution of our spectra.

 It should be noticed that in Paper I we were forced to limit
  the analysis of the metallicity effects, because of the narrow
  [Fe/H] range spanned by the IRTF stars. We decided to adopt the same
  library in this second paper to perform an homogeneous analysis in
  all the atmospheric windows. In a forthcoming paper, we will extend
the investigation of the metallicity effects to the wider [Fe/H] range
spanned by the new extended IRTF stellar library
\citep{2017ApJS..230...23V}.


\section{Main spectral features in Y, J, H, and L atmospheric 
windows}\label{sec:main_features}

In this section we give a brief review of the charachteristics of the
atmospheric windows considered in the paper and we describe their main
spectral features.

\subsection{Y atmospheric window}

The spectral region around 1\,$\mu$m has long been neglected, because
it falls in between the optical and IR regimens where for a long time
detector technology did not work well. The photometric $Y$ filter
centered at 1.035\,$\mu$m (with a full whidth half maximum (FWHM) of $\sim 0.13\,\mu$m) was
introduced by \citet{Hil02} to take advantage of the better
atmospheric transmission in that region. They also investigated which
features dominate the spectra of cool stars and brown dwarfs at
$\lambda$$\sim$0.97-1.07\,$\mu$m. A few low- and medium-resolution 
spectral libraries of late-type stars covered this spectral region
\citep{Leg96, Joy98, 2000A&AS..146..217L,
  McL03,2005ApJ...623.1115C,2005A&A...440.1061L}. More recently,
\citet{Sha10} published a high-resolution (R$\sim$25,000) atlas for
the full range of luminosity classes of F, G, K, and M stars. The XSL
also covers the Y window \citep{chen2014,2019A&A...627A.138A}. We note that
here and in the following we only mention works that include a few tens or more stars and that span 
many SpT and/or luminosity classes, because only these
libraries are actually of interest for stellar population studies.

The strongest absorption features in the Y atmospheric window
($\sim$0.94--1.10\,$\mu$m) are the H lines Pa$\gamma$, Pa$\delta$,
and Pa$\epsilon$. There are also many low and moderate excitation lines of
neutral metals, such as Ca{\sc i}, Cr{\sc i}, Fe{\sc i}, K{\sc i},
Mg{\sc i}, Na{\sc i}, Ni{\sc i}, Si{\sc i,} and Ti{\sc i} along with
singly ionized Sr{\sc ii} \citep{Leg96, Lyu04}. The spectra show
molecular bands, including the first-overtone band heads of CN
extending redward at 1.097$\,\mu$m \citep{McK88}. For the later SpTs,
the Y window contains TiO bands that appear at 0.982$\,\mu$m and
1.014$\,\mu$m, FeH band-head appearing at 0.989$\,\mu$m \citep{Leg96},
and VO bands at 1.06$\,\mu$m
\citep{Leg96,2005ApJ...623.1115C,2009ApJS..185..289R}.  Finally,
broad H$_{2}$O absorption bands appear on both sides of the Y
window. Generally, the H lines are stronger in F- and G-type stars,
while the metal and molecular features are more relevant in K- and
M-type stars.

\begin{figure*}[ht]
\includegraphics[width=16truecm,angle=0]{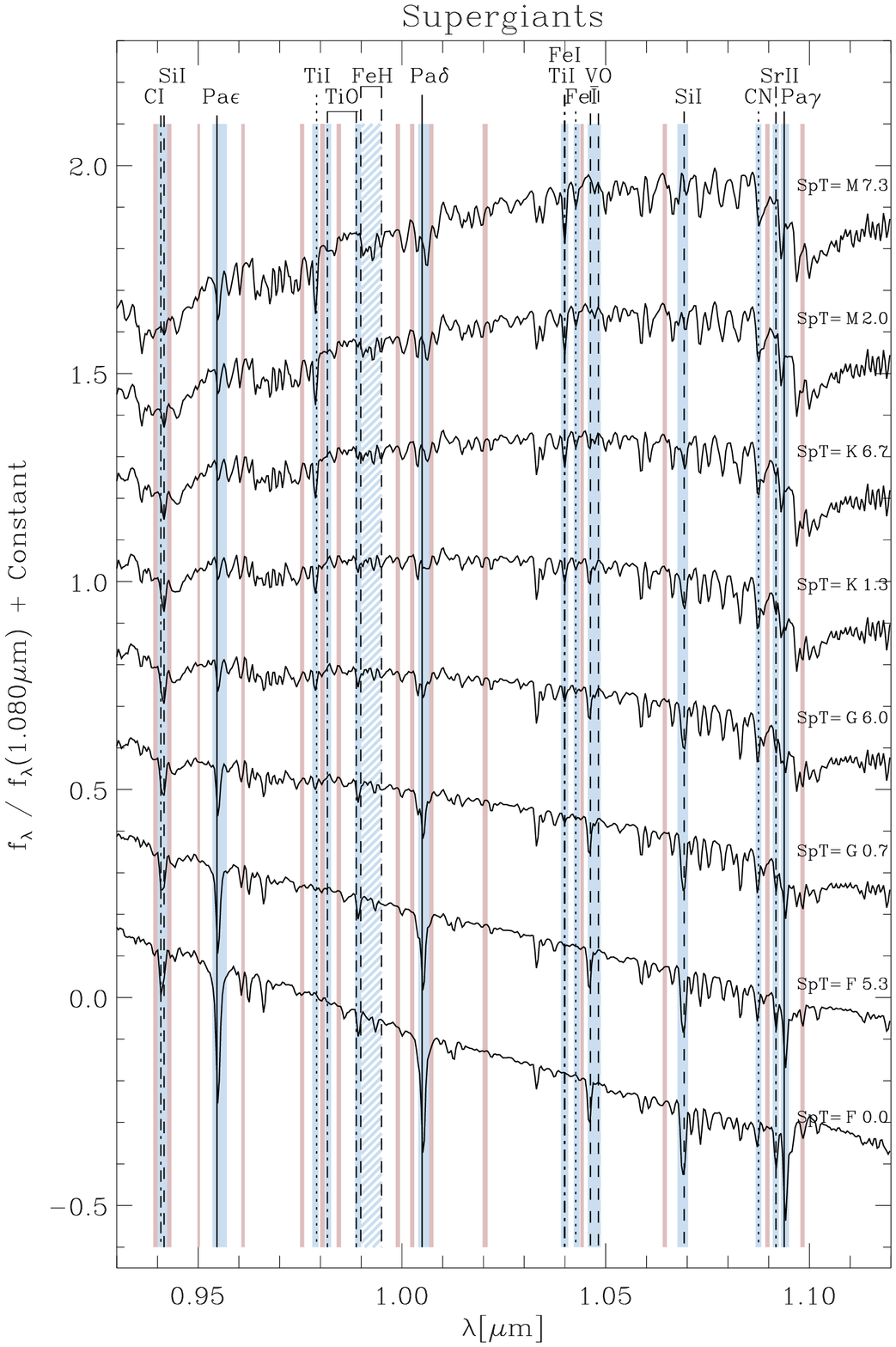}
\caption{Y atmospheric window  model spectra of supergiants 
obtained by fitting at each wavelength the flux-normalized (at 
1.08$\,\mu$m) sample spectra along SpTs. The model spectra for 
different SpTs are offset for displaying purposes and the SpTs 
are given on the right. The green regions mark the bandpasses of 
the newly defined indices, the gray regions mark their adjacent 
continuum as defined in this paper (see Table\,\ref{tab:indices}).  
Some relevant absorption features are marked. Here and for the 
other bands we did not consider features at the edges of the 
windows, because they can be affected by variable atmospheric 
transmission. The plots for stars of other luminosity classes and 
atmospheric windows are shown in 
Appendix\,\ref{app:plots}.}\label{fig:SGiant_fitted_Y}
\end{figure*}

The Y atmospheric window interpolated spectra for the different
SpTs (i.e., for different $T_{\rm eff}$) for supergiant, giant, and dwarf stars are
shown in Figs.\,\ref{fig:SGiant_fitted_Y}, \ref{fig:Giant_fitted_Y},
and \ref{fig:Dwarf_fitted_Y}, respectively.

\begin{table*}[ht]
\caption{Definition of the spectral indices. We have marked 
with the symbol $^{\ddag}$ the index definitions similar to \citet{riffel2019}, 
with the symbol $^{\triangle}$ those similar to \citet{origlia1993},
with the symbol $^{\Box}$ those similar to \citet{Iva04},
with the symbol $^{\dag}$ those similar to \citet{2012ApJ...760...71C} and \citet{2017ApJS..230...23V}, and
with the symbol $^{\diamondsuit}$ those similar to \citet{McL03} and \citet{2005ApJ...623.1115C}.
}\label{tab:indices}

\begin{tabular}{@{}l@{}c@{}c@{}c@{}@{}l@{}c@{}c@{ }c@{}}
\hline
\hline
Index       &\,Species                 & \multicolumn{2}{c}{Bandpass~~~Definitions~~~($\mu$m)}~&~~Index      &\,Species               & \multicolumn{2}{c}{Bandpass~~~Definitions~~~($\mu$m)}\\
Name        &                          & Central         & Blue and Red Continua                           ~&~~Name       &                        & Central          & Blue and Red Continua  \\               
\hline                                                                                                                                                  
            &                          &                 &                                    ~&~             &                        &                  &                                 \\
\multicolumn{4}{c}{\bf Y atmospheric window}                                                  & \multicolumn{4}{c}{\bf H atmospheric window (cont):}                                    \\
CSi         &\,C{\sc i}+Si{\sc i}      &\,0.9400-0.9424~&~~0.9390-0.9399\,\,0.9425-0.9434    ~&~~Br$_{16}$   &H{\sc i}(n=4)           &\,1.552-1.557~    &\,1.548-1.550\,\,1.593-1.594    \\
Pa$\epsilon$&\,H{\sc i}(n=3)           &\,0.9535-0.9570~&~~0.9498-0.9504\,\,0.9606-0.9614    ~&~~CO1$^{\ddag}$ &$^{12}$CO(2,0)          &\,1.557-1.5635~   &\,1.548-1.550\,\,1.593-1.594    \\
Ti          &\,Ti{\sc i}               &\,0.9780-0.9795~&~~0.9750-0.9760\,\,0.9800-0.9810    ~&~~Br$_{15}$   &H{\sc i}                &\,1.567-1.572~    &\,1.548-1.550\,\,1.593-1.594    \\
TiO\,A      &\,TiO                     &\,0.9812-0.9826~&~~0.9800-0.9810\,\,0.9840-0.9850    ~&~~Mg3$^{\ddag}$ &Mg{\sc i}               &\,1.573-1.580~    &\,1.548-1.550\,\,1.593-1.594    \\
TiO\,B      &\,TiO                     &\,0.9885-0.9900~&~~0.9840-0.9850\,\,0.9985-0.9995    ~&~~FeH1$^{\ddag}$$^{\triangle}$$^{\Box}$  &FeH                 &\,1.582-1.586~    &\,1.548-1.550\,\,1.593-1.594    \\
FeH$^{\dag}$$^{\diamondsuit}$  &\,FeH     &\,0.9900-0.9950~&~~0.9840-0.9850\,\,0.9985-0.9995    ~&~~Si$^{\ddag}$$^{\triangle}$$^{\Box}$ &Si{\sc i} &\,1.587-1.591~    &\,1.548-1.550\,\,1.593-1.594    \\
Pa$\delta$  &\,H{\sc i}(n=3)           &\,1.0040-1.0067~&~~1.0020-1.0030\,\,1.0067-1.0077    ~&~~CO2$^{\ddag}$  &$^{12}$CO(2,0)          &\,1.595-1.600~    &\,1.593-1.594\,\,1.616-1.618    \\
FeTi        &\,Fe{\sc i}+Ti{\sc i}     &\,1.0390-1.0408~&~~1.0198-1.0210\,\,1.0438-1.0446    ~&~~Fe2         &Fe{\sc i}               &\,1.605-1.609~    &\,1.593-1.594\,\,1.616-1.618    \\
Fe          &\,Fe{\sc i}               &\,1.0422-1.0436~&~~1.0198-1.0210\,\,1.0438-1.0446    ~&~~Br$_{13}$   &H{\sc i}(n=4)           &\,1.610-1.614~    &\,1.593-1.594\,\,1.616-1.618    \\
VO          &\,VO                      &\,1.0456-1.0488~&~~1.0438-1.0446\,\,1.0640-1.0650    ~&~~CO3$^{\triangle}$ &$^{12}$CO(2,0)          &\,1.618-1.622~    &\,1.616-1.618\,\,1.634-1.637    \\
Si$^{\ddag}$      &\,Si{\sc i}          &\,1.0676-1.0702~&~~1.0640-1.0650\,\,1.0892-1.0902    ~&~~FeH2        &FeH                     &\,1.624-1.628~    &\,1.616-1.618\,\,1.634-1.637    \\
CN          &\,CN                      &\,1.0868-1.0882~&~~1.0640-1.0650\,\,1.0892-1.0902    ~&~~CO4         &$^{12}$CO(2,0)          &\,1.639-1.647~    &\,1.634-1.637\,\,1.6585-1.6605  \\
Sr          &\,Sr{\sc ii}              &\,1.0913-1.0923~&~~1.0892-1.0902\,\,1.0978-1.0988    ~&~~Fe3         &Fe{\sc i}               &\,1.651-1.658~    &\,1.634-1.637\,\,1.6585-1.6605  \\
Pa$\gamma$  &\,H{\sc i}(n=3)           &\,1.0930-1.0950~&~~1.0892-1.0902\,\,1.0978-1.0988    ~&~~CO5         &$^{12}$CO(2,0)          &\,1.6605-1.6640   &\,1.6585-1.6605\,\,1.6775-1.679 \\
            &                          &                &                                    ~&~~Al1         &Al{\sc i}               &\,1.6705-1.6775   &\,1.6585-1.6605\,\,1.6775-1.679 \\
\multicolumn{4}{c}{\bf J atmospheric window}                                                 ~&~~Br$_{11}$   &H{\sc i}(n=4)           &\,1.6790-1.6825   &\,1.6775-1.679\,\,1.6825-1.6835 \\
Na$^{\ddag}$$^{\dag}$         &\,Na{\sc i}       &\,1.1358-1.1428~&~~1.1320-1.1340\,\,1.1430-1.1460    ~&~~COMg$^{\Box}$ &$^{12}$CO(2,0)    &\,1.705-1.713~    &\,1.692-1.696\,\,1.714-1.716    \\
FeCr$^{\ddag}$    &\,Fe{\sc i}+Cr{\sc i} &\,1.1600-1.1624~&~~1.1560-1.1585\,\,1.1716-1.1746    ~&~~            &\,+Mg{\sc i}            &                  &                                 \\
K1\,A$^{\dag}$$^{\diamondsuit}$ &\,K{\sc i} &\,1.1670-1.1714~&~~1.1560-1.1585\,\,1.1716-1.1746    ~&~~Br$_{10}$   &H{\sc i}(n=4)           &\,1.735-1.739~    &\,1.725--1.728\,\,1.744--1.748  \\
C           &\,C{\sc i}                &\,1.1746-1.1765~&~~1,1716-1.1746\,\,1.1805-1.1815    ~&~~            &                        &                  &                                 \\
K1\,B$^{\dag}$$^{\diamondsuit}$ &\,K{\sc i} &\,1.1765-1.1800~&~~1.1716-1.1746\,\,1.1805-1.1815    ~& \multicolumn{4}{c}{\bf L atmospheric window}                                              \\
Mg$^{\ddag}$       &\,Mg{\sc i}               &\,1.1820-1.1840~&~~1.1805-1.1815\,\,1.1855-1.1875    ~&~~Mg1$^{\ddag}$  &Mg{\sc i}               &\,3.388-3.406~    &\,3.360-3.376\,\,3.425-3.440    \\
Si$^{\ddag}$       &\,Si{\sc i}               &\,1.1977-1.2004~&~~1.1910-1.1935\,\,1.2050-1.2070    ~&~~Mg2$^{\ddag}$  &Mg{\sc i}               &\,3.454-3.464~    &\,3.425-3.440\,\,3.476-3.496    \\
SiMg        &\,Si{\sc i}+Mg{\sc i}     &\,1.2070-1.2095~&~~1.2050-1.2070\,\,1.2150-1.2180    ~&~~P$_{16}$    &OH(1-0)\,P$_{16}$       &\,3.514-3.532~    &\,3.476-3.496\,\,3.540-3.560    \\
K2\,A$^{\ddag}$$^{\diamondsuit}$&\,K{\sc i} &\,1.2415-1.2455~&~~1.2350-1.2380\,\,1.2460-1.2490    ~&~~P$_{14}$    &OH(2-1)\,P$_{14}$       &\,3.564-3.582~    &\,3.540-3.560\,\,3.604-3.612    \\
K2\,B$^{\diamondsuit}$       &\,K{\sc i}  &\,1.2495-1.2540~&~~1.2460-1.2490\,\,1.2560-1.2580    ~&~~P$_{17}$    &OH(1-0)\,P$_{17}$       &\,3.582-3.597~    &\,3.540-3.560\,\,3.604-3.612    \\
Pa$\beta$$^{\ddag}$   &\,H{\sc i}(n=3)         &\,1.2795-1.2840~&~~1.2755-1.2780\,\,1.2855-1.2873    ~&~~P$_{15}$    &OH(2-1)\,P$_{15}$       &\,3.628-3.648~    &\,3.604-3.612\,\,3.665-3.673    \\
Al$^{\ddag}$$^{\dag}$ &\,Al{\sc i}               &\,1.3115-1.3168~&~~1.3050-1.3075\,\,1.3230-1.3250    ~&~~Pf$\gamma$  &H{\sc i}(n=5)           &\,3.736-3.745~    &\,3.718-3.728\,\,3.793-3.803    \\
            &                          &                &                                    ~&~~Mg3$^{\ddag}$ &Mg{\sc i}               &\,3.862-3.870~    &\,3.840-3.850\,\,3.870-3.880    \\
\multicolumn{4}{c}{\bf H atmospheric window}                                                 ~&~~Hu$_{15}$   &H{\sc i}(n=6)           &\,3.900-3.914~    &\,3.870-3.880\,\,3.914-3.922    \\
Mg1$^{\ddag}$ &\,Mg{\sc i}               &\,1.485-1.491  ~&~~1.483-1.485\,\,1.491-1.500        ~&~~SiO1        &SiO(2,0)                &\,3.998-4.012~    &\,3.990-3.998\,\,4.025-4.035    \\
Mg2$^{\ddag}$$^{\Box}$ &\,Mg{\sc i}       &\,1.500-1.508  ~&~~1.491-1.508\,\,1.510-1.512        ~&~~SiO2        &SiO(3,1)                &\,4.035-4.050~    &\,4.025-4.035\,\,4.060-4.070    \\
K1          &\,K{\sc i}                &\,1.515-1.520  ~&~~1.510-1.512\,\,1.523-1.525        ~&~~SiO3        &SiO(4,2)                &\,4.080-4.090~    &\,4.060-4.070\,\,4.095-4.105    \\
Fe1         &\,Fe{\sc i}               &\,1.528-1.533  ~&~~1.523-1.525\,\,1.548-1.550        ~&~~            &                        &                  &                                 \\
\hline
\end{tabular}
\end{table*}

\subsection{J atmospheric window}

Photometrically, the first practical exploration of the J window
($\lambda\sim1.11-1.33\,\mu$m) dates back to
\citet{1951MNRAS.111..537F}.  \citet{1966ARA&A...4..193J} defined the
precursors of the $JHK$ filters used today. The J window is relatively
free of telluric features compared with the other parts of the IR, and
this makes it more valuable for quantitative spectroscopy. The first
spectroscopic atlas of stars observed in the IR was published by
\citet{1970AJ.....75..785J}. A number of other spectral libraries
covered it before the IRTF library
\citep{Leg96,Joy98,2000A&AS..146..217L,McL03,2005A&A...440.1061L,Lan07,2007BASI...35..359R,2012A&A...539A.109L}.
The work of \citet{Wal00} stands out because their spectra have
higher resolving power ($R\sim 3000$) and better S/N. The XSL also
covers the J window \citep{chen2014}.

A prominent feature of F and G stars in the J window is the Pa$\beta$
at 1.282\,$\mu$m, whose strength decrease toward late sub-types. It
becomes very weak in K stars and is absent for M stars whose spectra
are dominated by lines of neutral metal species of lower ionization
potential. Notable atomic features in the spectra of K and M stars are
the Na{\sc i} doublet at 1.14\,$\mu$m, the K{\sc i} doublet at
1.18\,$\mu$m and 1.25\,$\mu$m, and the Al{\sc i} doublet at
1.313\,$\mu$m and 1.318\,$\mu$m. They are strongest in late-M dwarfs
and weaker in corresponding giants and supergiants. Therefore, they
are excellent luminosity class (or surface gravity) indicators. The
Si{\sc i} features appear in early F stars and fade in early M stars. We
also see some Fe{\sc i}, Mg{\sc i}, Ti{\sc i}, and Mn{\sc i} lines, but they are too
weak to be suitable for extragalactic work.

At the resolution of our data, a number of spectral features are 
blends of absorption lines from different chemical elements. For 
example, the feature at $\sim$1.169 $\mu$m is due primarily to 
C{\sc i} in F and G stars. However, for later spectral types, K{\sc i} 
absorption plays the dominant role in determining the feature 
strength. The Al, Mn, Fe, and Mg lines behave similarly to Si, but 
persist through the latest observed types. The continuum shape in 
the J window is redder toward later spectral type. The J atmospheric window interpolated spectra for the different
SpTs (i.e., for different $T_{\rm eff}$) for supergiant, giant, and dwarf stars are shown in
Figs.\,\ref{fig:SGiant_fitted_J}, \ref{fig:Giant_fitted_J}, and
\ref{fig:Dwarf_fitted_J}, respectively.

\subsection{H atmospheric window}

The photometric use of the H window ($\lambda \sim$1.48-1.78\,$\mu$m) also 
goes back to \citet{1951MNRAS.111..537F} and \citet{1966ARA&A...4..193J}, 
but the spectroscopic treatment is more difficult than the J window
because it contains more telluric absorptions. Furthermore, the 
background emission is due to numerous O$_2$H and OH lines that 
vary on a timescale of a few minutes, making the data processing 
more challenging\footnote{For further information see the ISAAC user 
manual: 
\url{http://www.eso.org/sci/facilities/paranal/decommissioned/isaac/doc/VLT-MAN-ESO-14100-0841_v92.pdf}.}.

Although the K window ($\lambda \sim 1.92- 2.40\, \mu$m) is the most
studied, the presence of circumstellar dust or active galactic nuclei (AGN) often results in
a significant excess of continuum emission at $\lambda$$\geq$2\,$\mu$m
\citep{Sch97,Iva00}. This excess weakens or even renders invisible the
photospheric absorption features, making the K window unsuitable for
stellar population studies. Yet, the radiation from the stellar
photospheres still dominates shorter wavelengths in the H window. This
motivated the first studies of star spectra in the H window
\citep[\eg][]{1970AJ.....75..785J,Lan92,Ori93,Dall96,1996ApJ...470..597M,Blu97,1998ApJ...508..397M,Blu97,1998ApJ...508..397M}.
More recently, a number of spectral libraries have been reported for wide varieties of
stars
\citep[\eg][]{Iva04,2004BASI...32..311R,2005A&A...440.1061L,2005ApJS..161..154H,2008MNRAS.385.1076V,2011JKAS...44..125H,Coo13,2019MNRAS.484.4619G}.
The spectral libraries of \citet{2012A&A...539A.109L} and
\citet{2018ApJS..238...29P} stand out because of their significantly
higher resolving power of $R\sim 100,000$ and $45,000$,
respectively. However, they contain less stars and are not really
suitable for galaxy population modeling.

A large effort was devoted to abundance studies of red giants and
supergiants because of the rich concentration of both molecular and
metal lines in the H window
\citep{Ori93,Ori02,Ori03,Ori04,Cun07,Dav09a,Dav09b}. High-resolution
spectra in the H-band domain enabled the Apache Point Observatory
Galactic Evolution Experiment survey to determine the chemical
abundances of 16 elements for all their stars
\citep{2013ApJ...765...16S,2016AJ....151..144G,2019ApJ...886L...8F}.

The identification of features in the H window is challenging because
there are many relatively weak metal absorption features and
high-order H lines from the Br series and the second CO overtone
vibrational-rotational band is present in cooler
objects. Fortuitously, a few metal absorptions of MgI (at
$\sim$1.49--1.51\,$\mu$m and 1.71\,$\mu$m), Si ($\sim$1.59\,$\mu$m),
and Al ($\sim$1.67\,$\mu$m) are clearly separated especially for hot
stars.
The Mg and Al features are the deepest in K and early M dwarfs and
they are significantly weaker in giants and supergiants for a given
SpT. For this reason, these lines appear to be luminosity class
indicators.

H$_2$O bands are present on both sides of the H atmospheric window for
spectral types later than M4 \citep[see also][]{2009ApJS..185..289R}.
Some FeH absorption features are also evident, as pointed out by
\citet{Cus03}. The second overtone vibrational-rotational band of CO
is another good surface gravity indicator in K and M stars. Many
  OH band features are present in this spectral range. They are often
  in a region crowded with spectral features as shown by
  \citet{2012A&A...539A.109L} and \citet{2013ApJ...765...16S}, and
  could contaminate close spectral features. Therefore, we consider
  here only the most isolated ones to try to avoid confusion as much as possible.
The H atmospheric window  interpolated spectra for the different 
SpTs (i.e., for different $T_{\rm eff}$) for supergiant, giant, and 
dwarf stars are shown in 
Figs.\,\ref{fig:SGiant_fitted_H}, \ref{fig:Giant_fitted_H}, and  \ref{fig:Dwarf_fitted_H} respectively.

\subsection{L atmospheric window}

The L atmospheric window ($\lambda \sim 3.35-4.15\, \mu$m) was first
explored by \citet{1951MNRAS.111..537F} and
\citet{1966ARA&A...4..193J} too and a review of other early works can
be found in \citet{1979ARA&A..17....9M}.  Other spectral libraries in
this range were presented by \citet{1984ApJS...54..177R},
\citet{Wal02}, \citet{2009PASP..121..125M}, and
\citet{2017A&A...598A..79N}.  Some space-based libraries of $\sim$300
stars are also available from the Infrared Space Observatory
\citep{2002A&A...390.1033V,2002A&A...394..539H,2003ApJS..147..379S}
and from the Akari mission \citep[][albeit with resolving power
  of $R\sim$20 only]{2013AJ....145...32S}.

The most prominent metal features belong to Mg at $\lambda \sim$3.39\,$\mu$m
and $\lambda \sim$3.46\,$\mu$m, while the strongest molecular bands belong to the 
SiO ($\lambda \sim$4.00\,$\mu$m, $\lambda \sim$4.06\,$\mu$m,  and $\lambda \sim$4.08\,$\mu$m).
Some weaker OH bands are also present. All these features are stronger 
in cooler stars. The molecular lines are stronger in  
supergiants and giants, than in dwarf stars. The blue edge of the L window can be 
affected by broad H$_2$O absorption for M6-M9 giant stars. We 
exclude from our analysis the spectral region with
$\lambda$$\lesssim$3.4\,$\mu$m because it is affected by poor
atmospheric transmission.

Importantly, the L window can be significantly affected by dust
emission that is circumstellar, when individual stars are considered, 
or in the galactic environment when galaxies are considered. The L atmospheric window interpolated spectra for different SpTs
(i.e., for different $T_{\rm eff}$) for supergiant, giant, and dwarf
stars are shown in Figs.\,\ref{fig:SGiant_fitted_L},
\ref{fig:Giant_fitted_L}, and \ref{fig:Dwarf_fitted_L}, respectively.

\section{Sensitivity maps}\label{sec:sensitivity}

The sensitivity maps for the different SpTs and surface gravity
according to their class and spectral type are shown in 
Fig.\,\ref{fig:SupGian_SpT_Y} and in Appendix\,\ref{app:sens}. 
The different plots are shown with a constant vertical offset for 
display purposes and the position of the promisingly sensitive 
elements is marked. L stars were excluded from the surface gravity 
sensitivity maps because of their limited gravity coverage. At the 
edge of the atmospheric window, the derivative was very noisy and we 
exclude the lines in this range from the analysis.

\subsection{Y atmospheric window}
\begin{figure*}[ht]
\includegraphics[width=16truecm,angle=0]{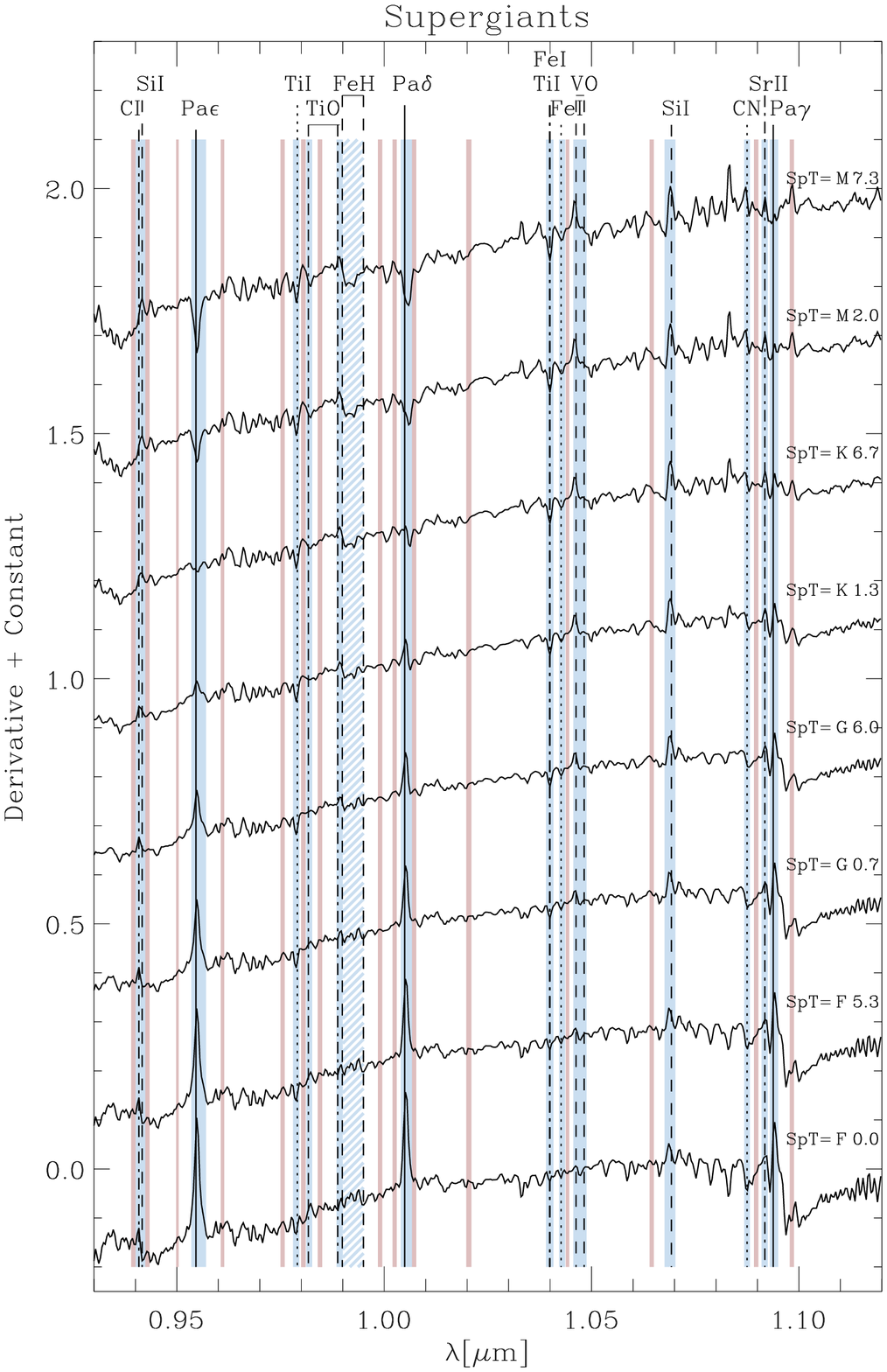}
\caption{Y atmospheric window  sensitivity map for SpT of supergiants.
Details are given in Fig.\,\ref{fig:SGiant_fitted_Y}. The plots for other
luminosity classes and atmospheric windows are shown in
Appendix\,\ref{app:plots}.}\label{fig:SupGian_SpT_Y}
\end{figure*}

The Y atmospheric sensitivity maps with respect to the SpTs are shown
in Figs.\,\ref{fig:SupGian_SpT_Y}, \ref{fig:Gian_SpT_Y}, and
\ref{fig:Dwarf_SpT_Y} for supergiant, giant, and dwarf stars,
respectively.
The regions corresponding to the Pa series of the H are sensitive to
the spectral type, showing a positive peak for the F stars that
gradually transforms into a negative peak for the M stars. A similar
trend is also shown in the FeH region and, even if with less
intensity, in the VO band.  This behavior is more evident for dwarf
than for supergiant and giant stars. The other elements show much
weaker features in the sensitivity map (i.e., much weaker change in the
spectrum). Si and Sr seem to be sensitive only for K and M stars, while
the CN is very close to a Fe line and they are probably contaminating each other, introducing some instability into the trend.

\begin{figure*}[ht]
\includegraphics[width=16truecm,angle=0]{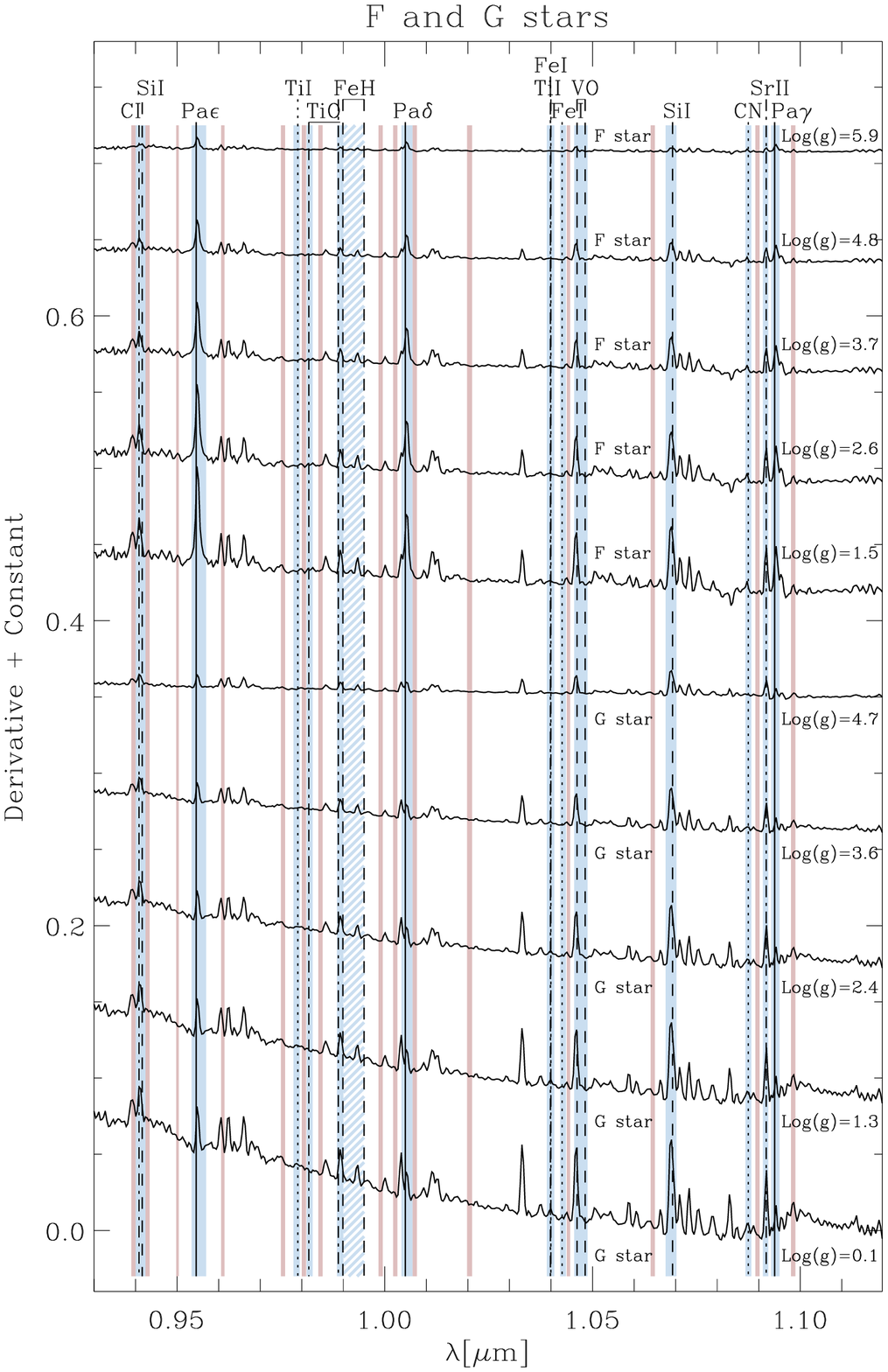}
\caption{Y atmospheric window sensitivity map for surface gravity
of F (top) and G-type stars (bottom). The sensitivities for different
gravity values are offset for display purposes and the central
values of the corresponding \logg bins are given. Figure\,\ref{fig:SupGian_SpT_Y} gives more details.}\label{fig:FG_Logg_Y}
\end{figure*}

\begin{figure*}[ht]
\includegraphics[width=16truecm,angle=0]{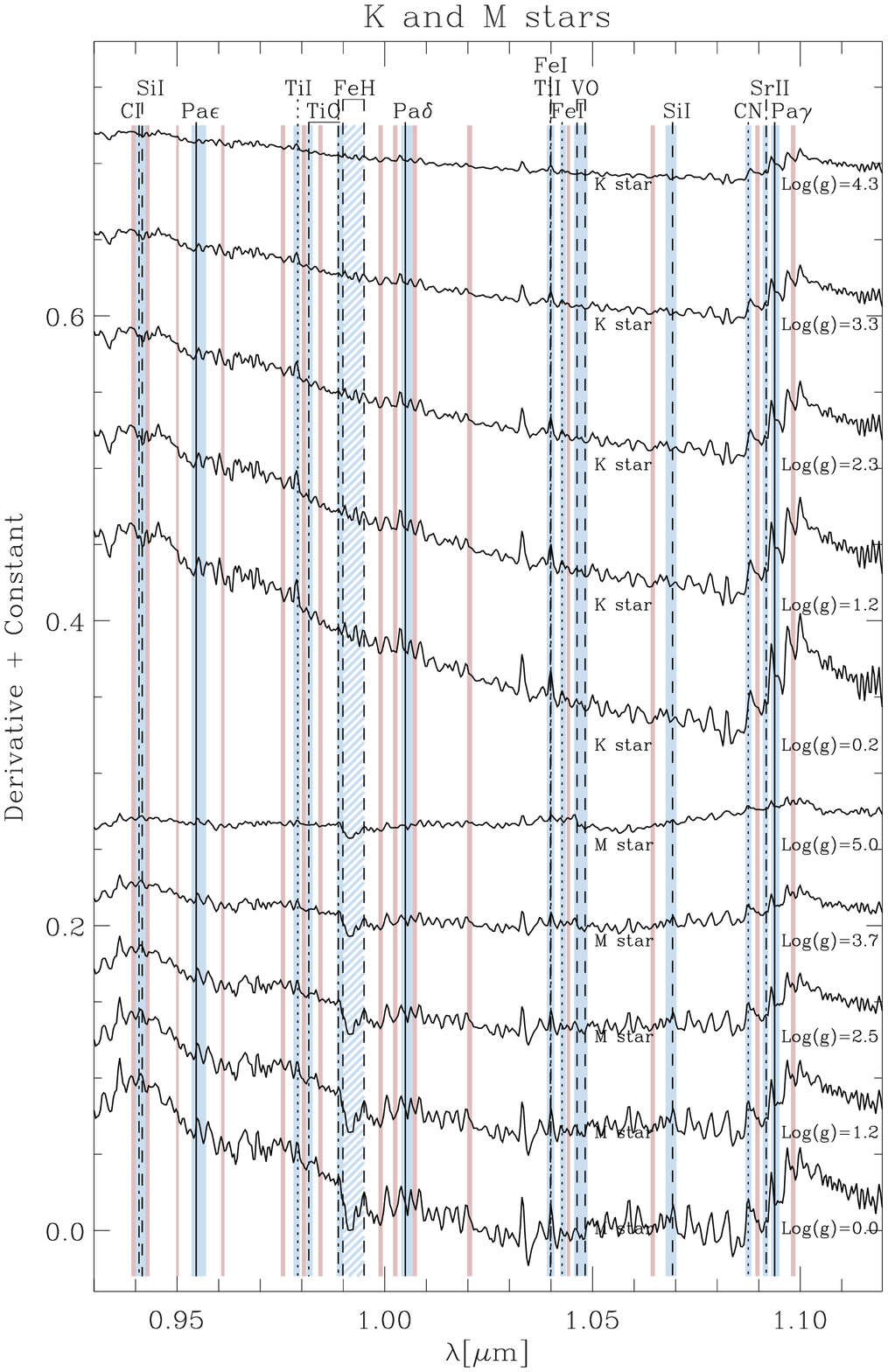}
\caption{Y atmospheric window  sensitivity map for surface gravity
of K (top) and M-type stars (bottom).  Figure\,\ref{fig:FG_Logg_Y} gives more
details.}\label{fig:KM_Logg_Y}
\end{figure*}

The Y atmospheric sensitivity maps with respect to the surface gravity
are shown in Figs.\,\ref{fig:FG_Logg_Y} and \ref{fig:KM_Logg_Y} for the F
and G stars and K and M stars, respectively.
The lines of the Pa series of the H show a positive peak
increasing from the low-gravity to high-gravity stars and being
evident only for the G and F type stars.  Some minor but significant
changes have been found in the derivative plot of the Si and
the Sr lines.

\subsection{J atmospheric window}

The J atmospheric sensitivity maps with respect to the SpTs are shown in
Figs.\,\ref{fig:SupGian_SpT_J}, \ref{fig:Gian_SpT_J}, and 
\ref{fig:Dwarf_SpT_J} for supergiant, giant, and dwarf stars,
respectively.
In this atmospheric window the behavior of supergiant and giant
stars is different from that of the dwarf stars.  The Na doublet, K, Mg, Si, and
to a small extent Fe and Al show a very strong positive peak that
decreases to very negative values moving from the early- to late-type dwarf stars.
This trend is not visible for the supergiants and giant stars.
In contrast, for supergiant and giant stars the most prominent features are
Pa$~\beta$ and C with values that are positive for the F stars
and decrease moving to the M stars. These changes are, instead, almost negligible for
dwarf stars.

The J atmospheric sensitivity maps with respect to the surface gravity
are shown in Figs.\,\ref{fig:FG_Logg_J} and \ref{fig:KM_Logg_J} for the F
and G stars and K and M stars, respectively.
Decreasing positive values from low-gravity to high-gravity stars for the C and Si elements
were found in F and G stars, while Na and K elements show a dependence
with gravity only in K and M stars.

\subsection{H atmospheric window}

The H atmospheric sensitivity maps with respect to the SpTs are shown
in Figs.\,\ref{fig:SupGian_SpT_H}, \ref{fig:Gian_SpT_H}, and
\ref{fig:Dwarf_SpT_H} for supergiant, giant, and dwarf stars,
respectively.  This is the atmospheric window with the strongest
sensitivity, no matter which class of stars we consider. However, the
elements showing sensitivity in giant and supergiant stars are
different from the elements showing sensitivity for dwarf stars.
For the latter the most evident changes are in the Mg, Si, Fe, and Al
lines, while the Br lines do not display any
significant peak. On the other hand, in sensitivity maps of the giant and supergiant stars
the Br lines have a large range of variations, going from positive
values in F stars to negative values in M stars.  Even if less
evident than in dwarf stars, the Mg, Si, Fe, and Al lines still maintain a weak
sensitivity trend with the spectral type also in these classes.

The H atmospheric sensitivity maps with respect to the surface gravity
are shown in Figs.\,\ref{fig:FG_Logg_H} and \ref{fig:KM_Logg_H} for F
and G stars and K and M stars, respectively.  In the F and G stars
the sensitivity maps show strong variations for the Br lines. Their
peaks are always very positive indicating strong changes in the line
width. The other elements have smaller variations not comparable
with those of H lines but still detectable. Regarding the F and G stars,
the dominant variations are detected in the Mg and CO lines, while the
Br lines show minor variations.

\subsection{L atmospheric window}

The L atmospheric sensitivity maps with respect to the SpTs are shown
in Figs.\,\ref{fig:SupGian_SpT_L}, \ref{fig:Gian_SpT_L}, and
\ref{fig:Dwarf_SpT_L} for supergiant, giant, and dwarf stars,
respectively.  The most prominent feature is Mg, which
decreases from the F to the M stars. For the Mg at $\lambda \simeq 3.4\,\mu$m we
found a possible contamination 
that needs to be considered.  The SiO also shows a clear peak in the
sensitivity map decreasing from F stars to M stars.  The P lines show
medium to weak sensitivity to the changes along SpT.

The L atmospheric sensitivity maps with respect to the surface gravity
are shown in Figs.\,\ref{fig:FG_Logg_L} and \ref{fig:KM_Logg_L} for F
and G stars and K and M stars, respectively.  F and G stars show
a weak dependency of the SiO lines with the surface gravity and almost
no sensitivity has been detected in the sensitivity map of F and K stars.

\section{Indices definition and equivalent width measurements}
\label{sec:ind_mesure}
We defined an index for all the features that display a significant variation in the
sensitivity maps (at least over some range of parameters), and that 
are therefore promising tools to measure \tef (or Spt) and
\logg.
For each index, we have a central bandpass covering the feature of
interest and two adjacent bandpasses tracing the local
``continuum'' \footnote{We note that at the IRTF library resolution
  these bands do not actually measure the stellar continuum in cool
  stars. } at the red and blue sides of the feature, as done in Paper
I. The equivalent width for each spectral feature has been derived
following Eq. 1 in Paper I. The central bandpass was designed to
include the peak of the particular feature at the sensitivity map and
the continuum bandpasses were placed on spectral regions where the
sensitivity map is, as much as possible, constant.

We defined 14 indices in Y, 12 in J, 22 in H, and 12 in the L atmospheric
windows. The index definitions are listed in Table\,\ref{tab:indices}
and their bandpasses are plotted in
Figs.\,\ref{fig:SGiant_fitted_Y}, \ref{fig:KM_Logg_Y}, and
\ref{fig:Giant_fitted_Y}--\ref{fig:Dwarf_fitted_L}.  We measured
the equivalent widths of all the indices. Their values are listed in
Tables\,\ref{tab:index_mis_Y_1} and \ref{tab:index_mis_L_1} and
plotted versus the star effective temperature and surface gravity in
Figs.\,\ref{fig:Ind_Spt_Y}, \ref{fig:Ind_Gr_Y}, and
\ref{fig:Ind_Spt_J}--\ref{fig:Ind_Gr_L}, respectively.

The precision of the measured indices depends on the S/N ratio of the
spectra, which varies widely because of the sky emission and telluric
absorption features.  To obtain a robust evaluation of the
uncertainties of the indices, we simulated some indices of the Y band,
choosing as example CSi, Pa$\epsilon$, and Pa$\delta$ for a FII star,
Ti, CN, and Pa$\delta$ for a KIII star, and FeH, VO, and Pa$\delta$
for a MI star. We generated spectra with eight different S/N levels,
ranging from 20 to 120, by adding artificial noise to the three test
stars from the IRTF library.

\begin{figure}
\includegraphics[angle=-90,width=0.5\textwidth]{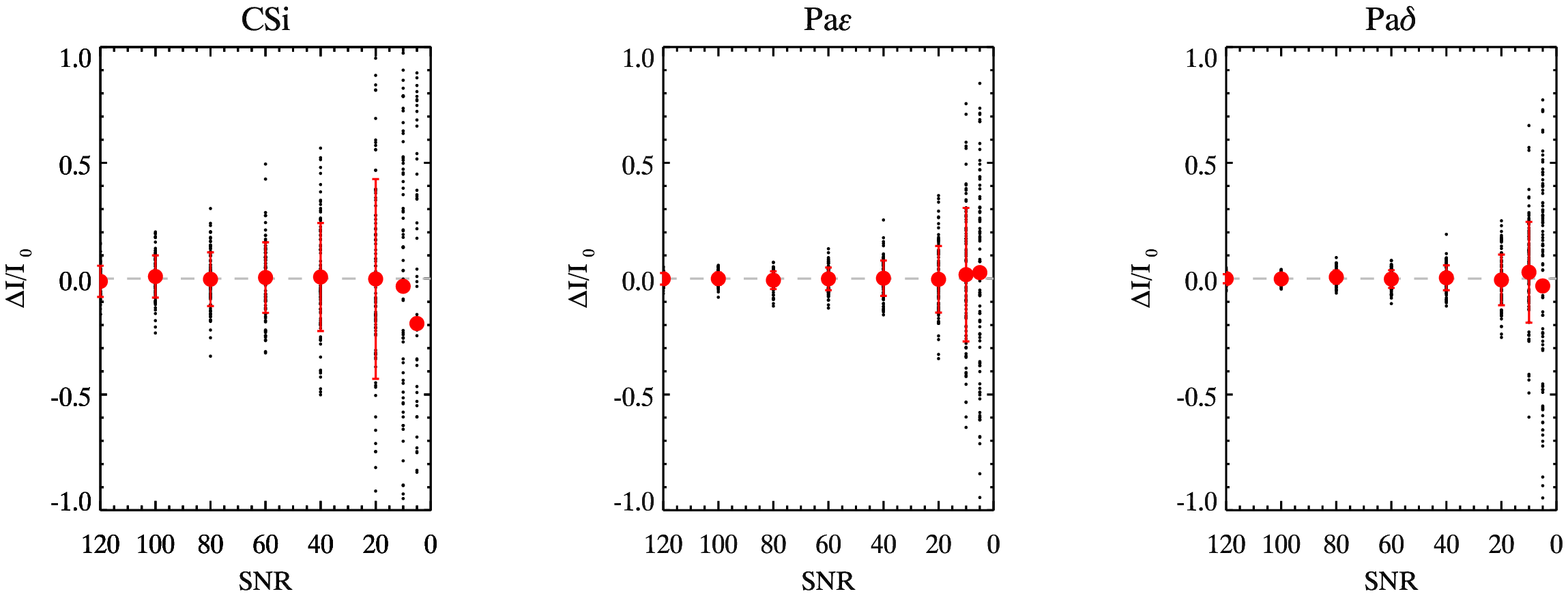}
\includegraphics[angle=-90,width=0.5\textwidth]{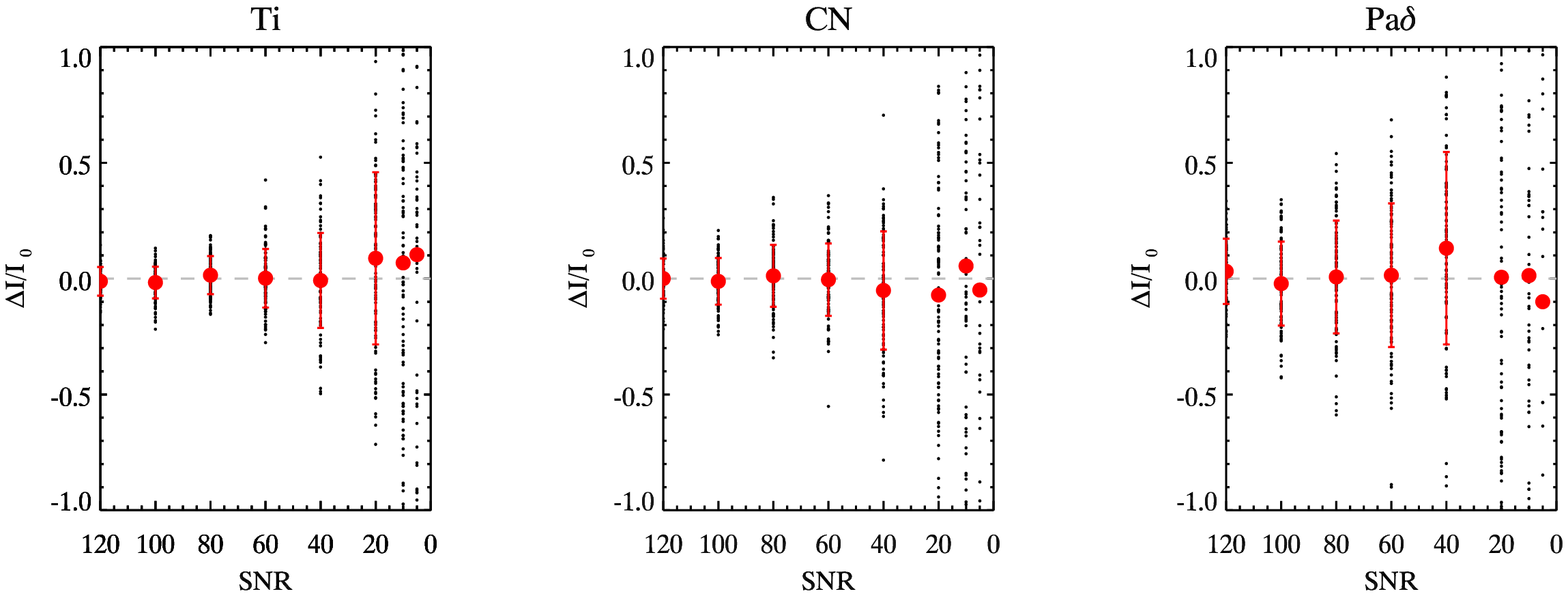}
\includegraphics[angle=-90,width=0.5\textwidth]{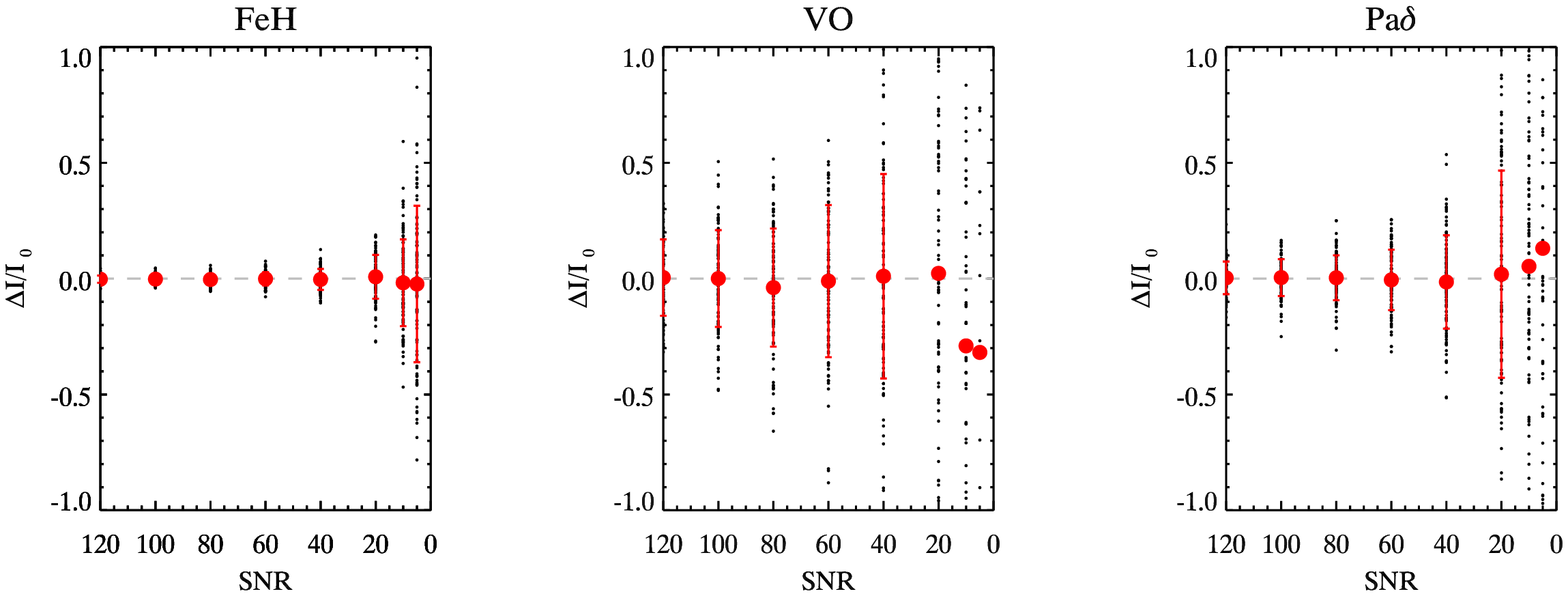}
\caption{Deviation of the index (measured on the noisy artificial spectrum)
from the index value in the ``noiseless'' spectrum, normalized 
by the latter, plotted versus the S/N of the noisy spectrum 
for Y band spectral features in MI (lower panels), KIII (central panels), 
and FII (upper panels) stars. The S/N is given per \AA.}\label{fig:SNR_plots.ps}
\end{figure}

Figure \ref{fig:SNR_plots.ps} shows the effect of the S/N on the
precision of the measured indices.  Generally, it is negligible except
for very low S/N (<10).  For some features (e.g., Pa$\epsilon$ in FII
star or FeH in MI star), we can obtain a precision of 10\% for
S/N$\sim$10-20, while for others (\eg\/ Pa$\delta$ in MI star or CN in
KIII star) we can reach a precision of 30\% for S/N$\sim$20-40. Only
for a few weaker features, the precision remains at a 50\% level even
for higher S/N.

\section{Sensitivity of indices to galaxy velocity dispersion broadening}
\label{sec:sigma_broadening}

The application of spectroscopic index systems to study stellar 
populations of unresolved galaxies 
\citep[\eg][]{1987ApJ...313...42D,1994ApJS...94..687W,Morelli2017} is affected 
by the intrinsic velocity dispersion that broadens the lines in the 
integrated spectra of the galaxies. This could render useless some 
weaker or close to each other indices that we defined in this work. 
It is therefore important to investigate how our measurements are 
affected by the internal kinematics of galaxies.

As done in Paper I and in order to take this into account, we
broadened all the spectra by convolving them with a Gaussian of
$\sigma$ varying from 115 to 400$\rm\,km\,s^{-1}$ in steps of
25$\rm\,km\,s^{-1}$. The indices were measured for all broadened
spectra and a third-order polynomial fit was performed to the relative
changes of each index $\Delta(Index)/Index(\sigma_0)$ as a function of
the velocity dispersion.  As expected for our indices, we found a
strong correlation between the line strength and broadening. The
indices with a low equivalent width value show a large variation even for small values of
$\sigma$ whereas the strongest indices are less affected.

\begin{figure*}
\includegraphics[angle=90,width=\textwidth]{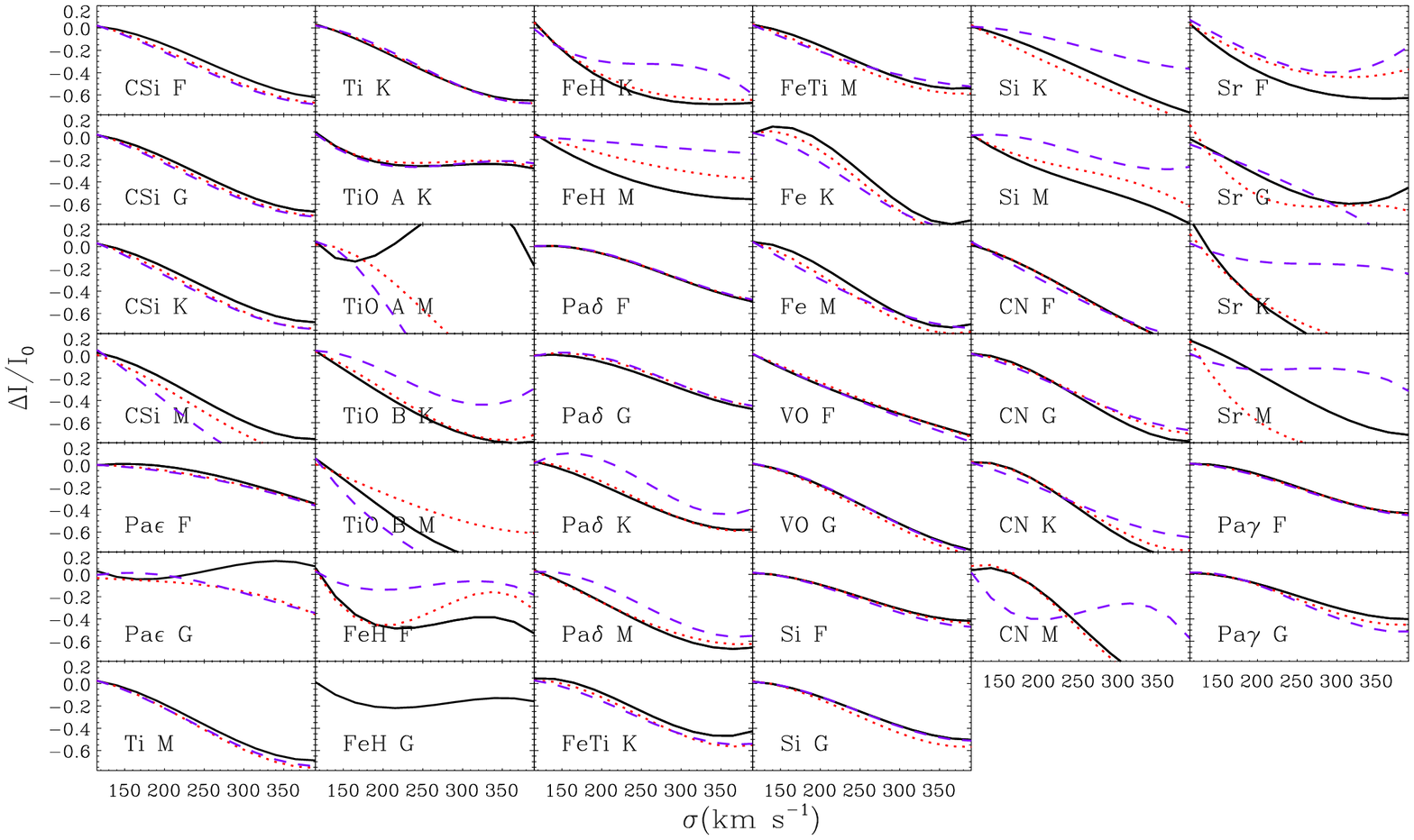}
\caption{Sensitivity of the indices in the Y atmospheric window 
to the velocity dispersion broadening. The relative variations of 
indices $\Delta(Index)/Index(\sigma_0)$ for different luminosity 
classes and spectral types are shown: supergiants with a black solid 
line, giants with a red dotted line, and dwarf stars with a violet 
dashed line.}\label{fig:Y-sigma}
\end{figure*}
Figure\,\ref{fig:Y-sigma} shows $\Delta(Index)/Index(\sigma_0)$ versus 
the broadened velocity dispersion for the indices in the Y 
atmospheric window. The faintest indices are omitted here and in 
the subsequent figures. The results for indices in the J, H, and L 
windows are shown in Figs.\,\ref{fig:J-sigma},\ref{fig:H-sigma}, and \ref{fig:L-sigma}, respectively.

To demonstrate the effect of broadening on the relations derived in
Figs.\,\ref{fig:Ind_Spt_Y}, \ref{fig:Ind_Spt_J}, \ref{fig:Ind_Spt_H},
and \ref{fig:Ind_Spt_L}, in Fig. 7 we show, as an example, the result of a
different broadening in for Ti and Si
versus the effective temperature (see Sect. \ref{sec:diagnostics}
for details). We choose Ti and Si since they have a small and large
scatter, respectively, in their correlation with the effective
temperature. As expected, as the velocity dispersion increases, the
broadening tends to reduce the trends and the equivalent widths
flatten for a large velocity dispersion ($\sigma \geq 350$ \kms),
especially for Ti.

\begin{figure*}
\includegraphics[angle=-90,width=\textwidth]{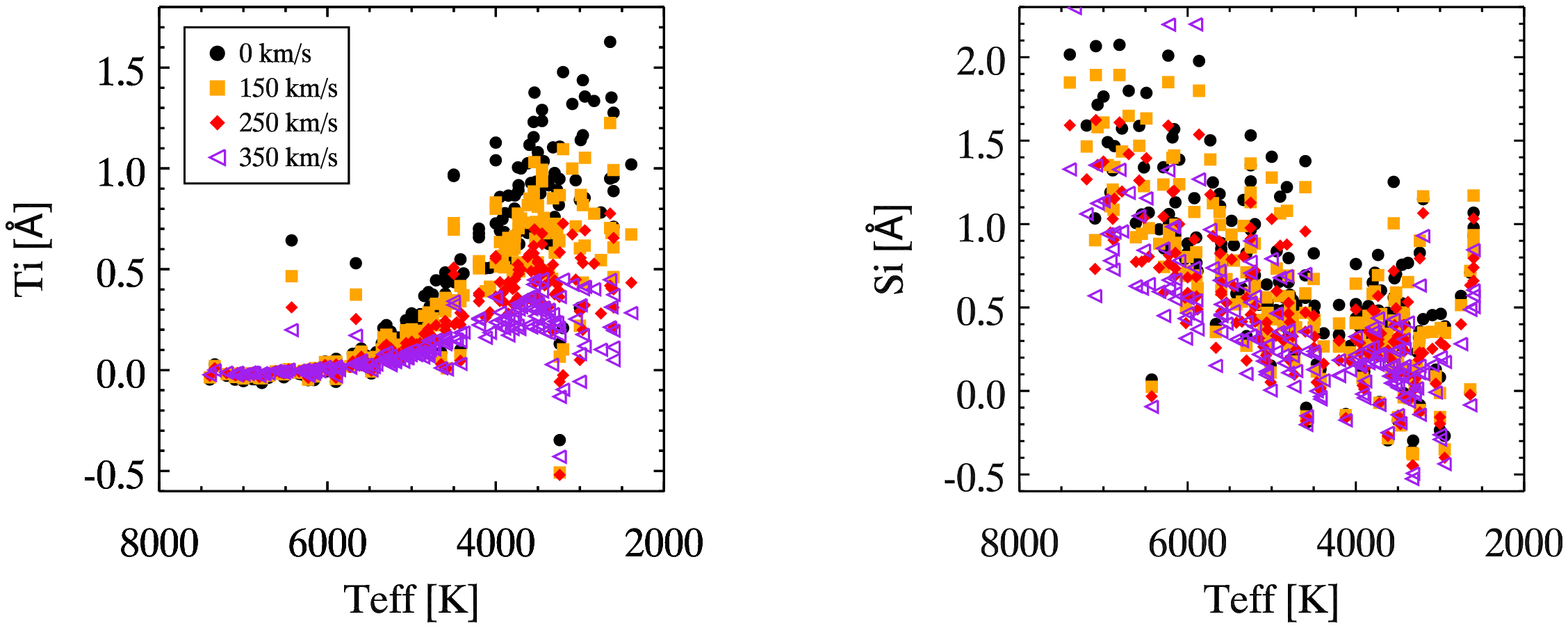}
\caption{ Effect of the velocity dispersion broadening on the $Ti - T_{eff}$ (left panel) 
and $Si - T_{eff}$ (right panel). With a black circle we show the original relation, 
and with a yellow square, red diamonds, and violet triangle the relation after a 
broadening of 150, 250, and 350 \kms, respectively.}
\label{fig:broad_disp_effect}
\end{figure*}

All indices and their correlation with the stellar properties are of
course affected by the broadening effect at a different level
depending on the spectral type, because lines may suffer from
cross-contamination by different species in stars with different
effective temperature.
Without exhaustive discussion, we point out that the most reliable and
stable indices over velocity dispersion ranging up to $\sim 200-250$
\kms are CSi, Ti, and Pa$\delta$ in the Y atmospheric window,  Na, FeCr, K1, Si,
K2 and Pa$\beta$ in the J atmospheric window, Mg2, Mg3, Si, COMg, and Br$_{10}$ in the H
atmospheric window, and Mg2 in the L atmospheric window.

\section{Spectral diagnostics}\label{sec:diagnostics}

The indices defined in Sect. \ref{sec:ind_mesure} have dependencies on
the stellar properties. In this section, we describe for
each atmospheric window the behavior of the most relevant features presents.

\subsection{Y atmospheric window}

Almost all indices in the Y atmospheric window show some dependence
on  \tef, although with a different level of significance 
and over different temperature ranges (Fig.\,\ref{fig:Ind_Spt_Y}).
For example, Pa$\epsilon$, Pa$\delta$, and Pa$\gamma$ indices are excellent 
temperature indicators for stars of types earlier than about G3. 
Other indices such as Ti, TiO\,A, TiO\,B, FeH, FeTi, Fe, and VO lines are 
complementary and work for the cooler stars. The Si, CN, 
and Sr indices are non-monotonic, which is also true to some extent for the 
Paschen indices. Finally, CSi has monotonic dependence over the 
almost entire range of interest, but this feature is relatively weak and shows 
some scatter.

The surface gravity is harder to infer from the Y window indices
(Fig.\,\ref{fig:Ind_Gr_Y}, upper panels). Some molecular features
related to TiO, FeH, VO, and CN do follow monotonic trends that are
not strong in comparison with the measurement errors. The trends
become more evident if the samples are constrained to a given
spectral type, for example, to G type stars for the VO index or to M stars for
the FeH index. A similar effect is seen with the metallicity. Indices such
as CSi and TiO\,B show trends only when a certain luminosity class (like supergiant stars)
is considered (Fig.\,\ref{fig:Ind_Gr_Y}, bottom panels).

\subsection{J atmospheric window}

Almost all the indices in the J window can serve as good $T_{\rm eff}$
indicators (Fig.\,\ref{fig:Ind_Spt_J}). However, in most cases the
trends in comparison with the scatter are relatively stronger than for
the indices in the Y window. Notably in some cases, such as the  K1A,
K1B, K2A, and K2B indices, the trends for stars of different luminosity
classes are identical. This allows us to estimate the stellar temperature
without information about the luminosity class. The dwarf stars cooler than
mid-M type show an extremely strong $T_{\rm eff}$ dependence for many
indices, including Na, K1A, C, K1B, Si, K2A, and K2B. However, this is due
to contamination by molecules such as methane. These indices reach values in the
dwarf stars that are not measured in the stars of other luminosity
classes, making it obvious that the stars are dwarf. There are no
indices that show degeneracy, with the exception of Mg and Al only for
dwarf stars.

Some J window indices can serve as luminosity indicators. They are
FeCr, K1A, K1B, and Si. Na, K2A, and Al are useful if only certain
spectral types are considered (Fig.\,\ref{fig:Ind_Gr_J}, upper
panels). The K1A index appears to be by far the best one, and more notably
it shows the same behavior over the entire range of spectral types.
Disappointingly, none of the J window indices seem to correlate with
[Fe/H], and Mg- and Fe-dominated indices need further investigation
(Fig.\,\ref{fig:Ind_Gr_J}, bottom panels).

\subsection{H atmospheric window}

The indices in the H atmospheric window in general show a more complex
behavior than those in the Y and J windows (Figs.\,\ref{fig:Ind_Tef_H}
and \ref{fig:Ind_Spt_H}). The dwarf stars show non-monotonic behavior
more often than not and there are no indices with identical trends for
supergiant, giant, and dwarf stars, except for some spectral classes,
for example, CO1, CO2, CO3, FeH1, and FeH2 for stars earlier than K0, and COMg for
stars earlier than K7. However, the different behavior of some indices
in stars of different luminosity classes can be used to our advantage
to separate the luminosity classes. For example, various CO-based
indices have very different values for dwarf stars, as long as they
are cooler than mid-K or early-M types. This diagnostic can be
brought to stars as hot as early-K for FeH2 and CO5 indices, and as
early as G with the help of the Mg1 and Mg2 indices provided
there are no metallicity effects. This behavior is clearly visible
in Fig.\,\ref{fig:Ind_Gr_H}.

Br$_{10}$, Si, and CO4 indices are shown to be very promising \logg indicators
in the H window (Fig.\,\ref{fig:Ind_Gr_H}). None of the investigated
indices here show significant metallicity trend
(Fig.\,\ref{fig:Ind_FeH_H}). Unfortunately, the Mg feature at $\lambda
\simeq 1.58\, \mu$m is severely contaminated with other lines and it
seems to be rather temperature sensitive at $\lambda \simeq 1.71\,
\mu$m, \citep{2011arXiv1108.1499L}. The passband of Si at $\lambda
\simeq 1.596\, \mu$m includes the strong line of Fe $\lambda \simeq
1.599\, \mu$m, which is also increasing with decreasing effective
temperature.  These effect requires further investigation but this
will be possible only after widening the metallicity range of the
library stars.

\subsection{L atmospheric window}

The L window is relatively devoid of spectral features in comparison
with the other windows. We see the same variety of behavior with
respect to the temperature as in the other windows 
(Fig.\,\ref{fig:Ind_Spt_L}).
The Pf$\gamma$ index appears to be the best temperature indicator 
for stars earlier than G2 and various indices encompassing the OH 
band for the cooler stars. The same OH-based indices together 
with the SiO-based indices can help to separate dwarfs from giants 
and supergiant stars for spectral types later than K3. 
However, the intrinsic scatter of the relations for L window 
indices is larger than for the indices in the other atmospheric 
windows -- not surprising given the stronger thermal background 
at $\lambda > $2.5\,$\mu$m.

The index with the most consistent behaviour with respect to
  the surface gravity across all the spectral types is
  Hu$_{15}$ (Fig.\,\ref{fig:Ind_Gr_L}, upper panels), although to be
  reliable diagnostics the S/N of the data is critical.
Like the previous cases, we found no significant correlations
with [Fe/H] (Fig.\,\ref{fig:Ind_Gr_L}, bottom panels).

\begin{figure*}
\includegraphics[width=9truecm,height=13truecm,angle=0]{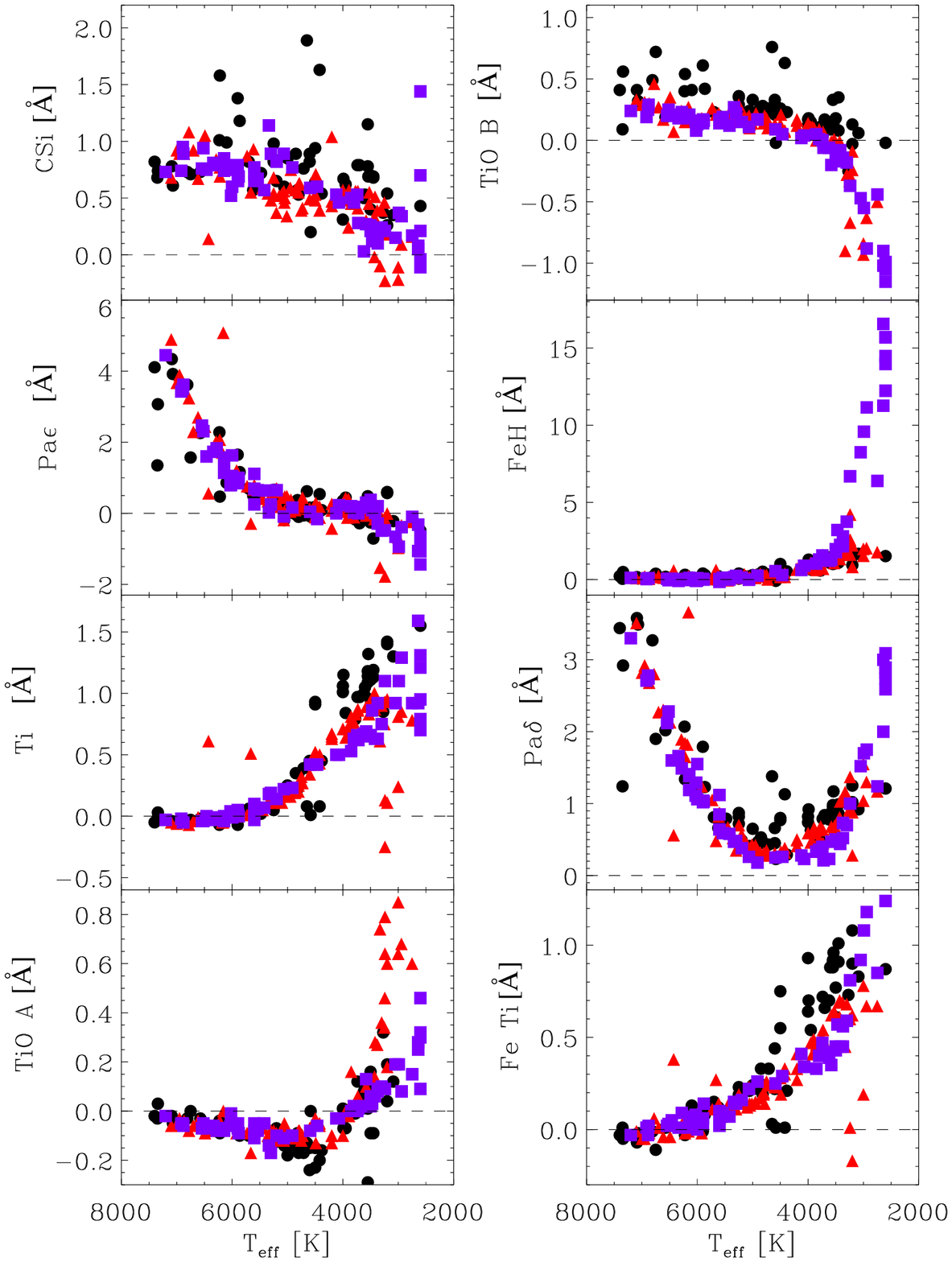}
\includegraphics[width=9truecm,height=13truecm,angle=0]{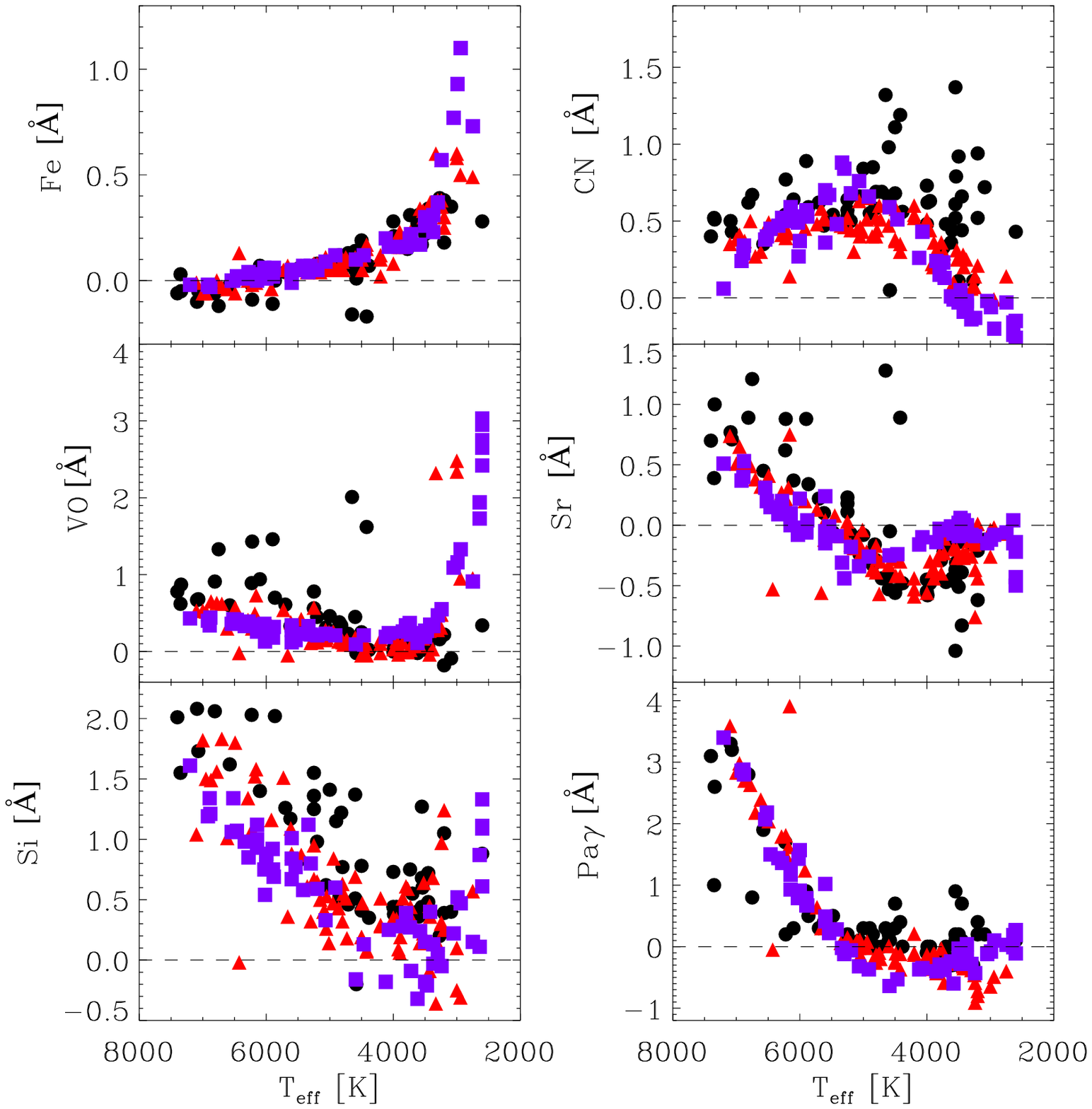}
\caption{Index measurements in the Y atmospheric window as
  functions of effective temperature. The different
  symbols correspond to the supergiants (circles), giants (triangles),
  and dwarf stars (squares), respectively. }
\label{fig:Ind_Spt_Y}
\end{figure*}

\begin{figure*}
\includegraphics[width=9truecm,height=11truecm,angle=0]{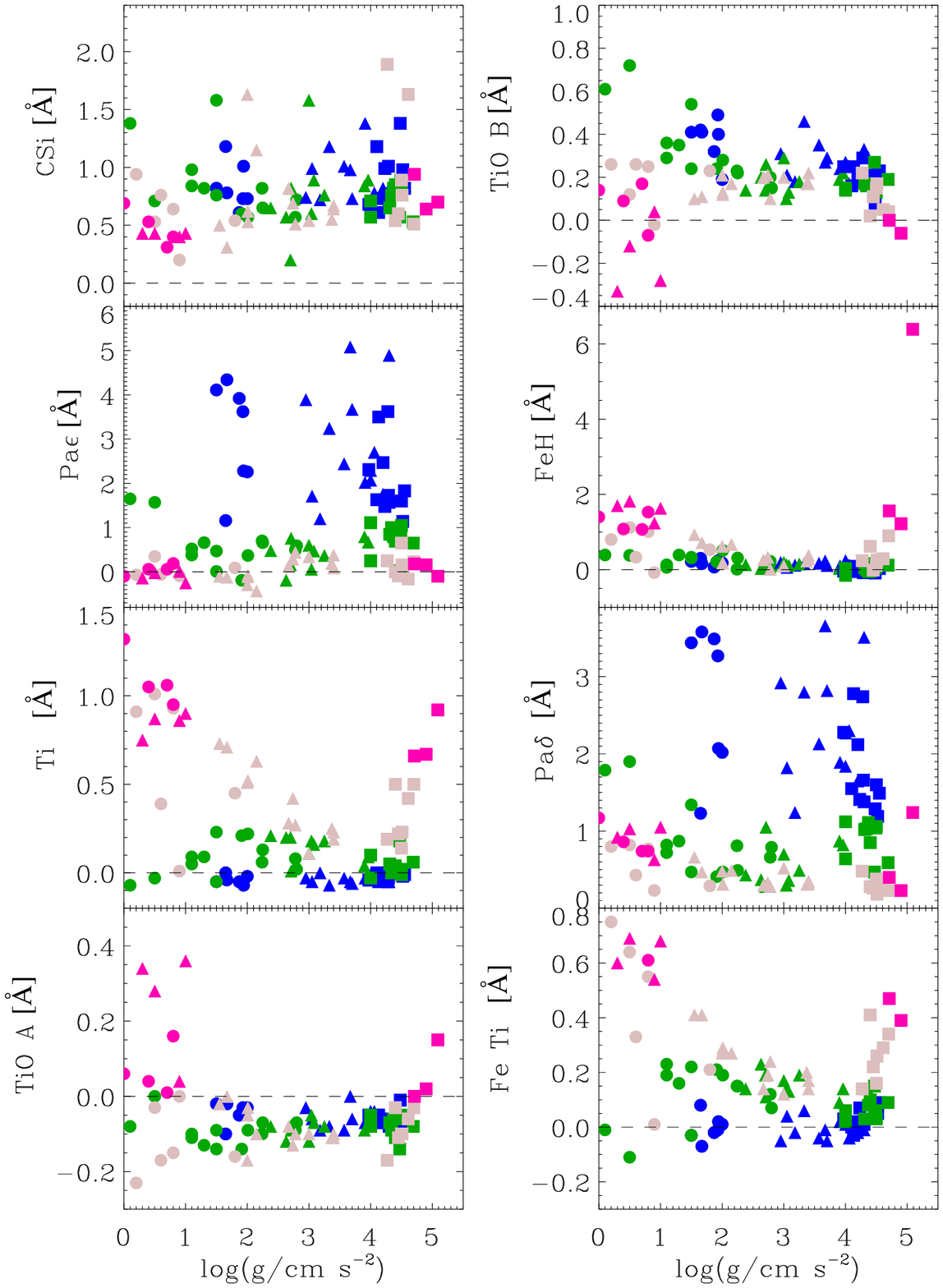}
\includegraphics[width=9truecm,height=11truecm,angle=0]{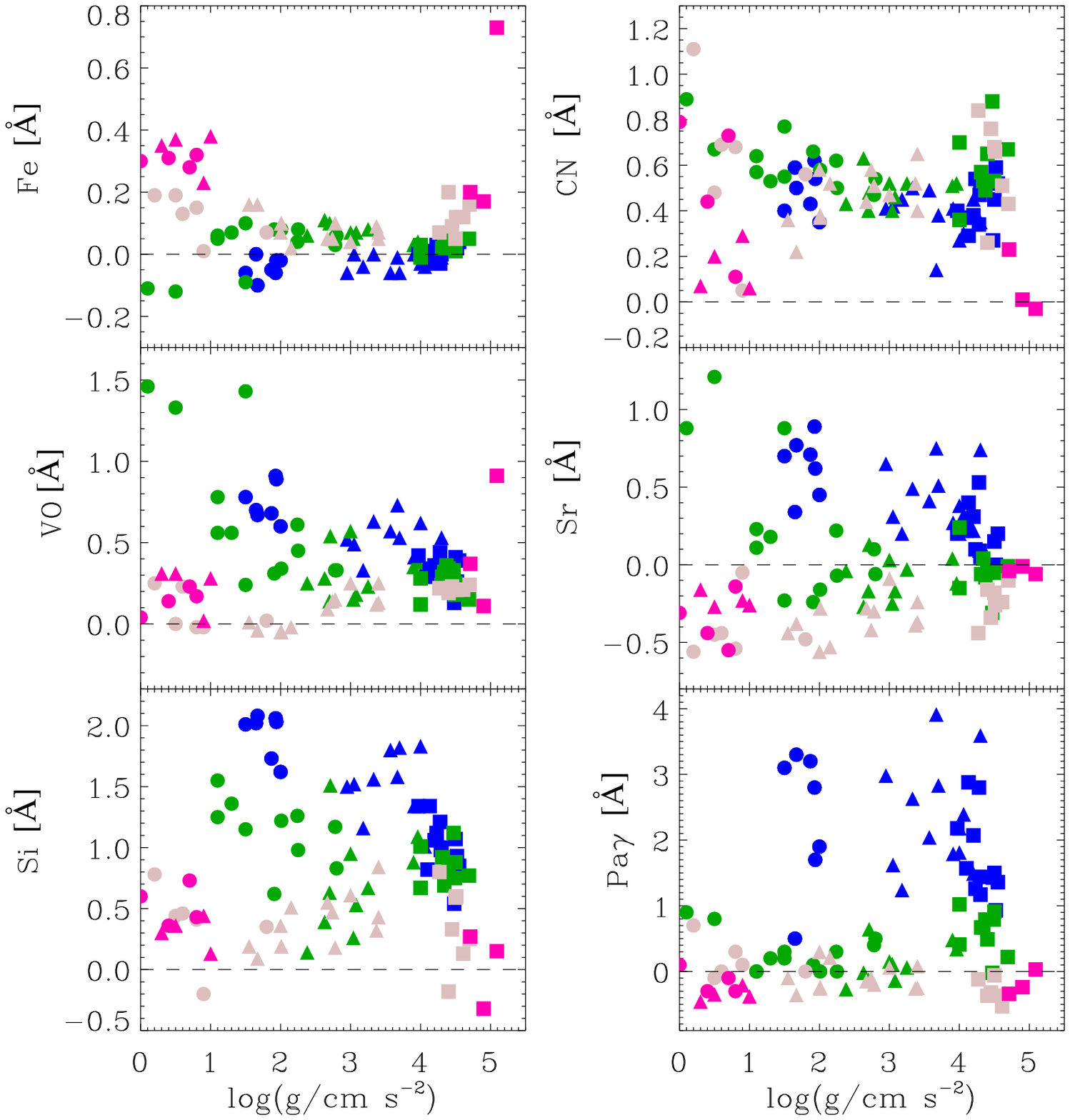}\\
\includegraphics[width=9truecm,height=11truecm,angle=0]{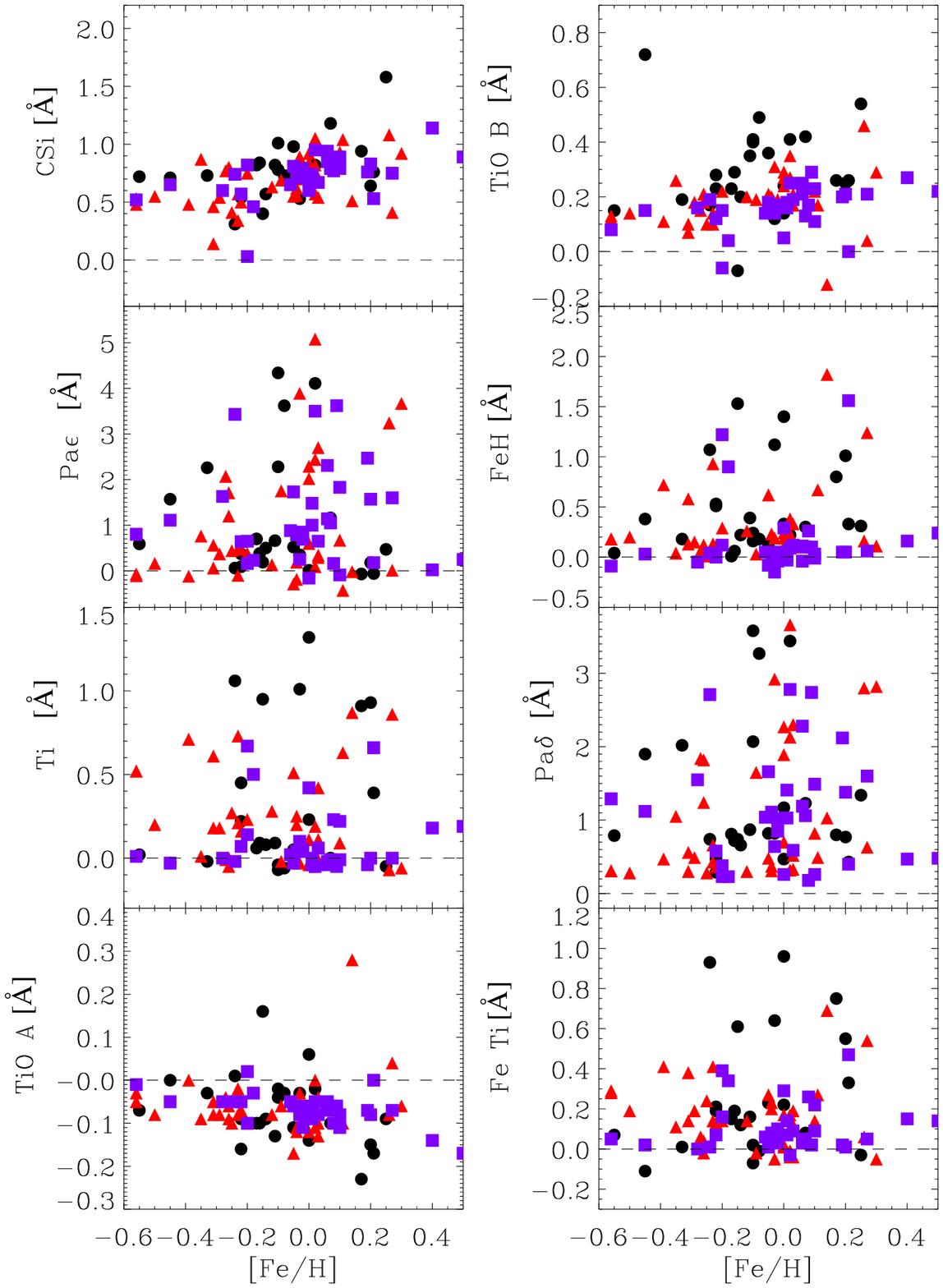}
\includegraphics[width=9truecm,height=11truecm,angle=0]{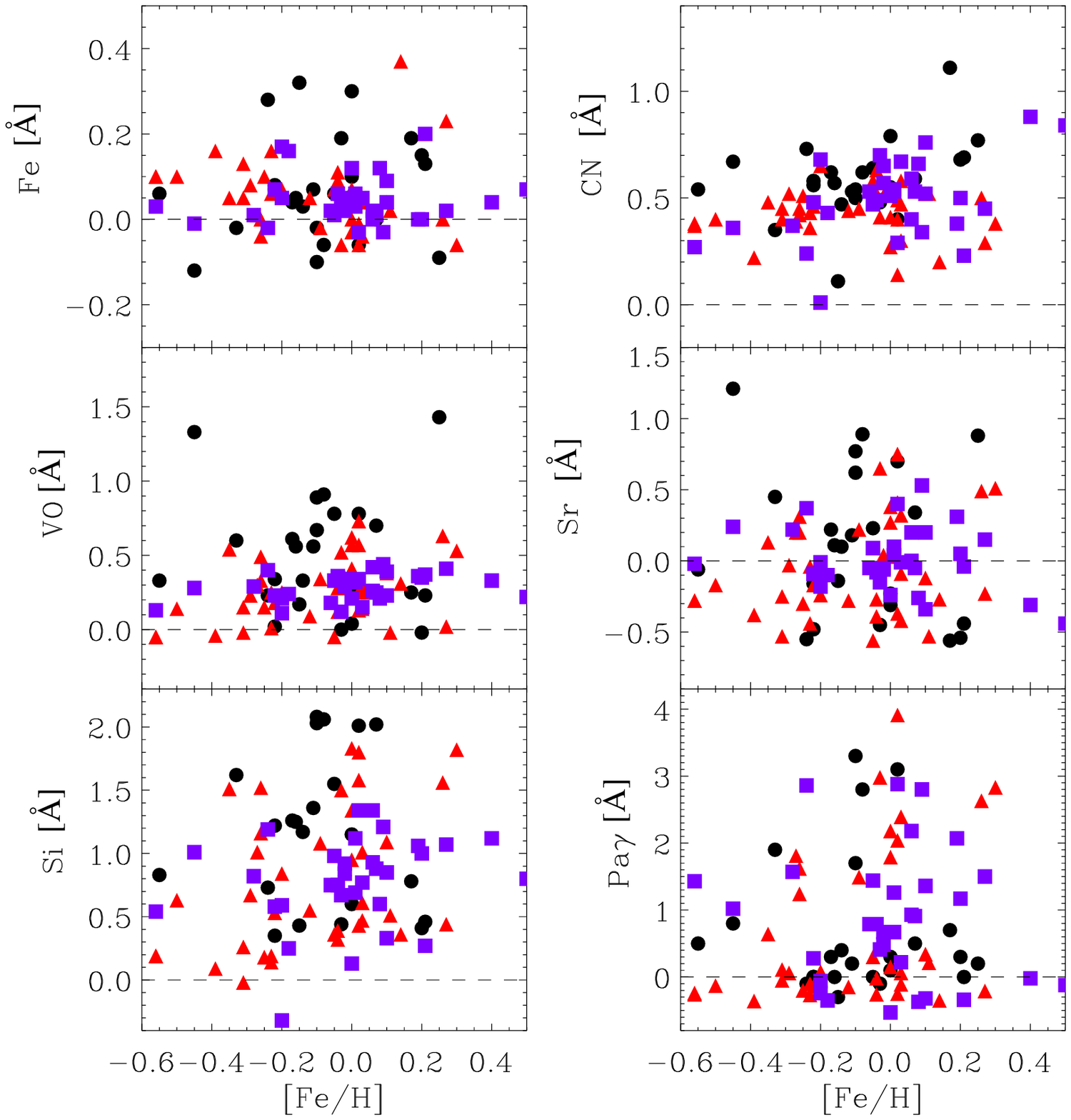}
\caption{Index measurements in the Y atmospheric window as 
functions of surface gravity (top panels) and metallicity (bottom
panels). Symbols for supergiants, giants, and dwarf stars
are as in Fig.\,\ref{fig:Ind_Spt_Y}. Different colors in the top panels
correspond to F (blue), G (green), K (gray), and M type stars (pink),
respectively.}
\label{fig:Ind_Gr_Y}
\end{figure*}

\section{Discussion and summary}\label{conclusions}

The analysis presented here is based on stellar spectra. We relate the
strengths of various spectral features with the physical properties of
the stars -- effective temperature \tef, surface gravity \logg, and
metallicity [Fe/H]. However, our main goal is to relate the strengths
of these spectral features to the physical parameters of stellar
populations in unresolved galaxies, and we adopt a simple and
straightforward mapping of one to the other. The temperature \tef is
related to the population age, because the luminosity-weighted
temperature decreases as the population passively evolves from being
dominated by younger main-sequence stars towards older red giant
stars. The surface gravity \logg tells us what is the dominant
luminosity class of a population. Finally, the metallicity [Fe/H]
translates into the cumulative abundance of the dominant population in
a galaxy. We will investigate these diagnostics in a forthcoming
  paper (Gasparri et al. 2020, submitted) by measuring them in the
  sample of galaxies observed with X-Shooter
  \citep{2019A&A...621A..60F}. Furthermore, we will also compare the
  observational behavior of our defined indices with the predictions
  from the theoretical models
  \citep[\eg][]{2016A&A...589A..73R,2018ApJ...854..139C}. 

We point out that the diagnostics described here can also be used for
spectral typing of individual stars. Last but not least, we stress
that the behavior of the indices is not necessarily the same as the
behavior of the particular species that dominate these indices,
because of the possible contamination by other species as noted also
by \citet{2012ApJ...760...71C} in their models analyzing the initial
mass function indicators. For example, some hydrogen and metal lines
are contaminated by molecular bands in the late-type giant and
supergiant stars.

To summarize, we present here an analysis of stellar spectra in the Y,
J, H, and L atmospheric windows, based on high-quality medium
resolution ($R\sim2000-2500$) spectra of more than 200 stars with
almost solar metallicity ($-0.5 \leq$ [Fe/H]$ \leq 0.5$ dex) and a
spectral S/N of about 100 \citep{2009ApJS..185..289R}.
We identify the best indicators for the different stellar physical
 parameters, and as discussed above, of some integrated parameters of
 the stellar populations in unresolved galaxies. The results discussed
 in Sect. \ref{sec:diagnostics} have been summarized in Table
 \ref{tab:results}.

%
\begin{table*}
\caption{Summary of the best indicators for the different stellar
  physical parameters. For the details, please refer to Sect.
  \ref{sec:diagnostics}. The columns show the following: (1) stellar
  physical parameters (2) spectral feature (3) quality of the
  index: strong (S) or weak (W) (4): notes on the index. }
\begin{center}
\begin{small}
\begin{tabular}{cccc}
\hline
\noalign{\smallskip}
\multicolumn{1}{c}{Stellar parameters} &
\multicolumn{1}{c}{Index} &
\multicolumn{1}{c}{Quality} &
\multicolumn{1}{c}{Notes}  \\ 
\multicolumn{1}{c}{(1)} &
\multicolumn{1}{c}{(2)} &
\multicolumn{1}{c}{(3)} &
\multicolumn{1}{c}{(4)}  \\ 
\hline
\hline
\noalign{\smallskip}
\noalign{\smallskip}
\multicolumn{4}{c}{ {\bf Y atmospheric windows}}  \\
\noalign{\smallskip}
\multicolumn{1}{c}{\multirow{3}{*}{ $T_{\rm eff}$}}&
\multicolumn{1}{c}{Pa$\epsilon$,Pa$\delta$,Pa$\gamma$} &
\multicolumn{1}{c}{S} &
\multicolumn{1}{c}{ F, G stars}  \\ 
\multicolumn{1}{c}{}&
\multicolumn{1}{c}{Ti,TiO\,A,TiO\,B,FeH,FeTi,Fe, VO} &
\multicolumn{1}{c}{S} &
\multicolumn{1}{c}{ G, K, M stars}  \\ 
\multicolumn{1}{c}{}&
\multicolumn{1}{c}{CSi} &
\multicolumn{1}{c}{W} &
\multicolumn{1}{c}{Stars of all spectral type}  \\ 
\noalign{\smallskip}
\multicolumn{1}{c}{\logg}&
\multicolumn{1}{c}{TiO,FeH,VO,CN} &
\multicolumn{1}{c}{S/W} &
\multicolumn{1}{c}{Very dependent on star spectral type}  \\ 
\noalign{\smallskip}
\multicolumn{1}{c}{[Fe/H]}&
\multicolumn{1}{c}{TiO\,B} &
\multicolumn{1}{c}{S} &
\multicolumn{1}{c}{Only in supergiant stars}  \\ 
\noalign{\smallskip}
\multicolumn{4}{c}{ {\bf J atmospheric windows}}  \\
\noalign{\smallskip}
\multicolumn{1}{c}{\multirow{2}{*}{ $T_{\rm eff}$}}&
\multicolumn{1}{c}{K1\,B,K2\,A,K2\,B} &
\multicolumn{1}{c}{S} &
\multicolumn{1}{c}{Similar trend in stars of all spectral types }  \\ 
\multicolumn{1}{c}{}&
\multicolumn{1}{c}{K1\,A,Na,Si} &
\multicolumn{1}{c}{S} &
\multicolumn{1}{c}{M stars, methane contamination}  \\ 
\noalign{\smallskip}
\multicolumn{1}{c}{\multirow{2}{*}{\logg}}&
\multicolumn{1}{c}{FeCr,K1\,B,Na,Si,Al} &
\multicolumn{1}{c}{S} &
\multicolumn{1}{c}{Dependent on star spectral type}  \\ 
\multicolumn{1}{c}{}&
\multicolumn{1}{c}{K1\,A} &
\multicolumn{1}{c}{S} &
\multicolumn{1}{c}{All spectral type stars }  \\ 
\noalign{\smallskip}
\multicolumn{4}{c}{ {\bf H atmospheric windows}}  \\
\noalign{\smallskip}
\multicolumn{1}{c}{$T_{\rm eff}$}&
\multicolumn{1}{c}{CO1,CO2,CO3,FeH1,FeH2,COMg} &
\multicolumn{1}{c}{S} &
\multicolumn{1}{c}{Mostly in  F,G stars}  \\ 
\noalign{\smallskip}
\multicolumn{1}{c}{\logg}&
\multicolumn{1}{c}{Br$_{10}$,Si,CO4} &
\multicolumn{1}{c}{S} &
\multicolumn{1}{c}{Stars of all spectral types}  \\ 
\noalign{\smallskip}
\multicolumn{4}{c}{ {\bf L atmospheric windows}}  \\
\noalign{\smallskip}
\multicolumn{1}{c}{$T_{\rm eff}$}&
\multicolumn{1}{c}{Pf$\gamma$} &
\multicolumn{1}{c}{W} &
\multicolumn{1}{c}{In stars earlier than G2}  \\ 
\noalign{\smallskip}
\multicolumn{1}{c}{\logg}&
\multicolumn{1}{c}{Mg3,Hu$_{15}$} &
\multicolumn{1}{c}{W} &
\multicolumn{1}{c}{Stars of all spectral type }  \\ 
\noalign{\smallskip}
\hline
\noalign{\medskip}
\end{tabular}
\end{small}
\label{tab:results}
\end{center}
\end{table*}

Features sensitive to the effective temperature are present and
measurable in all the investigated atmospheric windows. The surface
gravity is more challenging: the H window offers better tools for a
separation between dwarf and giant stars, followed by the J
window. This is almost impossible in the Y and L windows. The diagnostics
in the L window are the most challenging because the spectra are
typically noisier than in other windows due to the higher thermal
background emission. However, the next generation of both extremely
large ground-based telescopes and space-based facilities will work in
this regime, making even these L window diagnostic tools quite
valuable.

\begin{acknowledgements}   
We thank the anonymous referee for the useful discussion and
suggestions. We have made extensive use of the SIMBAD Database at
CDS (Center de Donn\'ees astronomiques) Strasbourg, the NASA/IPAC
Extragalactic Database (NED), which is operated by the Jet Propulsion
Laboratory, CalTech, under contract with NASA, and of the VizieR
catalog access tool, CDS, Strasbourg, France. EMC, AP, EDB, and LM are
supported by Padua University through grants DOR1715817/17,
DOR1885254/18, DOR1935272/19, and BIRD164402/16. EMC, AP, and EDB are
also supported by MIUR grant PRIN 2017 20173ML3WW\_001.

\end{acknowledgements}         

\bibliographystyle{aa}
\bibliography{Morelli_37505}

\clearpage
\begin{appendix}
\label{app:app}

\section{Main spectral features plots}\label{app:plots}

\begin{figure*}
\includegraphics[width=16truecm,angle=0]{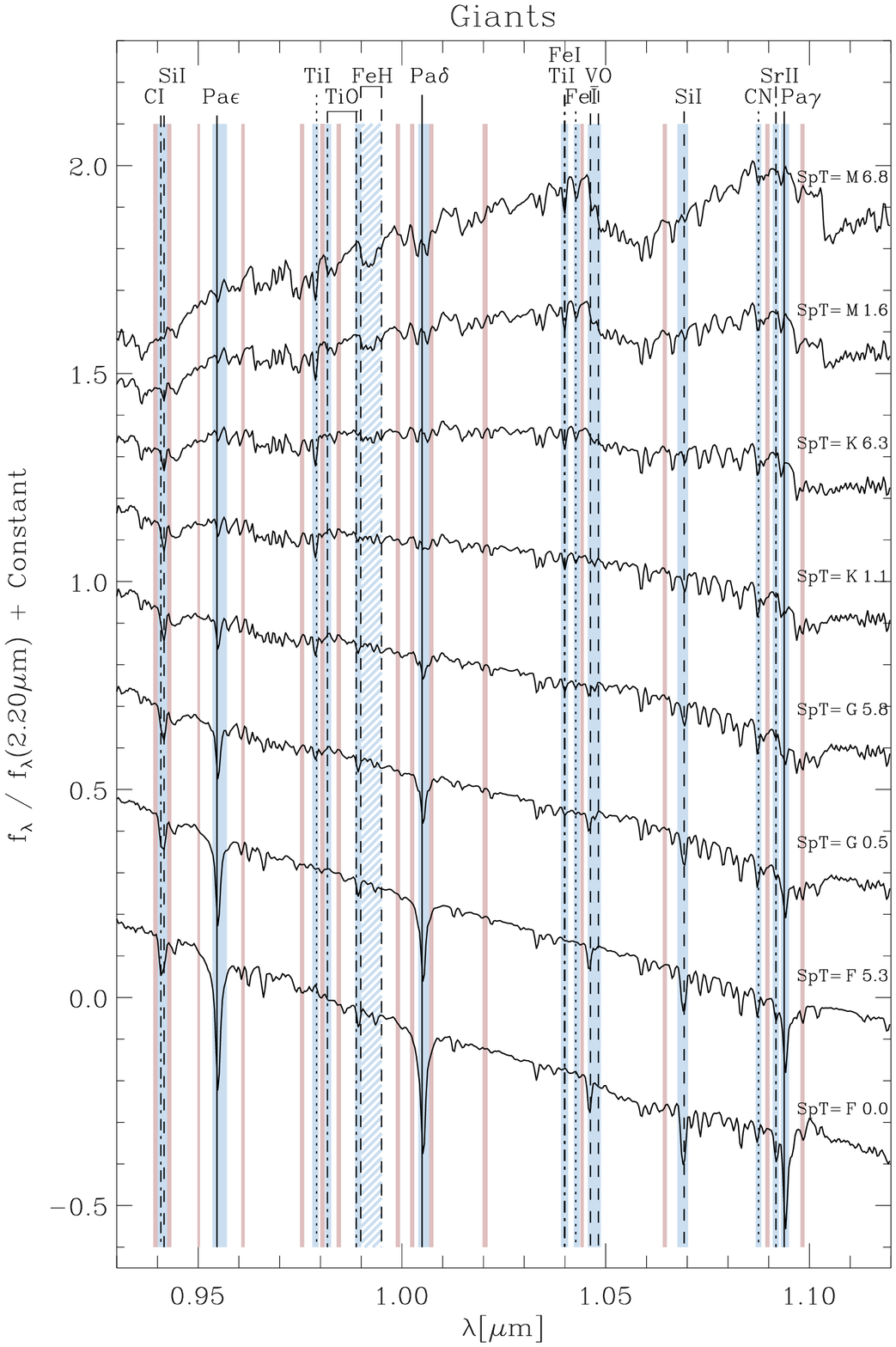}
\caption{Y atmospheric window model spectrum of giants. Figure\,\ref{fig:SGiant_fitted_Y} gives details.}\label{fig:Giant_fitted_Y}
\end{figure*}

\begin{figure*}
\includegraphics[width=16truecm,angle=0]{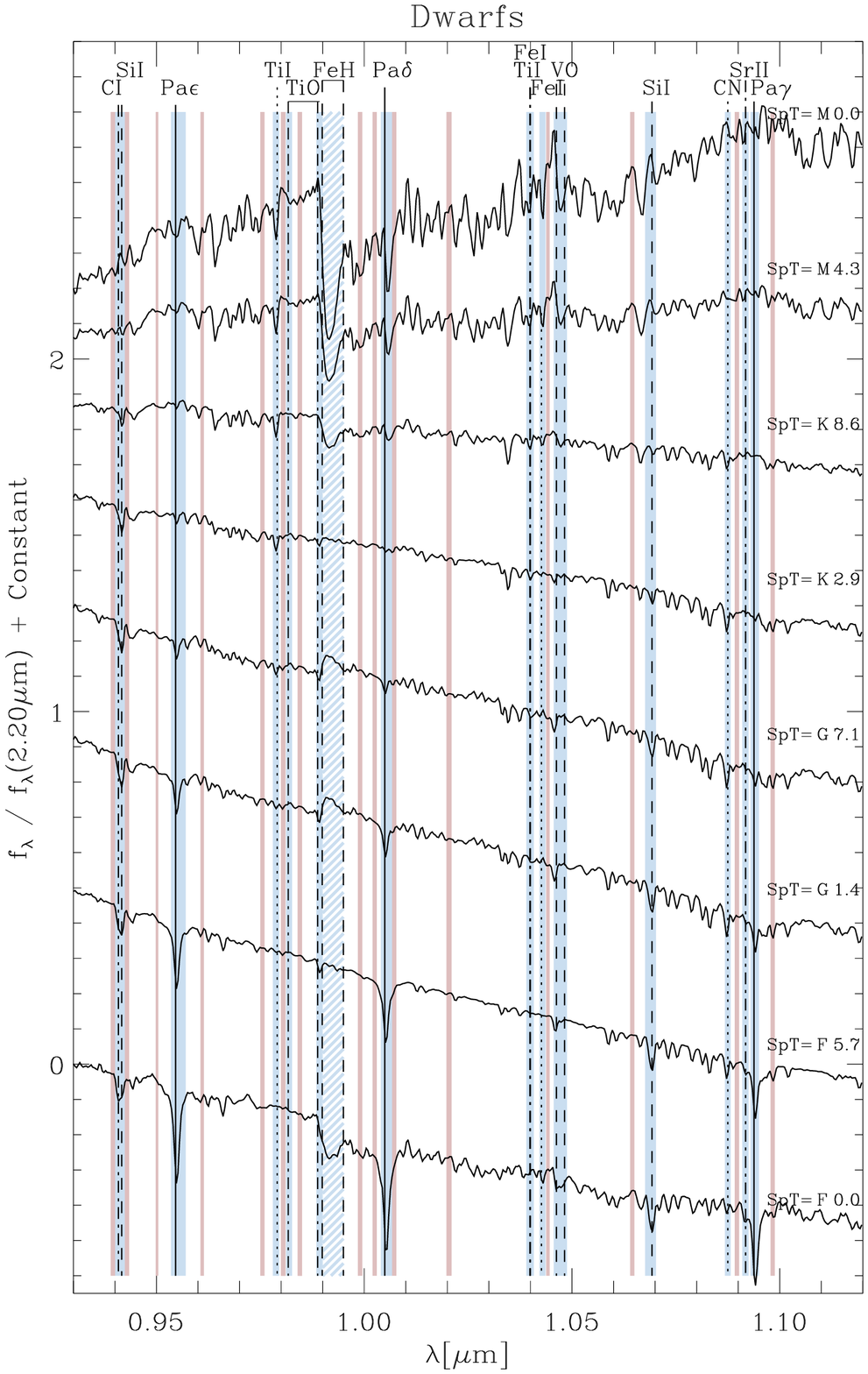}
\caption{Y atmospheric window model spectrum of dwarfs.  
Figure\,\ref{fig:SGiant_fitted_Y} gives details.}\label{fig:Dwarf_fitted_Y}
\end{figure*}

\begin{figure*}
\includegraphics[width=16truecm,angle=0]{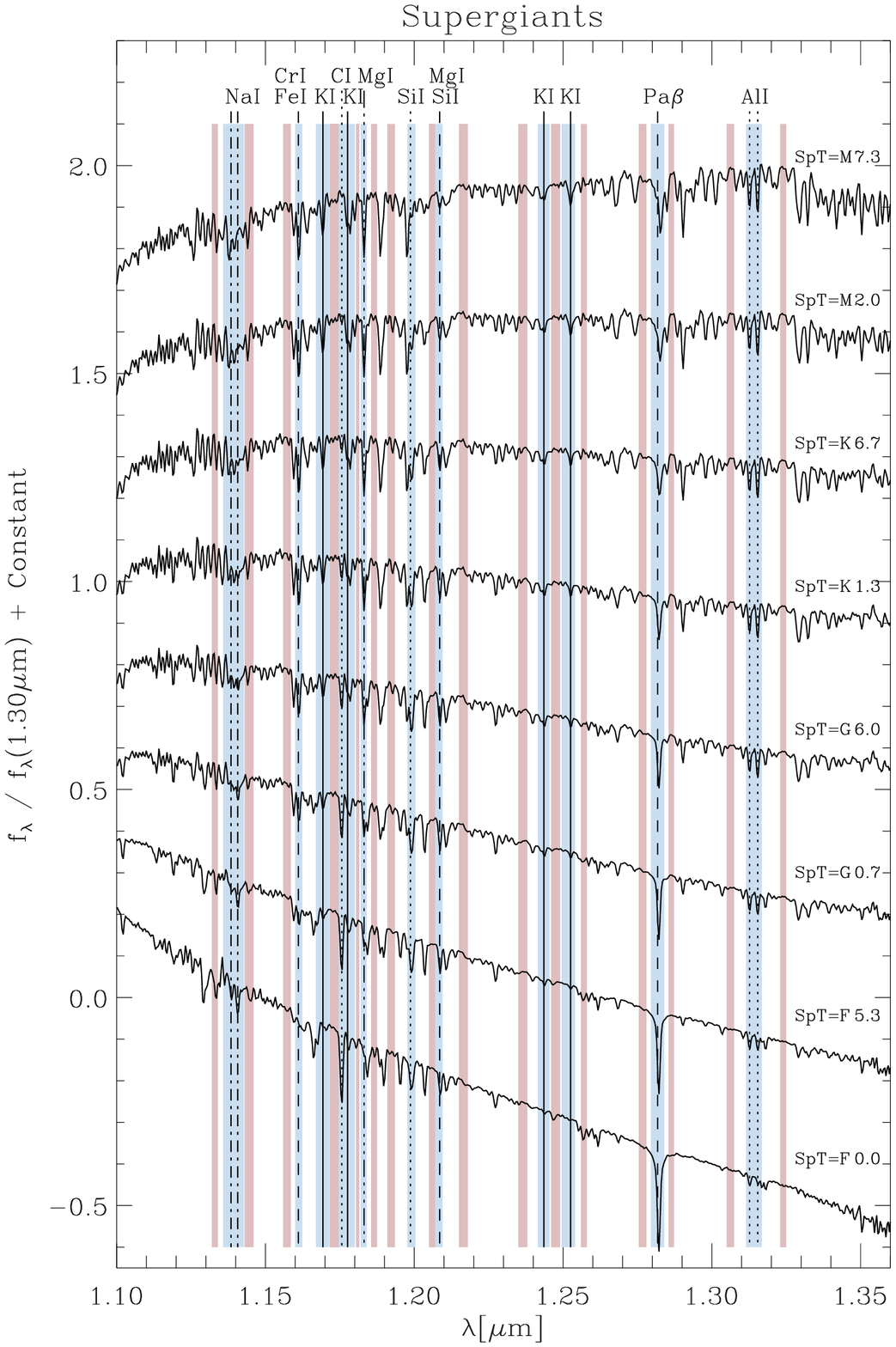}
\caption{$J$ atmospheric window  model spectrum of supergiants. The 
flux-normalization here is performed at 1.30$\,\mu$m.  
Figure\,\ref{fig:SGiant_fitted_Y} gives details.}\label{fig:SGiant_fitted_J}
\end{figure*}

\begin{figure*}
\includegraphics[width=16truecm,angle=0]{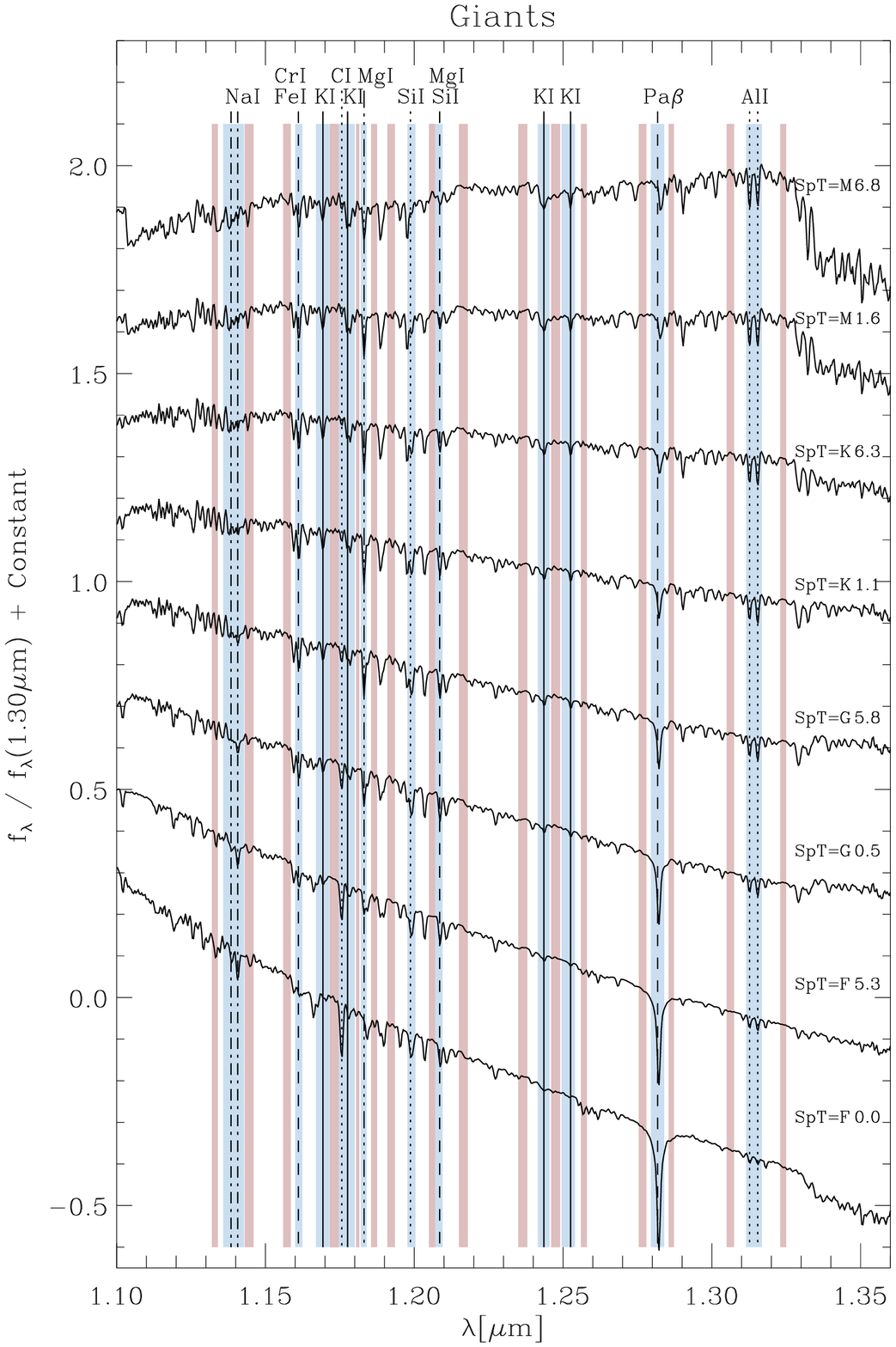}
\caption{$J$ atmospheric window  model spectrum of giants. 
Figure\,\ref{fig:SGiant_fitted_J} gives details.}\label{fig:Giant_fitted_J}
\end{figure*}

\begin{figure*}
\includegraphics[width=16truecm,angle=0]{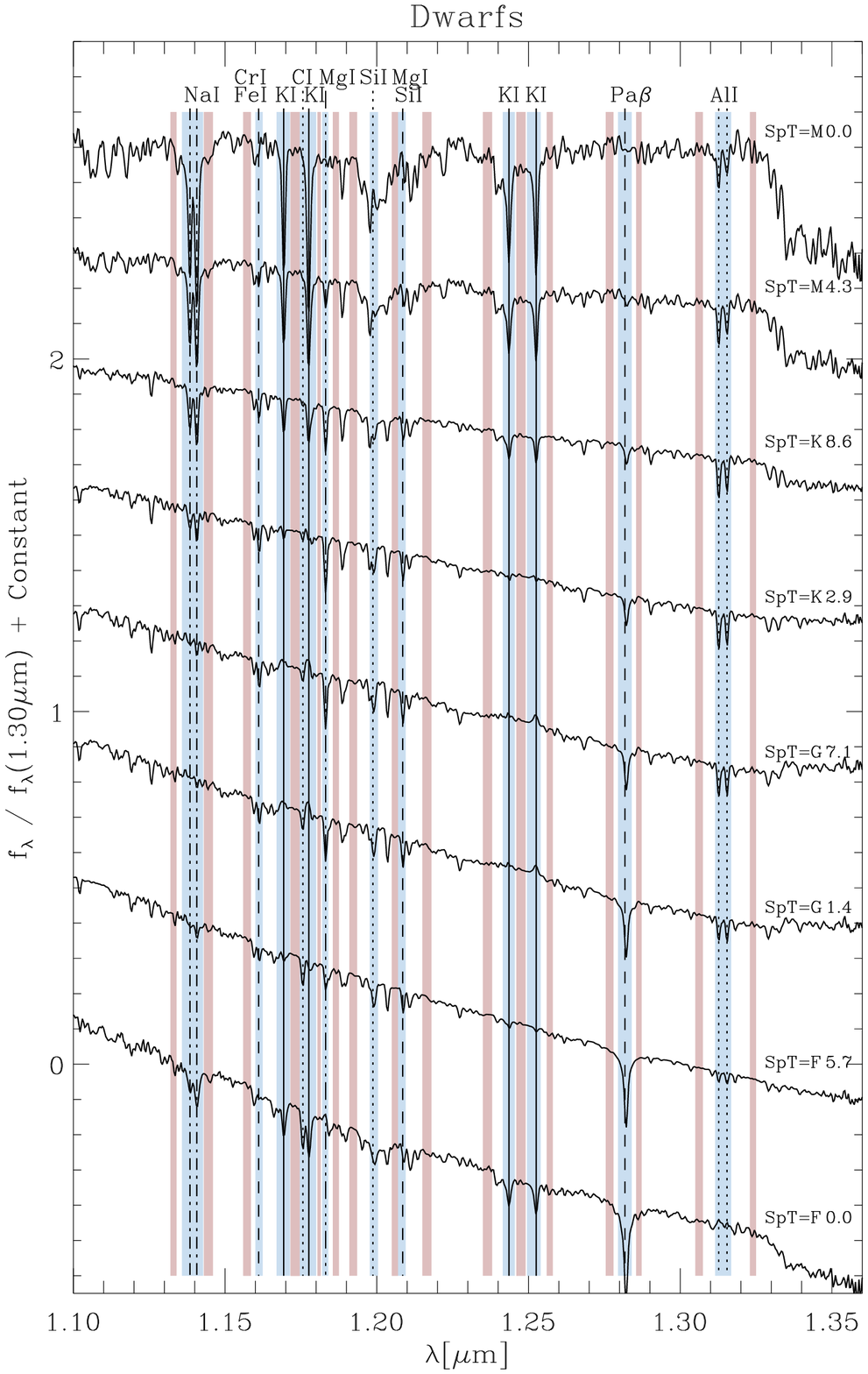}
\caption{$J$ atmospheric window model spectrum of dwarfs. 
Figure\,\ref{fig:SGiant_fitted_J} gives details.}\label{fig:Dwarf_fitted_J}
\end{figure*}

\begin{figure*}
\includegraphics[width=16truecm,angle=0]{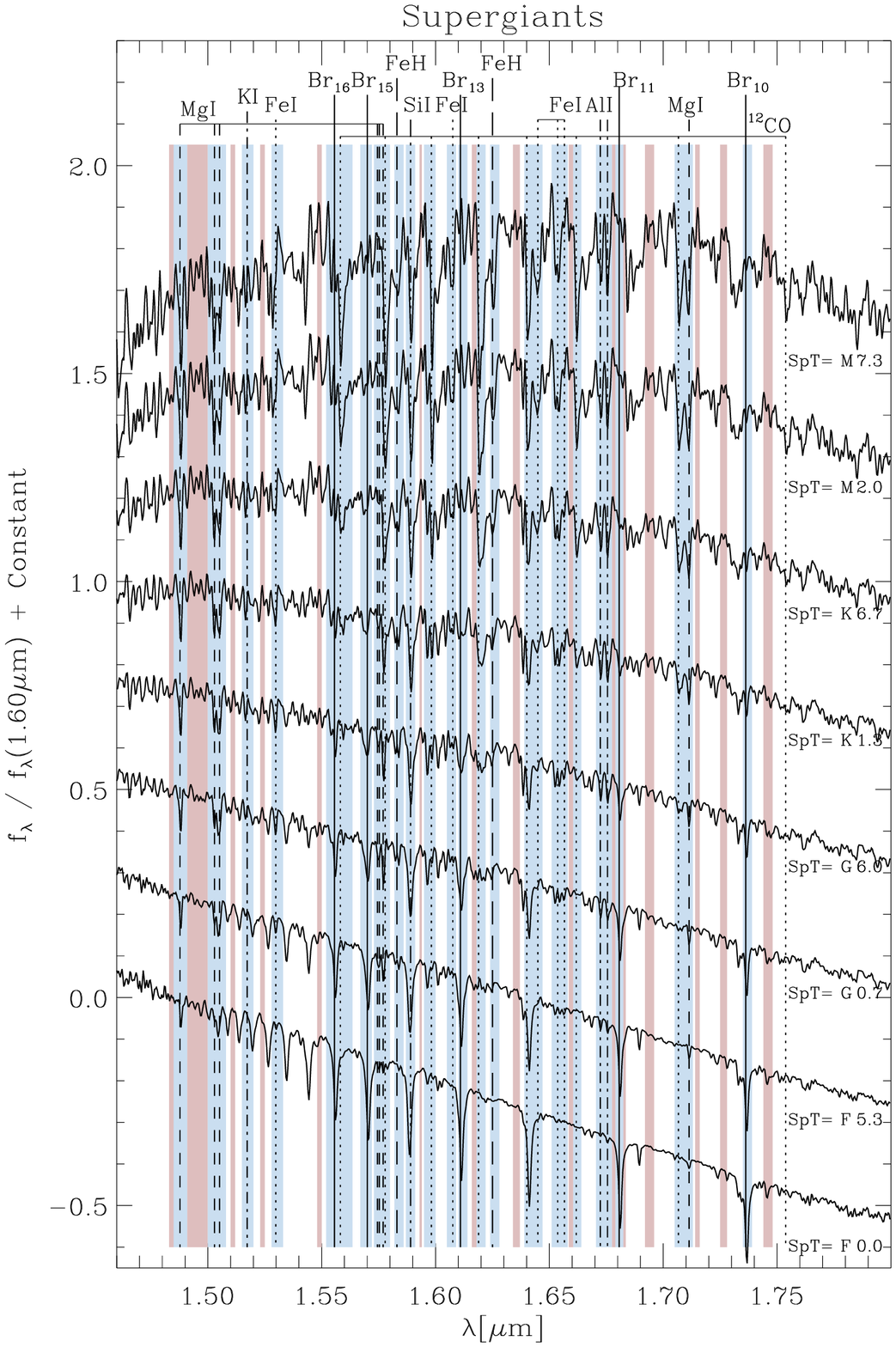}
\caption{H window  model spectrum of supergiants. The 
flux-normalization here is performed at 1.60$\,\mu$m.  
Figure\,\ref{fig:SGiant_fitted_Y} gives details.}\label{fig:SGiant_fitted_H}
\end{figure*}

\begin{figure*}
\includegraphics[width=16truecm,angle=0]{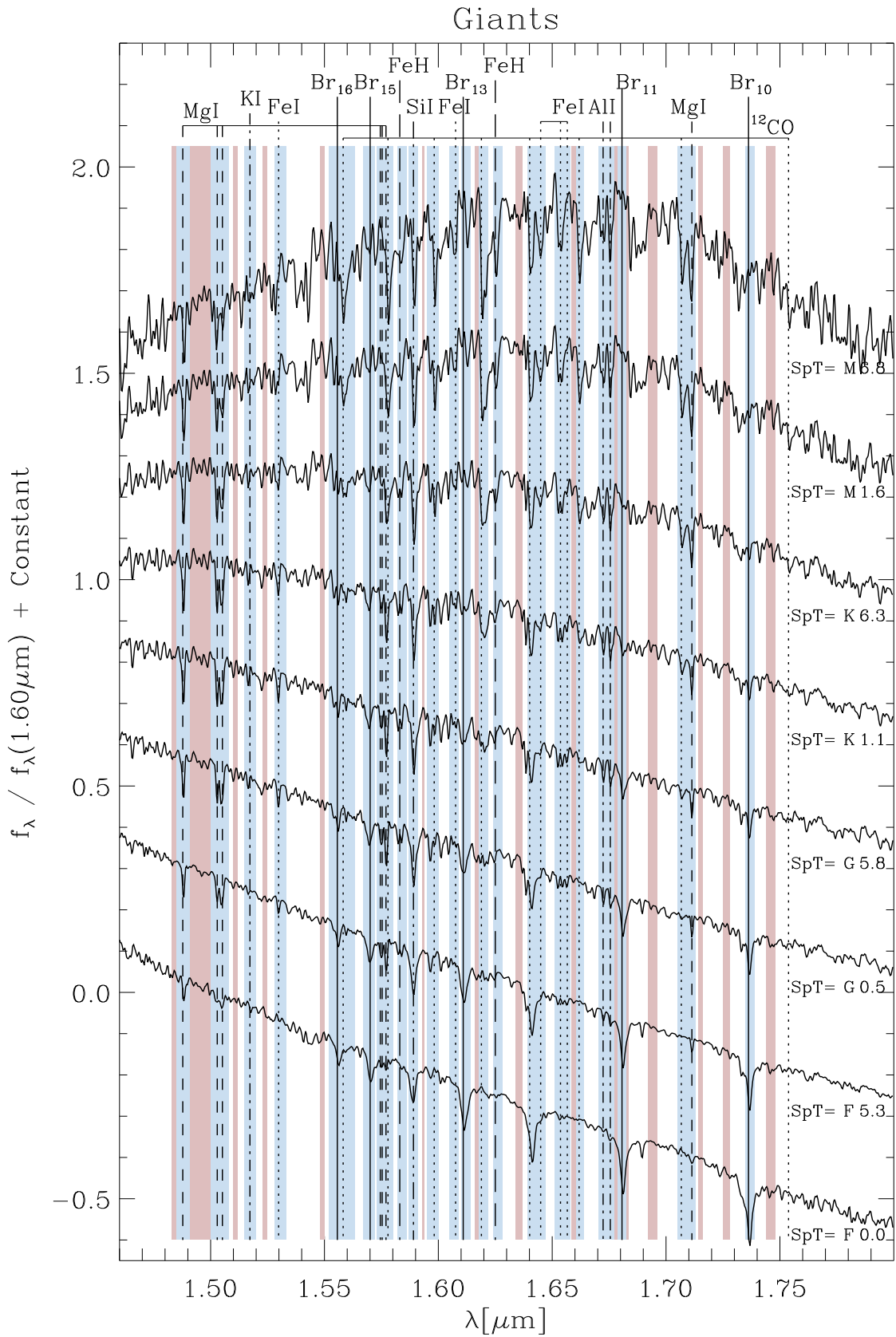}
\caption{H window  model spectrum of giants.  
Figure\,\ref{fig:SGiant_fitted_H} gives details.}\label{fig:Giant_fitted_H}
\end{figure*}

\begin{figure*}
\includegraphics[width=16truecm,angle=0]{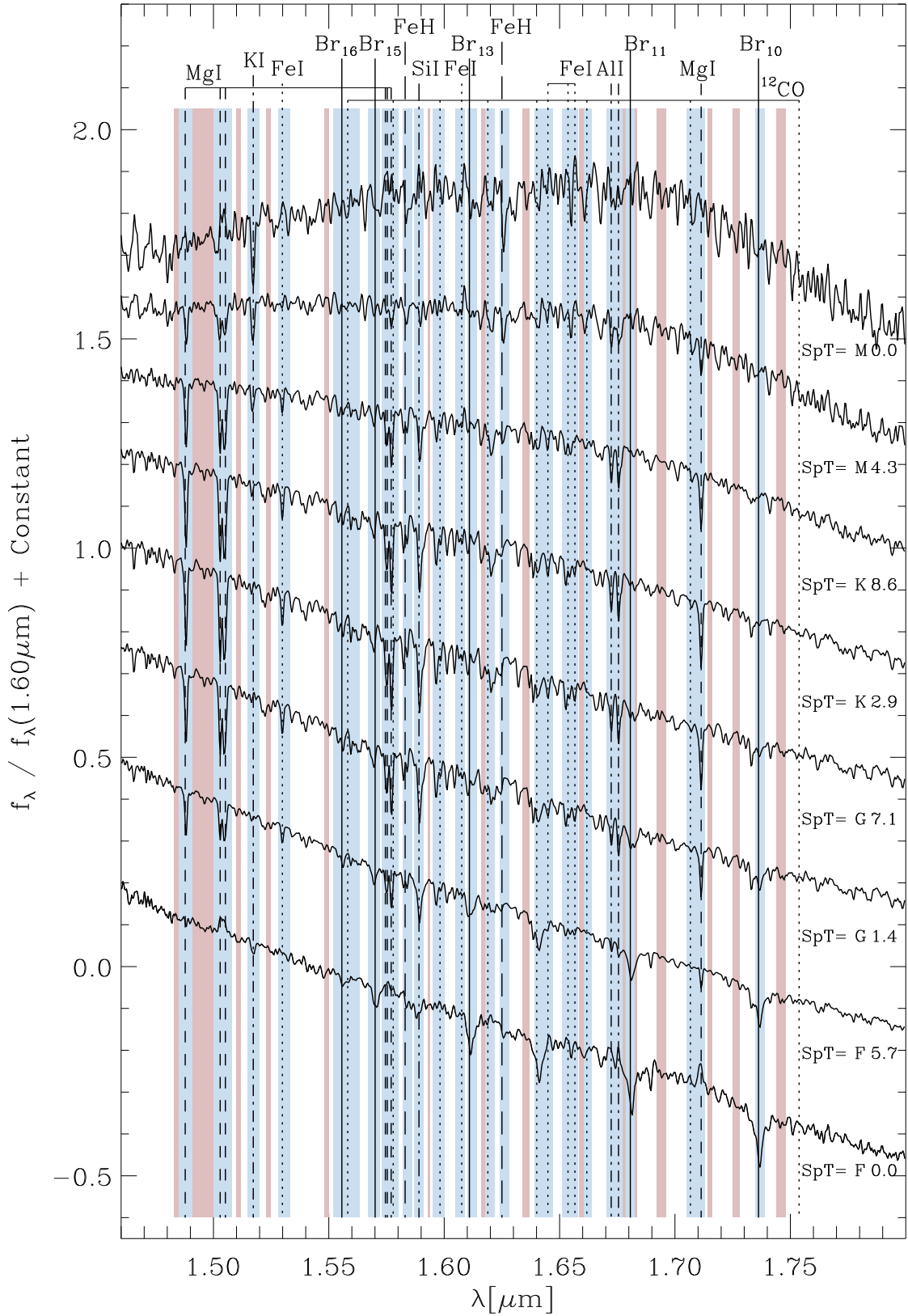}
\caption{H window model spectrum of dwarfs.  
Figure\,\ref{fig:SGiant_fitted_H} gives details.}\label{fig:Dwarf_fitted_H}
\end{figure*}

\begin{figure*}
\includegraphics[width=16truecm,angle=0]{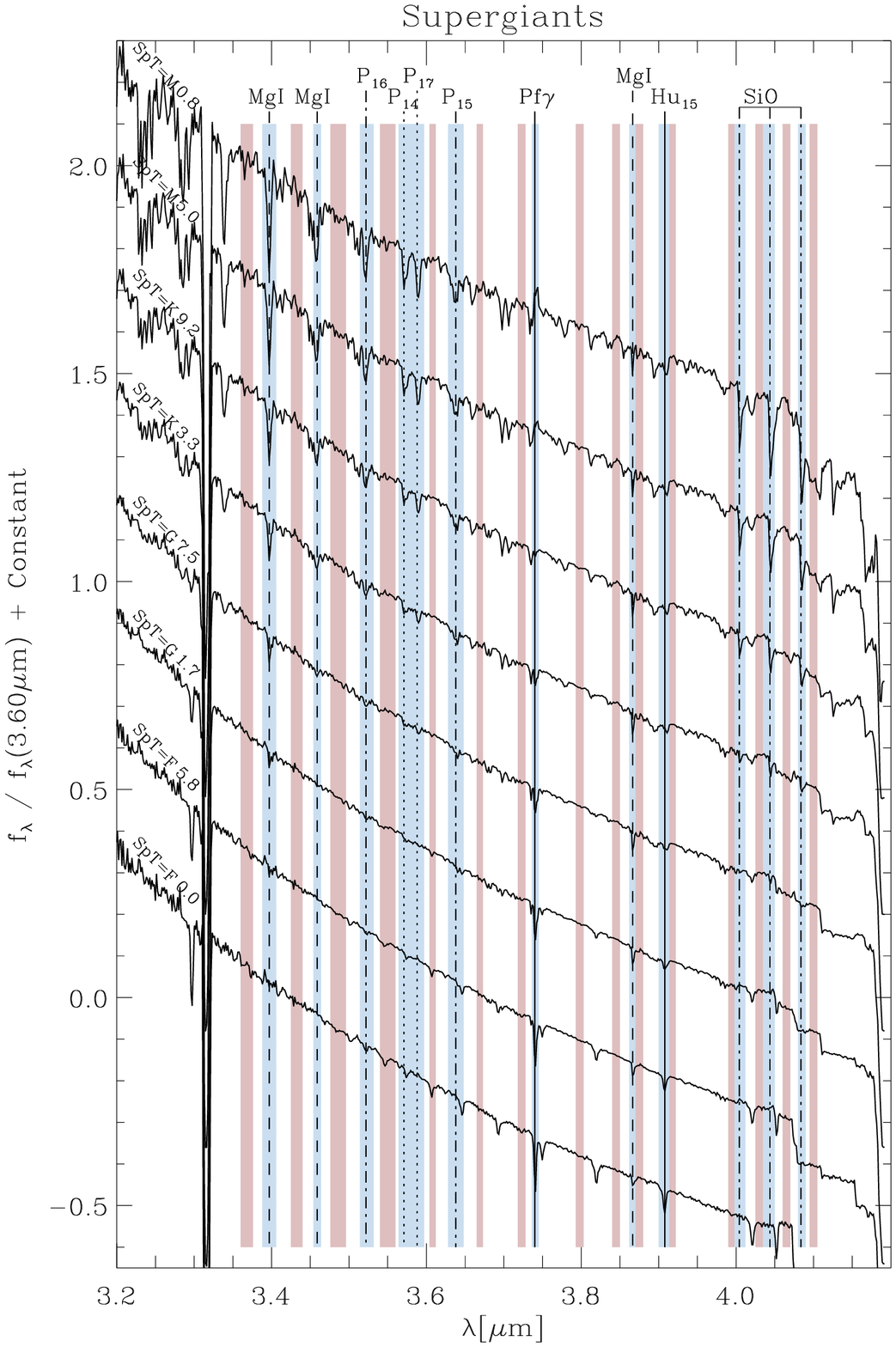}
\caption{L atmospheric window model spectrum of supergiants. The 
flux-normalization here is performed at 3.60$\,\mu$m.  
Figure\,\ref{fig:SGiant_fitted_Y} gives details.}\label{fig:SGiant_fitted_L}
\end{figure*}

\begin{figure*}
\includegraphics[width=16truecm,angle=0]{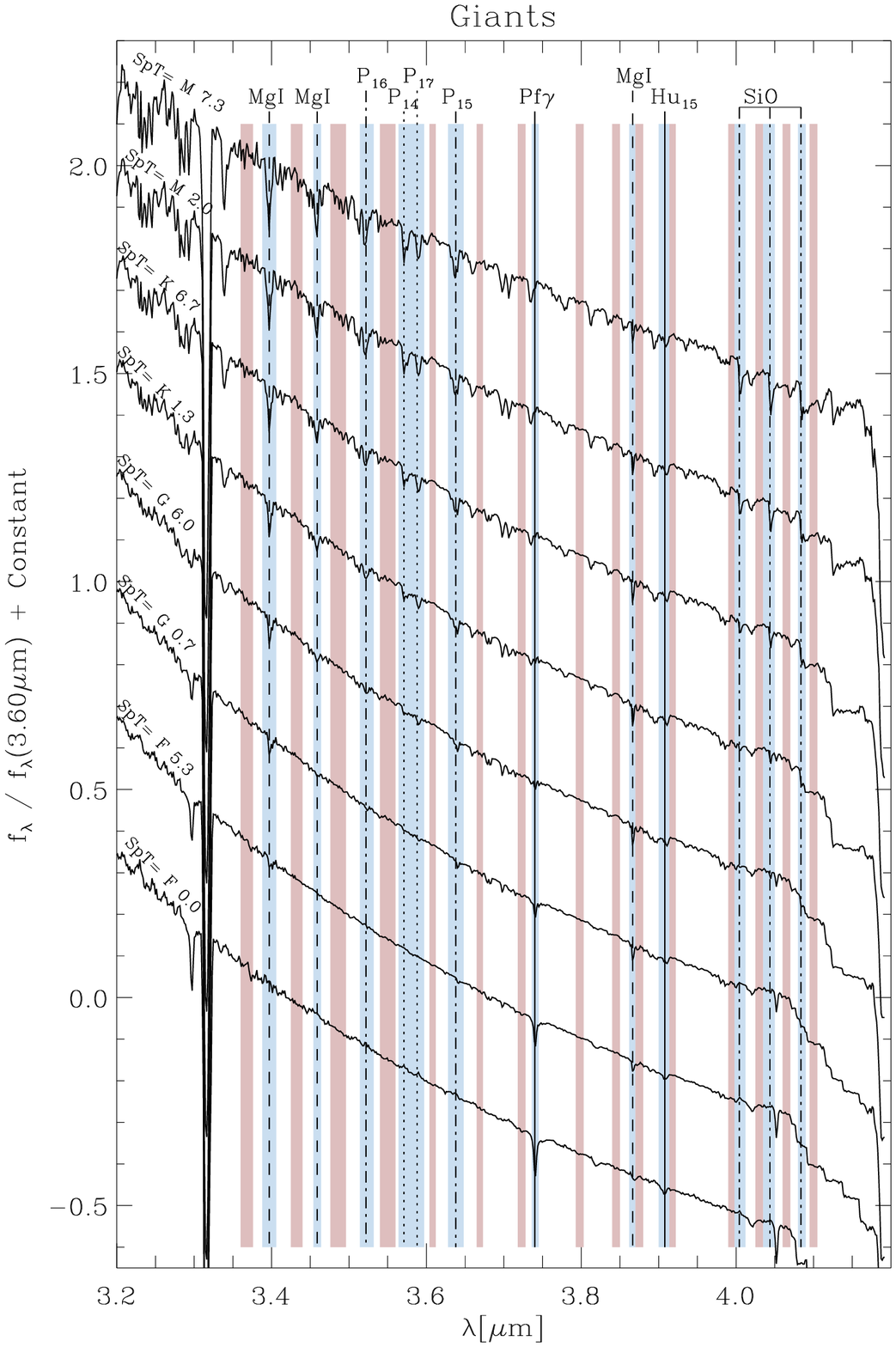}
\caption{L atmospheric window model spectrum of giants.  
Figure\,\ref{fig:SGiant_fitted_L} gives details.}\label{fig:Giant_fitted_L}
\end{figure*}

\begin{figure*}
\includegraphics[width=16truecm,angle=0]{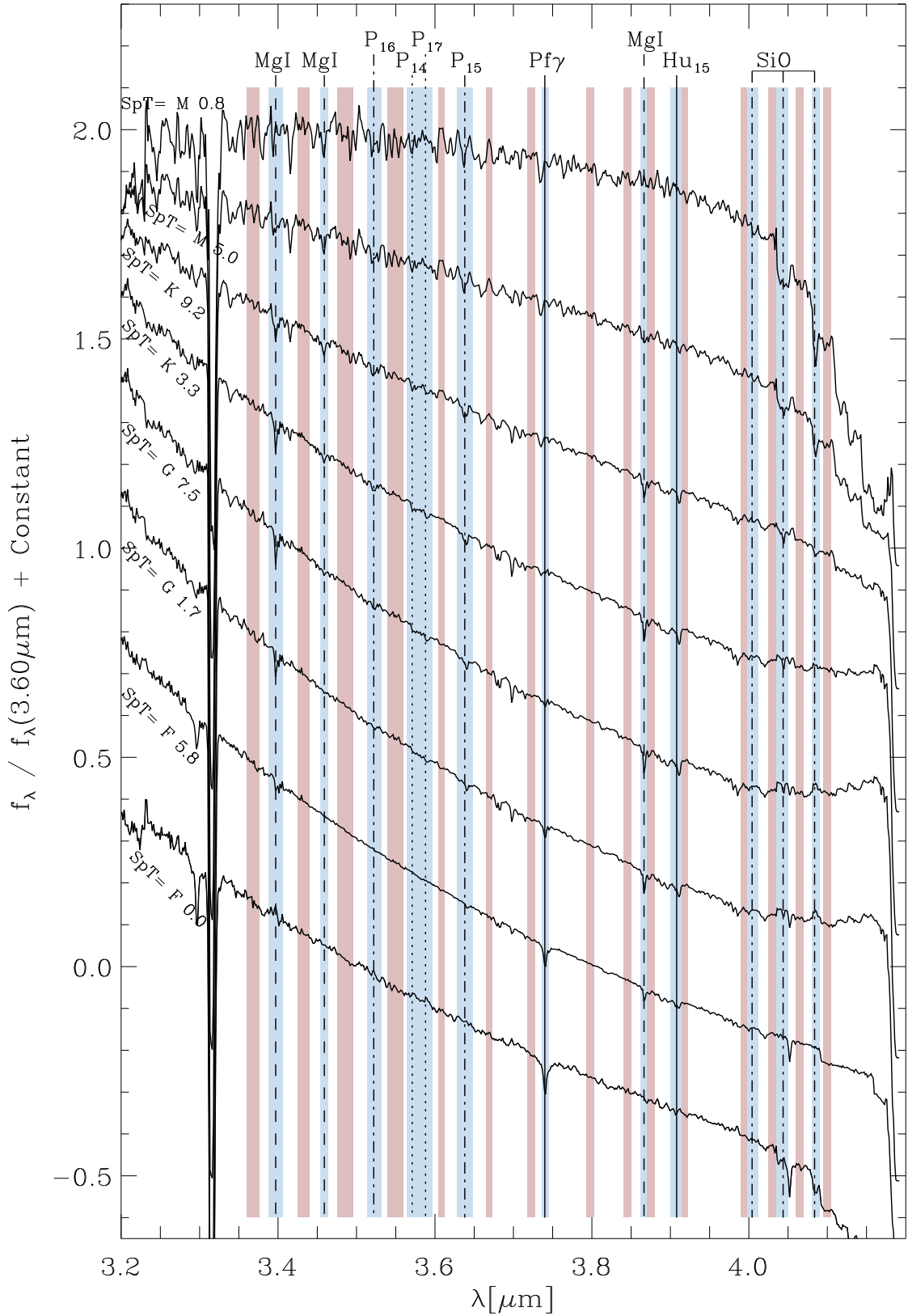}
\caption{L atmospheric window model spectrum of dwarfs.  
Figure\,\ref{fig:SGiant_fitted_L} gives details.}\label{fig:Dwarf_fitted_L}
\end{figure*}

The model spectra in Y, J, H and L atmospheric windows for stars of 
different luminosity classes are shown in Figs.\,\ref{fig:SGiant_fitted_Y} 
and \ref{fig:Giant_fitted_Y}--\ref{fig:Dwarf_fitted_L}.

\clearpage
\section{Sensitivity map plots}\label{app:sens}

\begin{figure*}
\includegraphics[width=16truecm,angle=0]{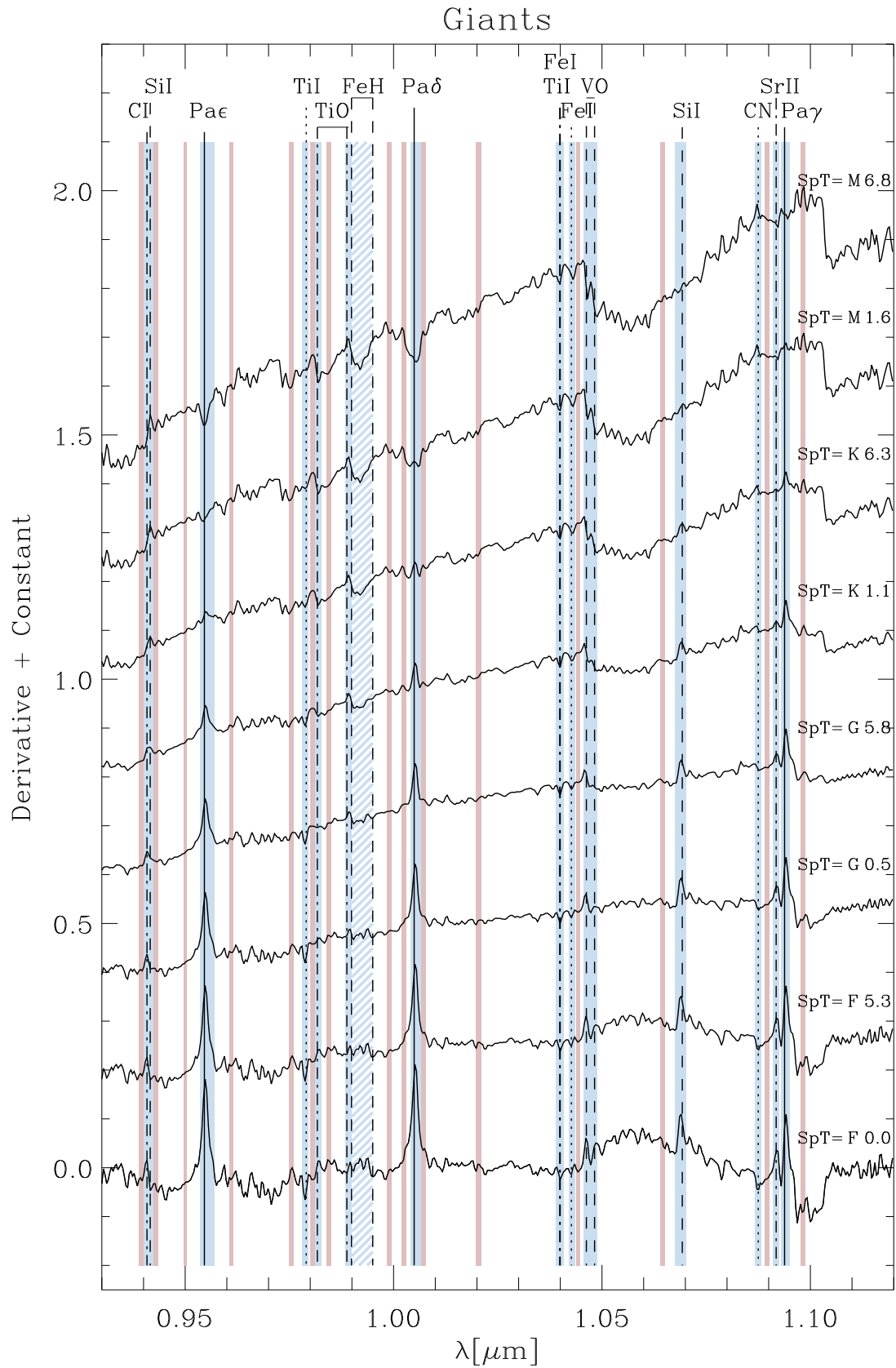}
\caption{Y atmospheric window sensitivity map for SpT of
  giants. Fig.\,\ref{fig:SupGian_SpT_Y} gives
  details.}\label{fig:Gian_SpT_Y}
\end{figure*}

The sensitivity maps for SpT in J, H, and L atmospheric windows 
for stars of different luminosity classes are shown in 
Figs.\,\ref{fig:SupGian_SpT_Y} and \ref{fig:Gian_SpT_Y}--\ref{fig:Dwarf_fitted_L}.
The corresponding sensitivity maps for surface gravity in Y, J, H, 
and L atmospheric windows for stars of different luminosity classes are 
shown in Figs.\,\ref{fig:SupGian_SpT_Y} and 
\ref{fig:Gian_SpT_Y}--\ref{fig:Dwarf_fitted_L}.

\begin{figure*}
\includegraphics[width=16truecm,angle=0]{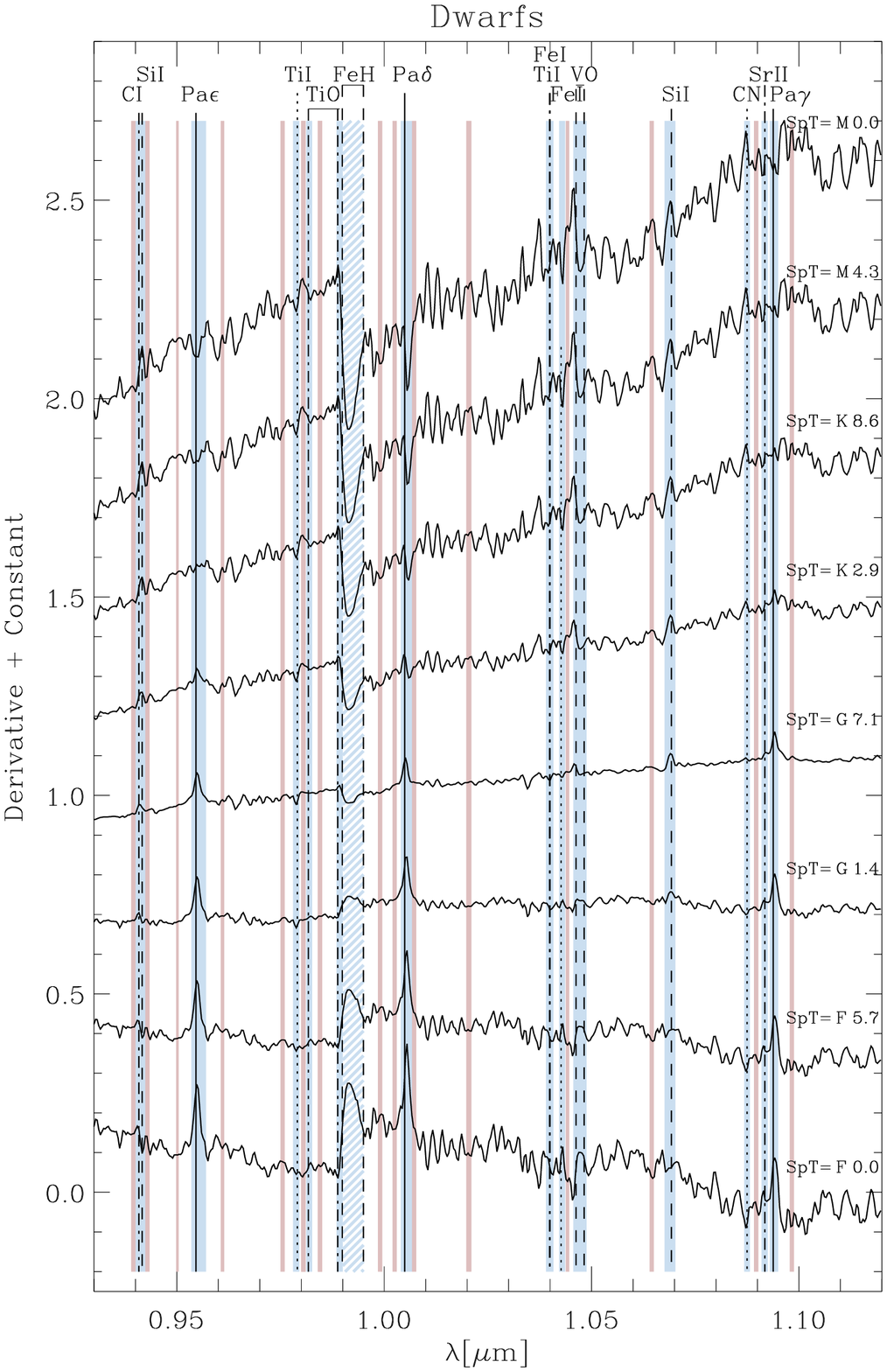}
\caption{Y atmospheric window sensitivity map for SpT of dwarfs.
Fig.\,\ref{fig:SupGian_SpT_Y} gives details.}\label{fig:Dwarf_SpT_Y}
\end{figure*}

\begin{figure*}
\includegraphics[width=16truecm,angle=0]{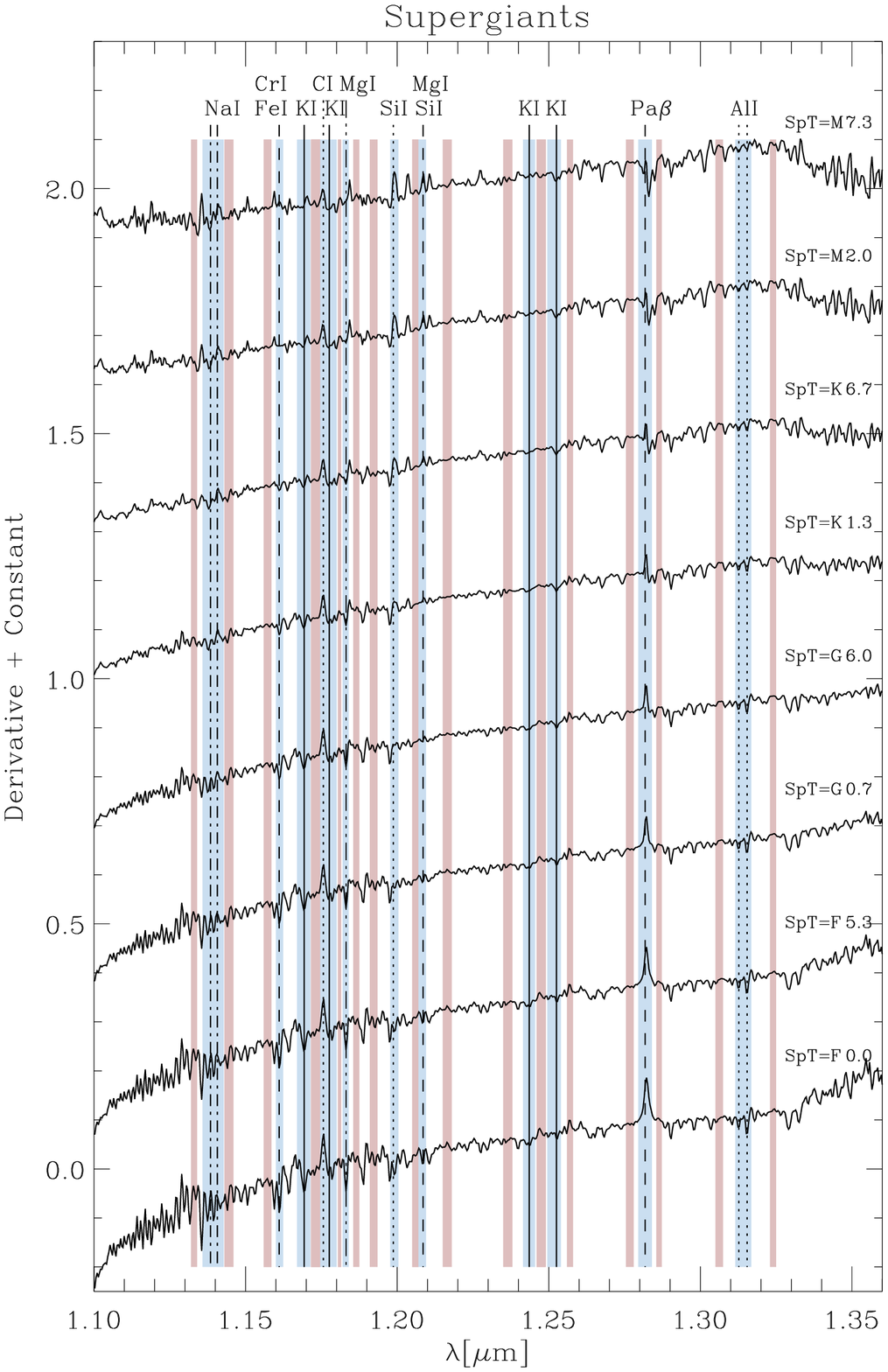}
\caption{J atmospheric window sensitivity map for SpT of supergiants.
Fig.\,\ref{fig:SupGian_SpT_Y} gives details.}\label{fig:SupGian_SpT_J}
\end{figure*}

\begin{figure*}
\includegraphics[width=16truecm,angle=0]{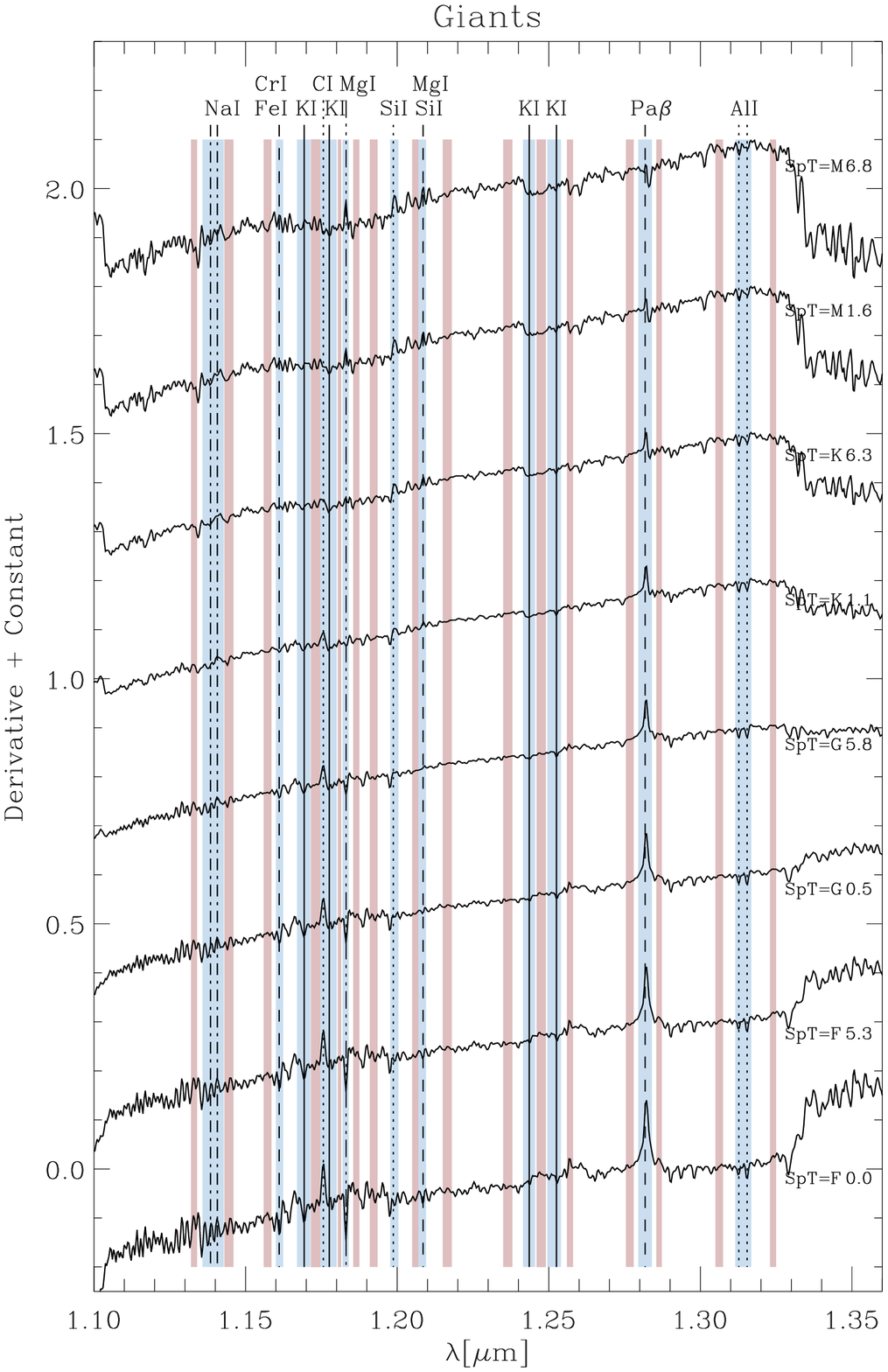}
\caption{J atmospheric window sensitivity map for SpT of giants.
Fig.\,\ref{fig:SupGian_SpT_J} gives details.}\label{fig:Gian_SpT_J}
\end{figure*}

\begin{figure*}
\includegraphics[width=16truecm,angle=0]{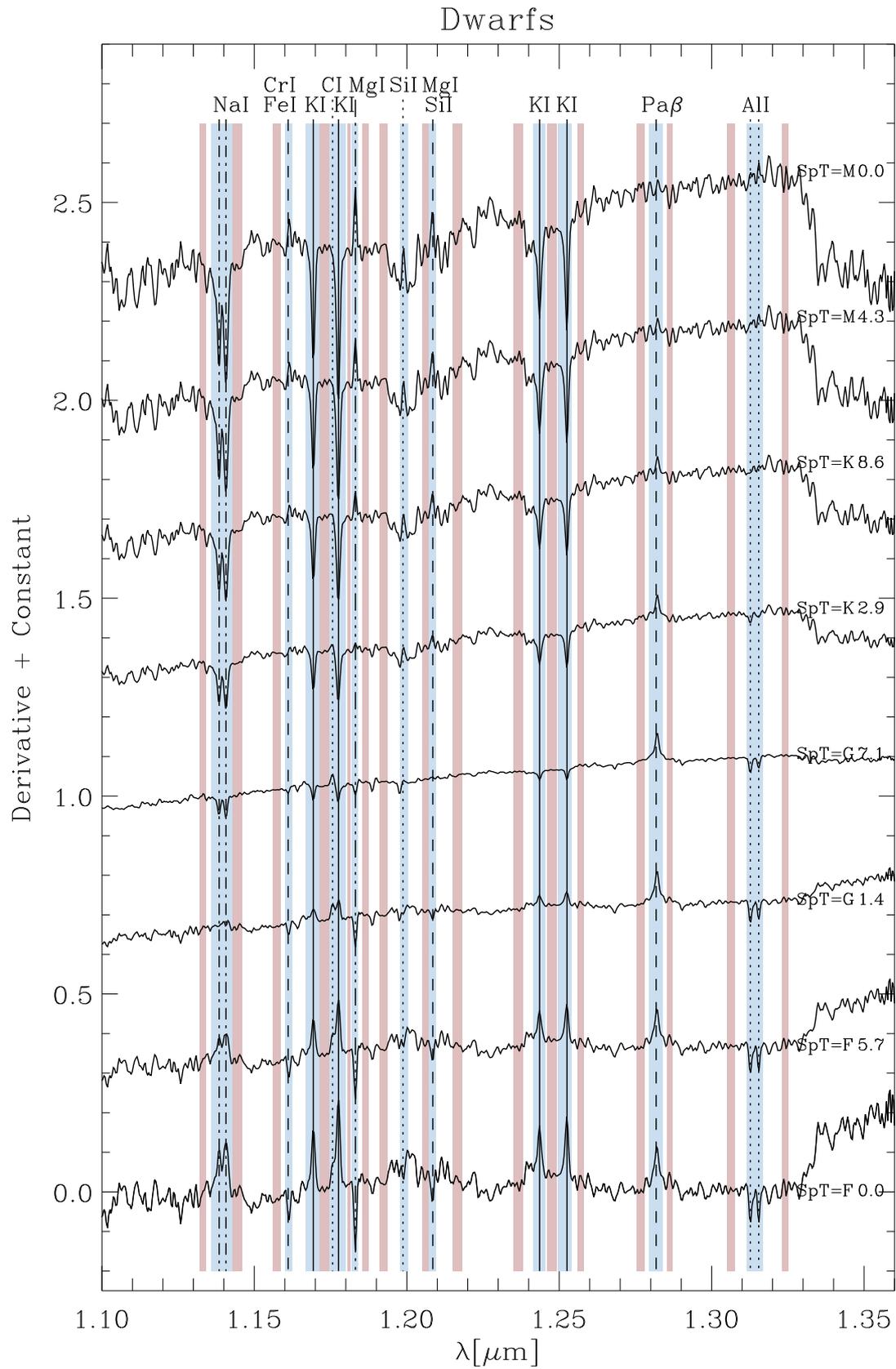}
\caption{J atmospheric window sensitivity map for SpT of dwarfs.
See Fig.\,\ref{fig:SupGian_SpT_J} for details.}\label{fig:Dwarf_SpT_J}
\end{figure*}

\begin{figure*}
\includegraphics[width=16truecm,angle=0]{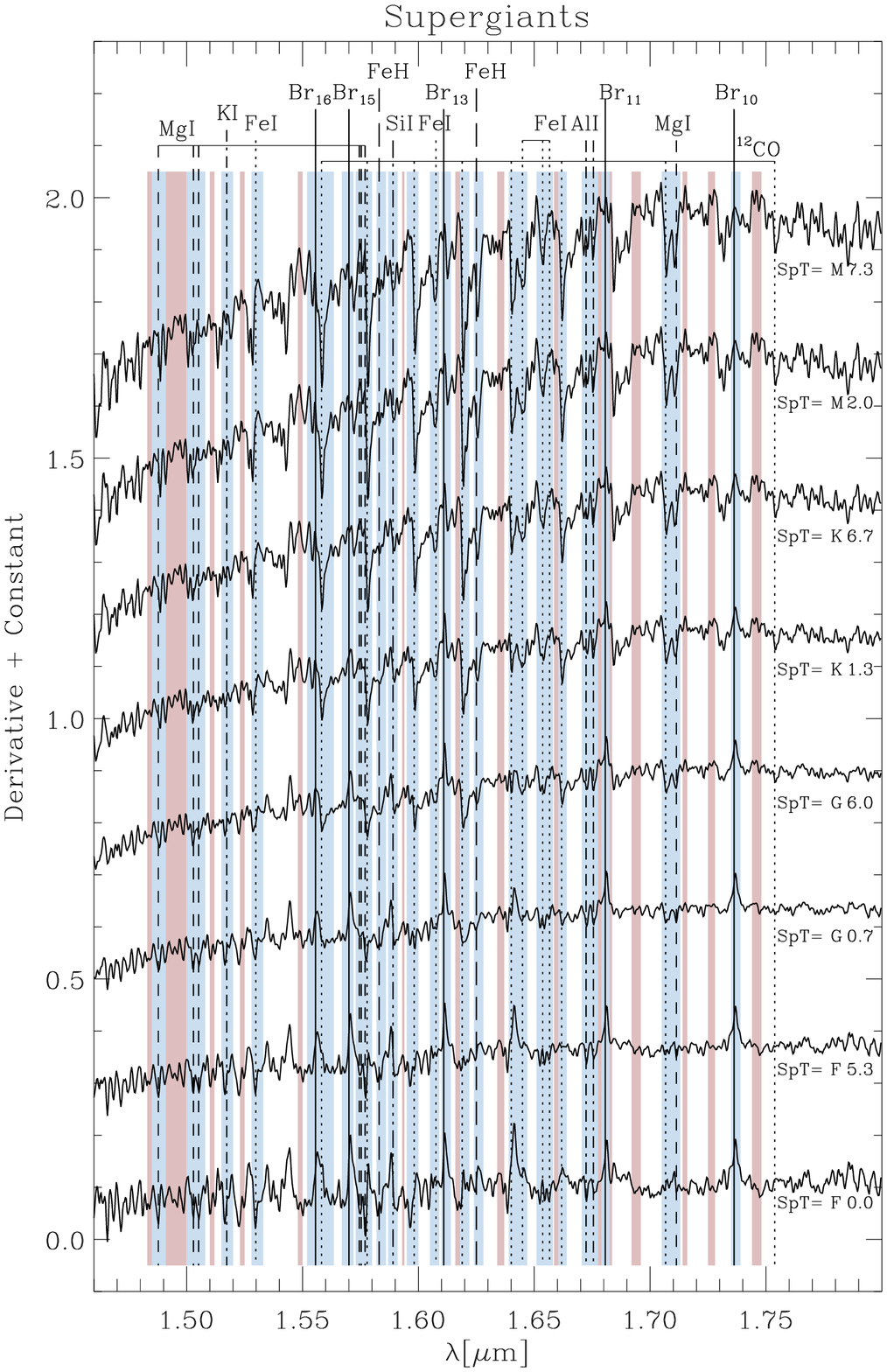}
\caption{H atmospheric window sensitivity map for SpT of supergiants.
ig.\,\ref{fig:SupGian_SpT_Y} gives details.}\label{fig:SupGian_SpT_H}
\end{figure*}

\begin{figure*}
\includegraphics[width=16truecm,angle=0]{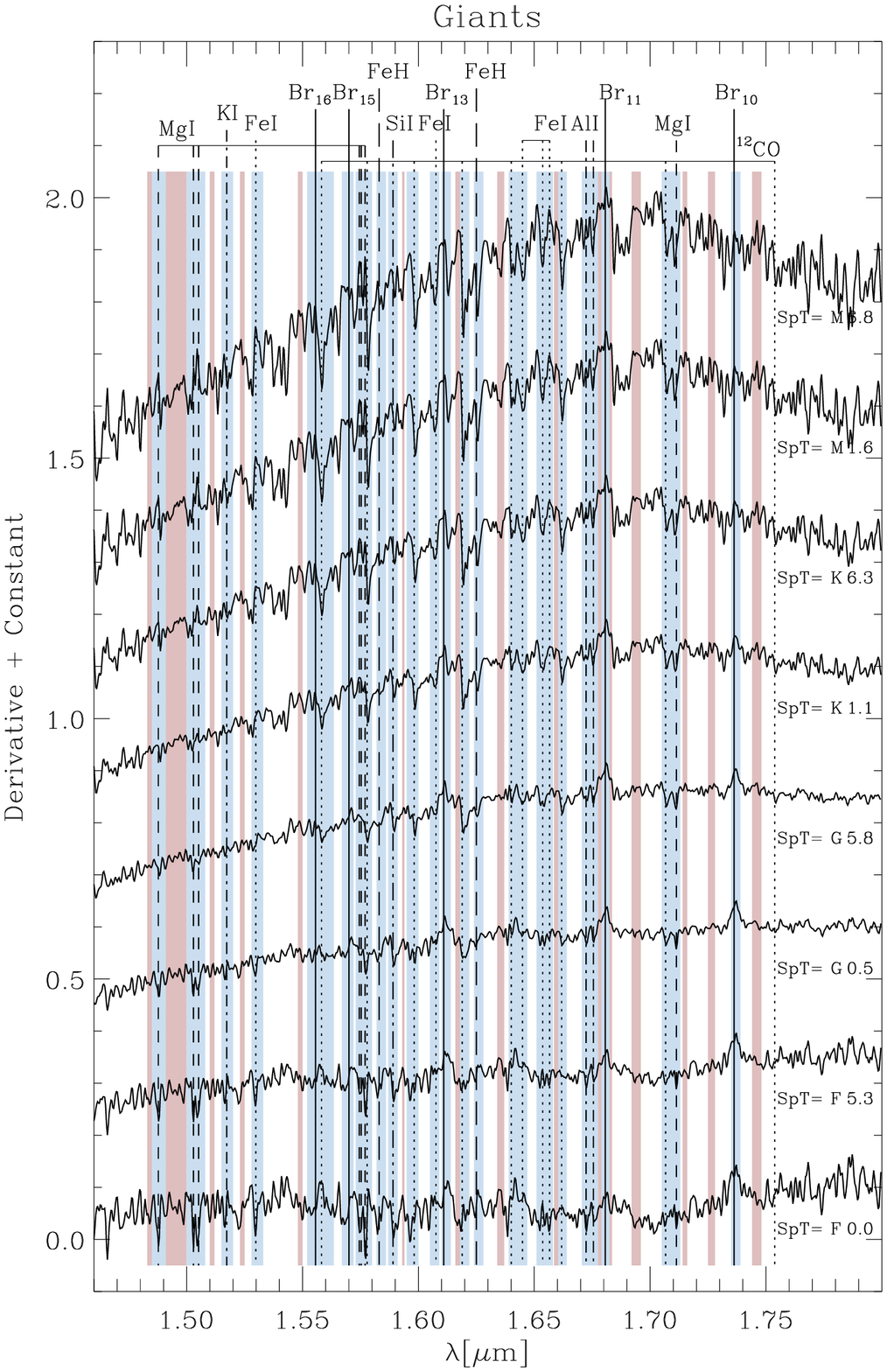}
\caption{H atmospheric window sensitivity map for SpT of giants.
Fig.\,\ref{fig:SupGian_SpT_H} gives details.}\label{fig:Gian_SpT_H}
\end{figure*}

\begin{figure*}
\includegraphics[width=16truecm,angle=0]{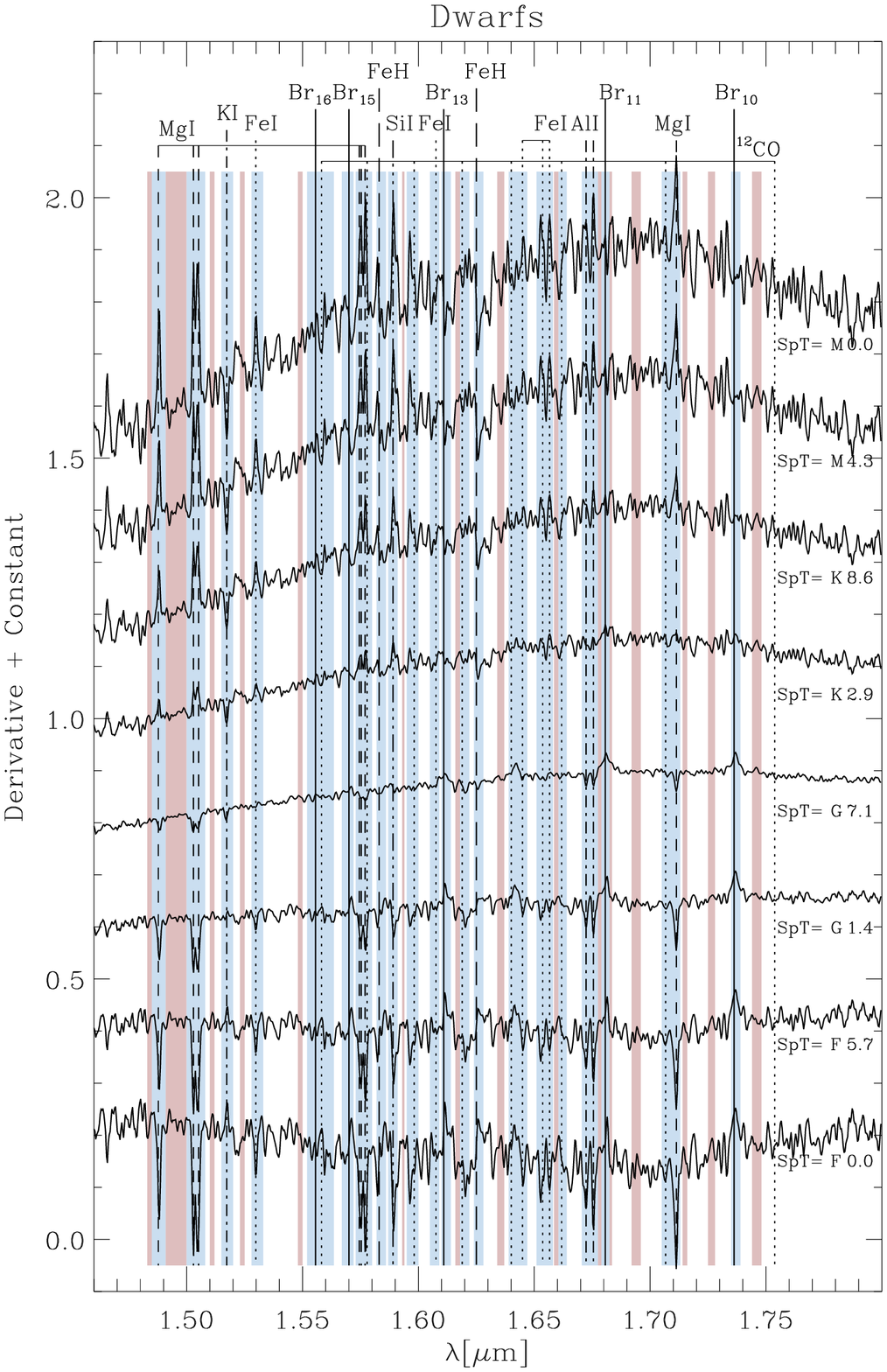}
\caption{H atmospheric window sensitivity map for SpT of dwarfs.
Fig.\,\ref{fig:SupGian_SpT_H} gives details.}\label{fig:Dwarf_SpT_H}
\end{figure*}

\begin{figure*}
\includegraphics[width=16truecm,angle=0]{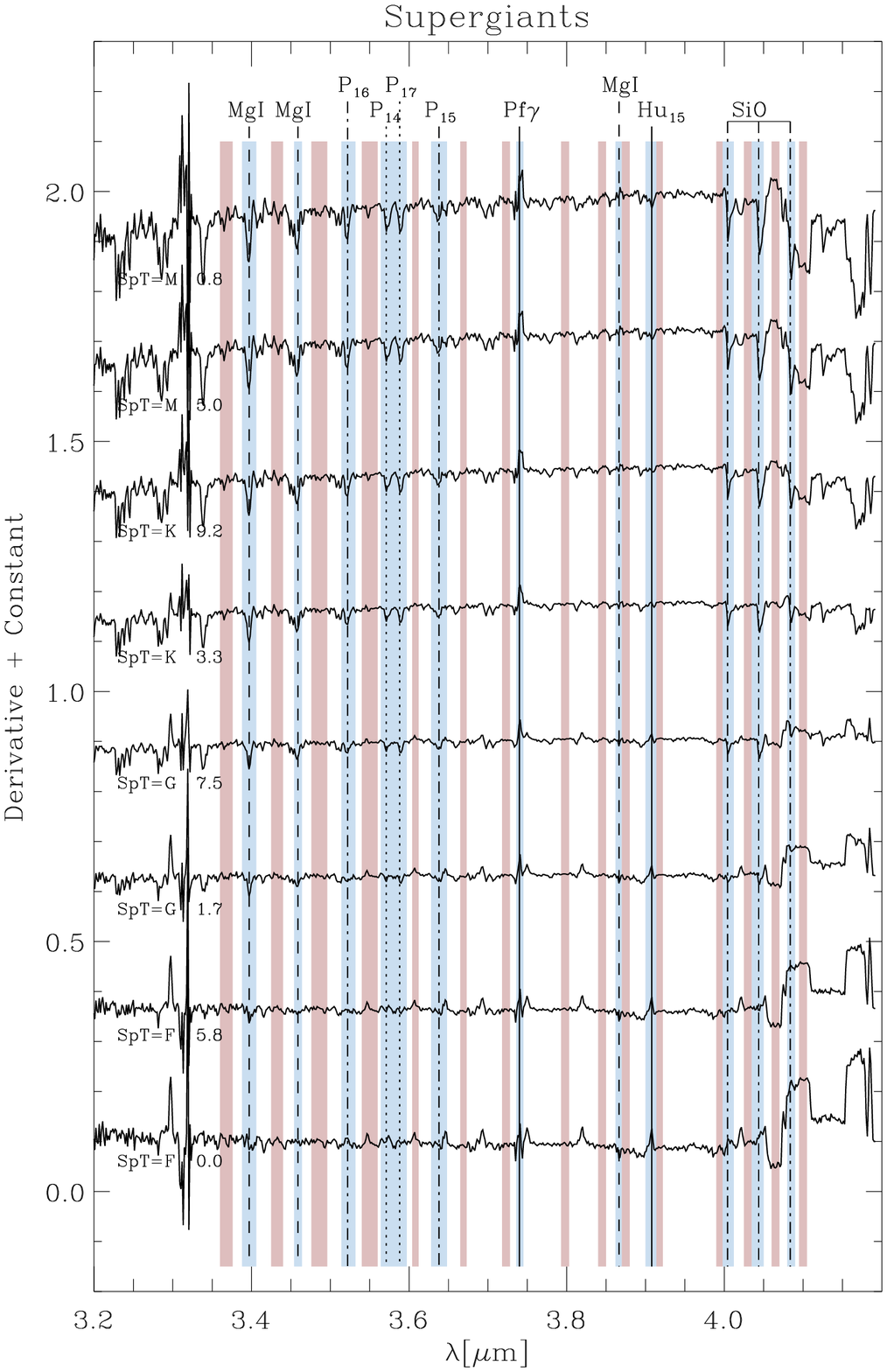}
\caption{L atmospheric window sensitivity map for SpT of supergiants.
See Fig.\,\ref{fig:SupGian_SpT_Y} for details.}\label{fig:SupGian_SpT_L}
\end{figure*}

\begin{figure*}
\includegraphics[width=16truecm,angle=0]{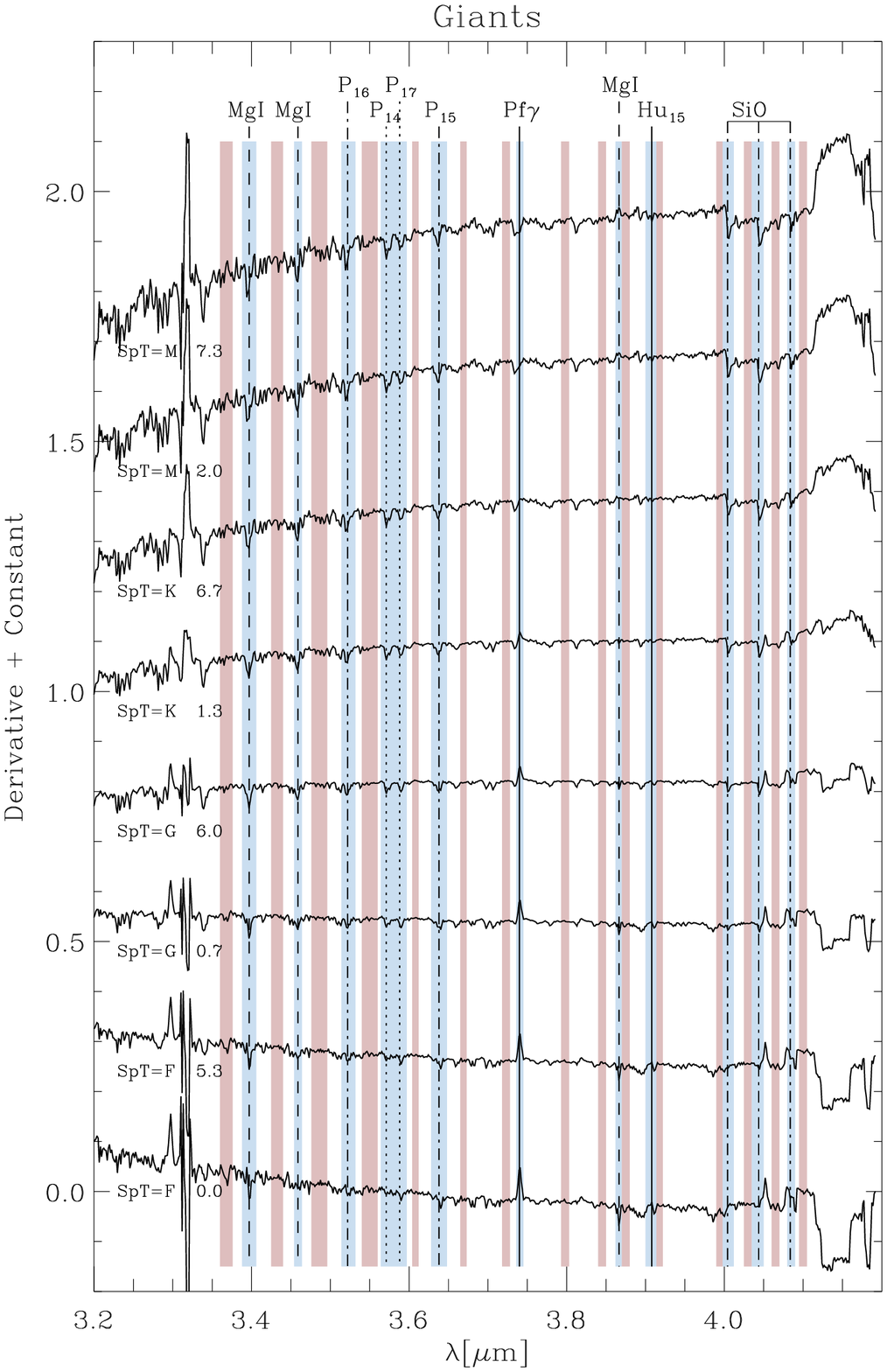}
\caption{L atmospheric window sensitivity map for SpT of giants.
Fig.\,\ref{fig:SupGian_SpT_L} gives details.}\label{fig:Gian_SpT_L}
\end{figure*}

\begin{figure*}
\includegraphics[width=16truecm,angle=0]{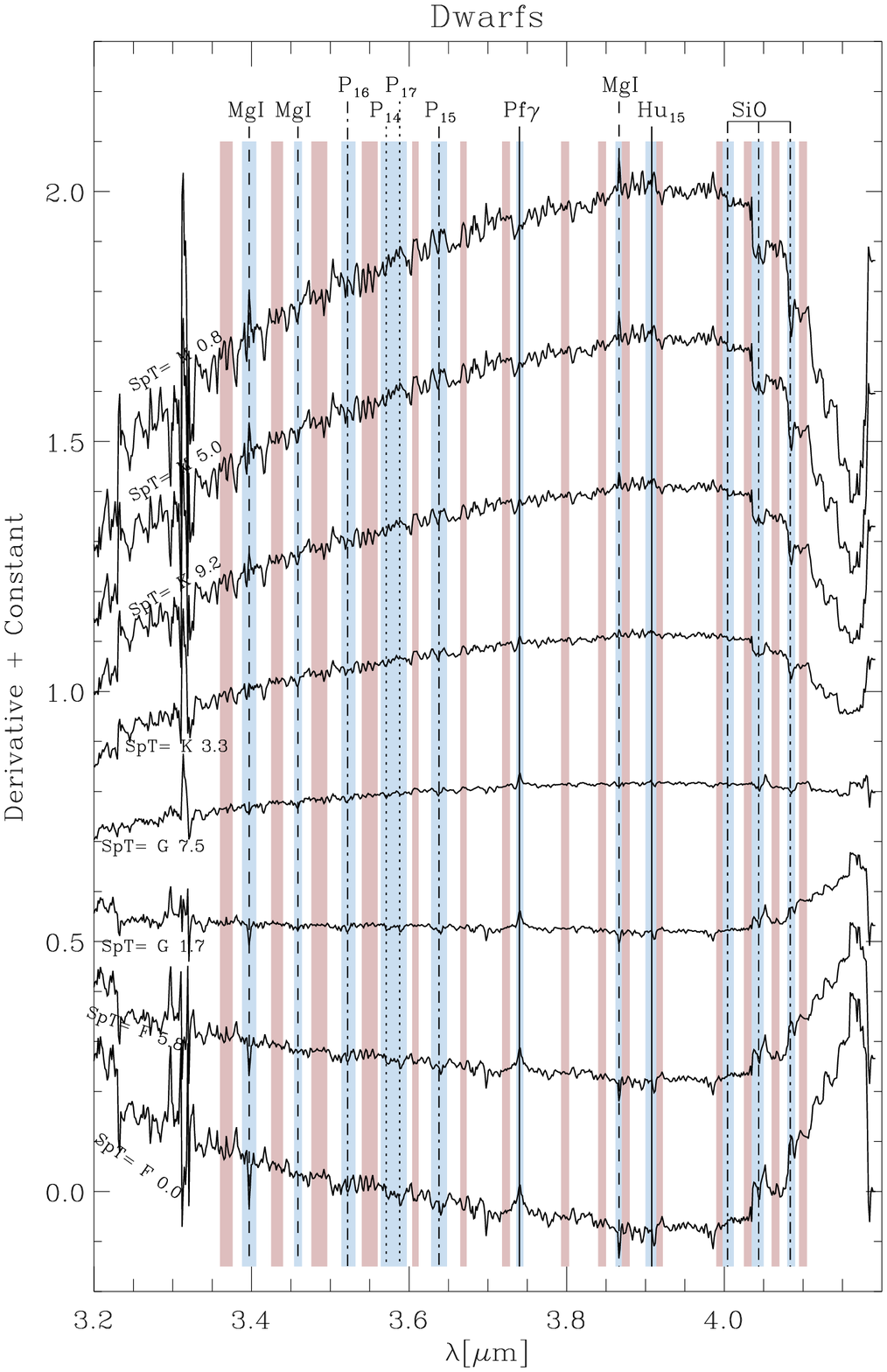}
\caption{L atmospheric window sensitivity map for SpT of dwarfs.
Fig.\,\ref{fig:SupGian_SpT_L} gives details.}\label{fig:Dwarf_SpT_L}
\end{figure*}

\clearpage

\begin{figure*}
\includegraphics[width=16truecm,angle=0]{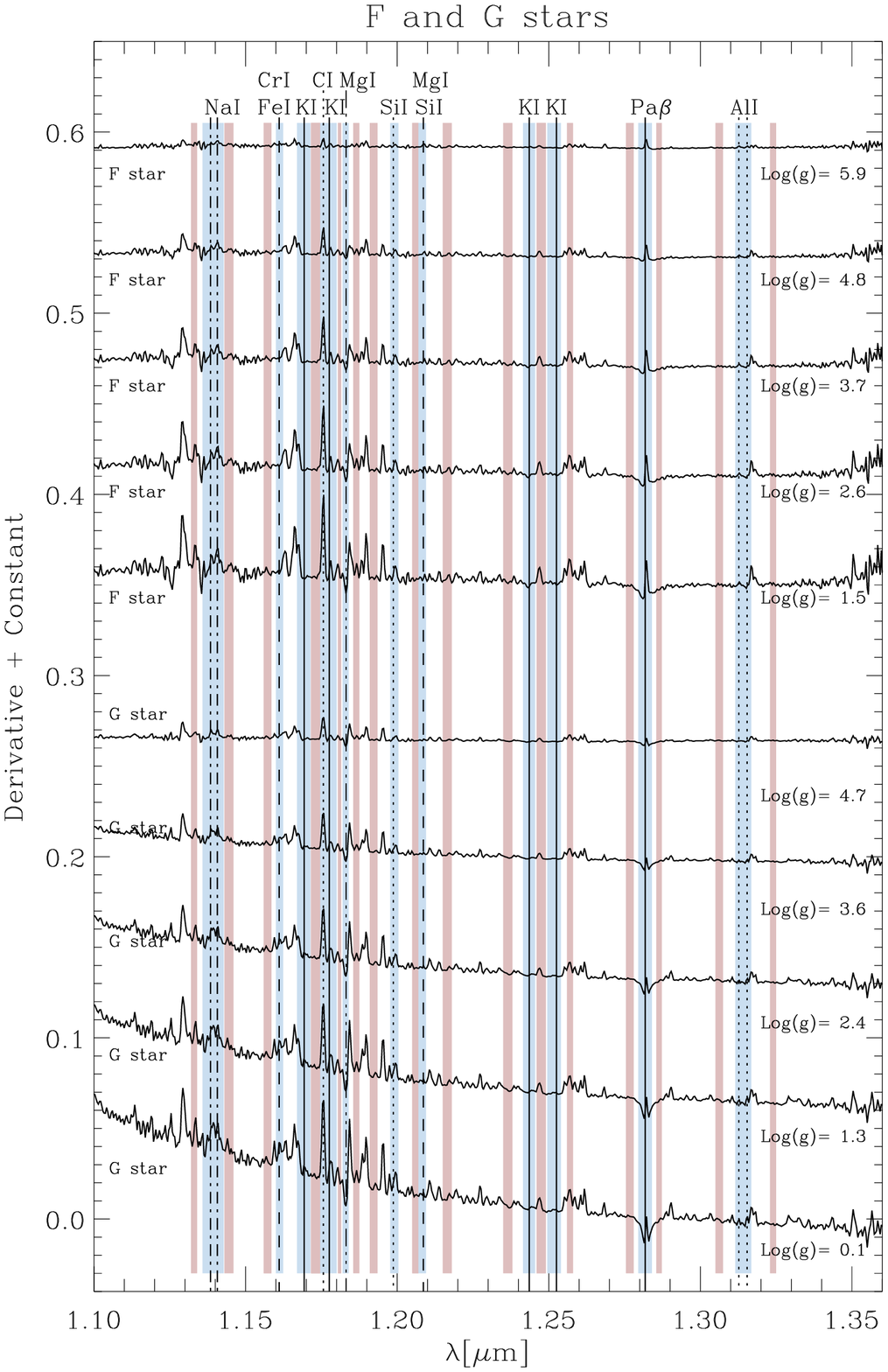}
\caption{J atmospheric window sensitivity map for surface gravity of
  F- (top) and G-type stars (bottom). Fig.\,\ref{fig:FG_Logg_Y} gives
  details.}\label{fig:FG_Logg_J}
\end{figure*}

\begin{figure*}
\includegraphics[width=16truecm,angle=0]{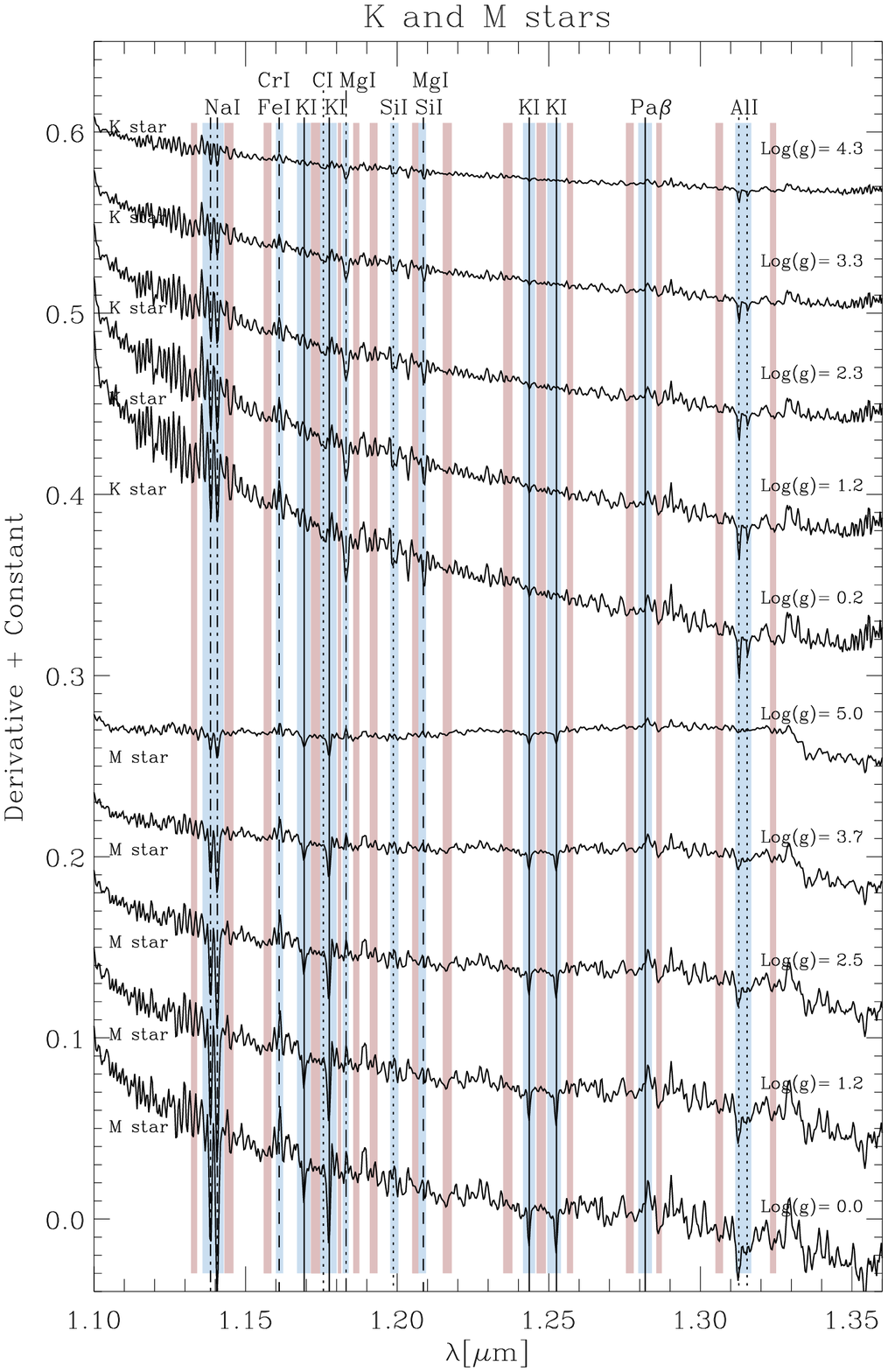}
\caption{J atmospheric window sensitivity map for surface gravity of
  K- (top) and M-type stars (bottom). Fig.\,\ref{fig:KM_Logg_Y} gives
  details.}\label{fig:KM_Logg_J}
\end{figure*}

\begin{figure*}
\includegraphics[width=16truecm,angle=0]{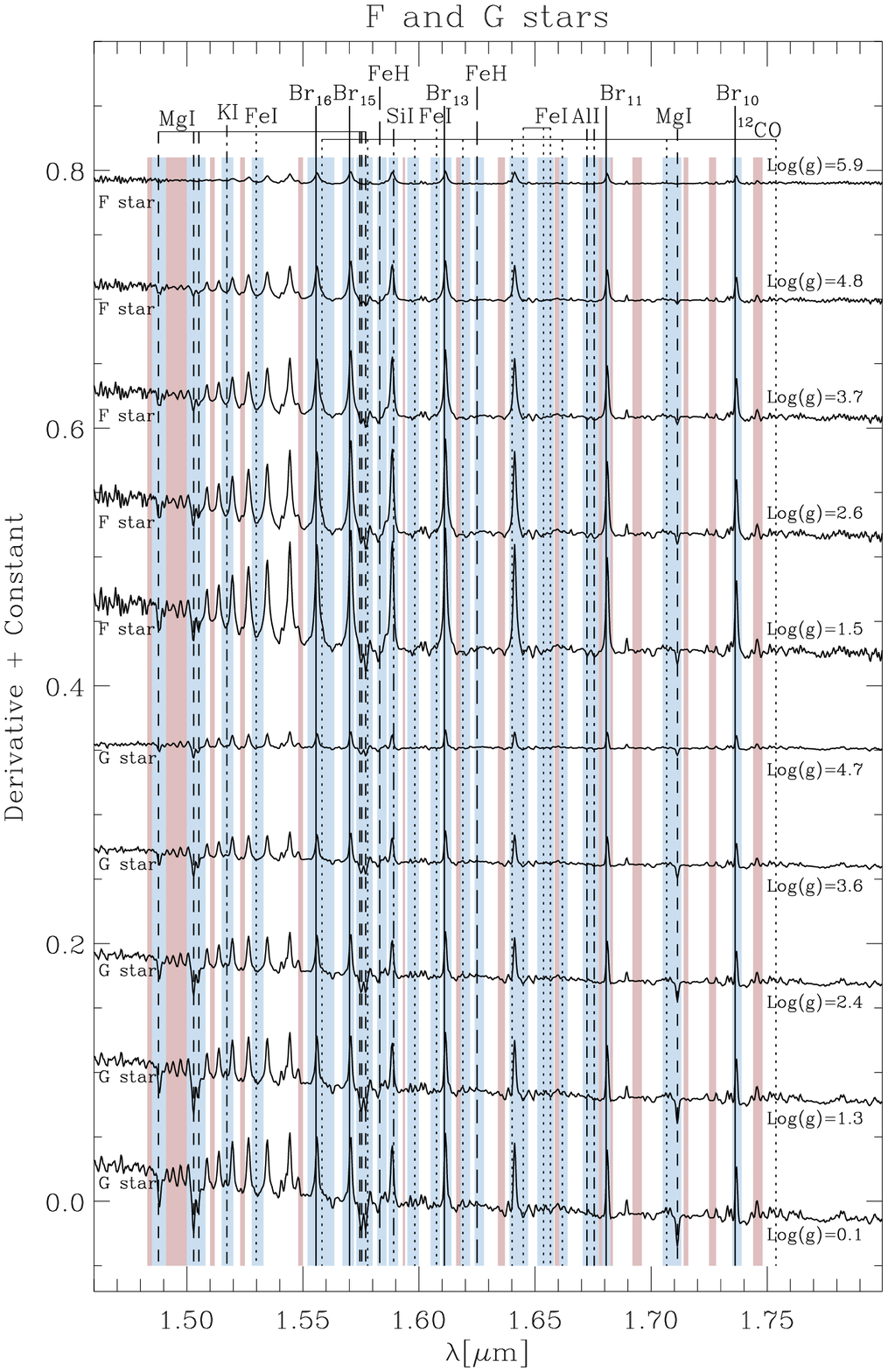}
\caption{H atmospheric window sensitivity map for surface gravity of
  F- (top) and G-type stars (bottom). Fig.\,\ref{fig:FG_Logg_Y} gives
  details.}\label{fig:FG_Logg_H}
\end{figure*}

\begin{figure*}
\includegraphics[width=16truecm,angle=0]{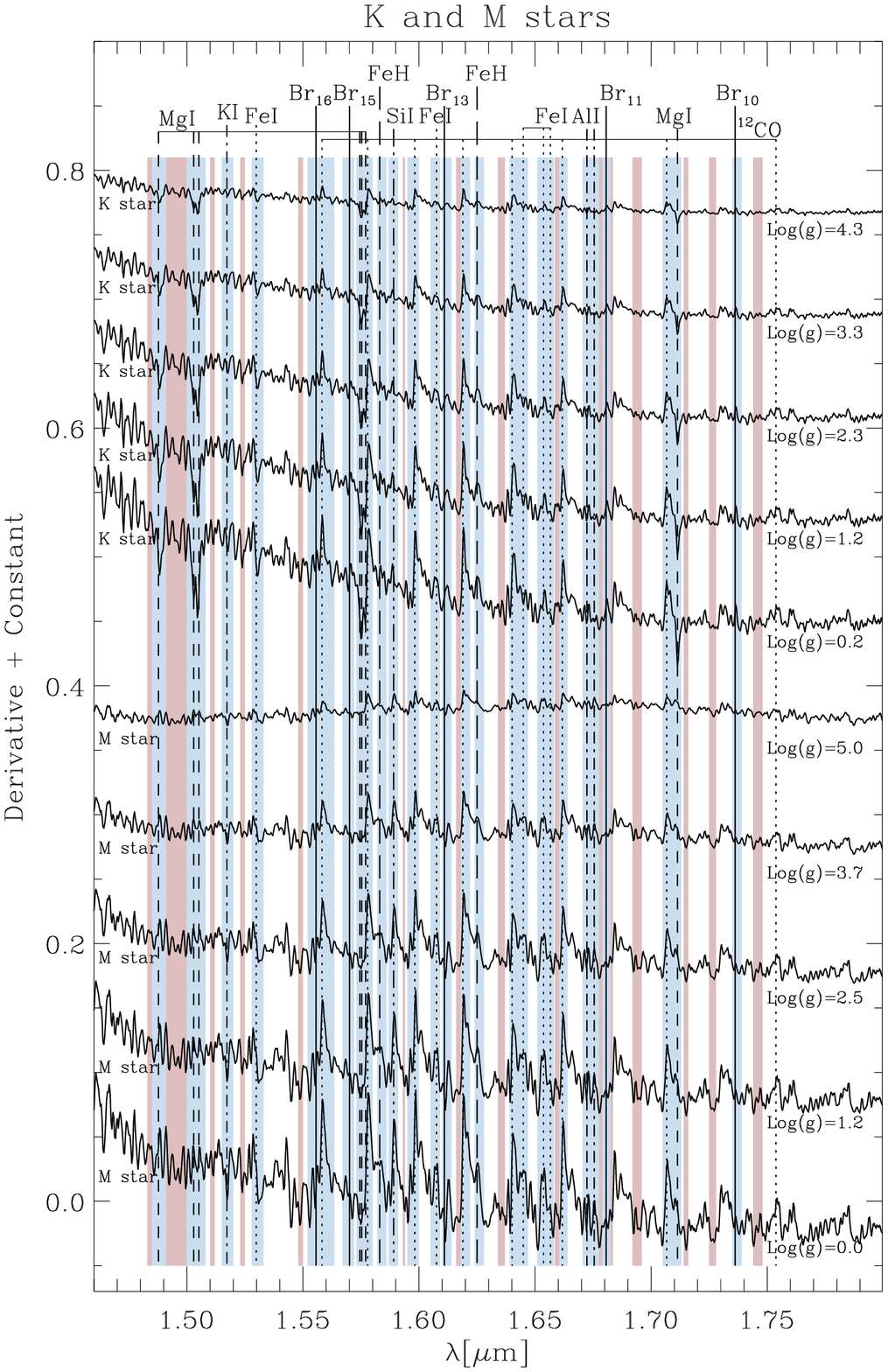}
\caption{H atmospheric window sensitivity map for surface gravity 
of K- (top) and M-type stars (bottom). Fig.\,\ref{fig:KM_Logg_Y} gives 
details.}\label{fig:KM_Logg_H}
\end{figure*}

\begin{figure*}
\includegraphics[width=16truecm,angle=0]{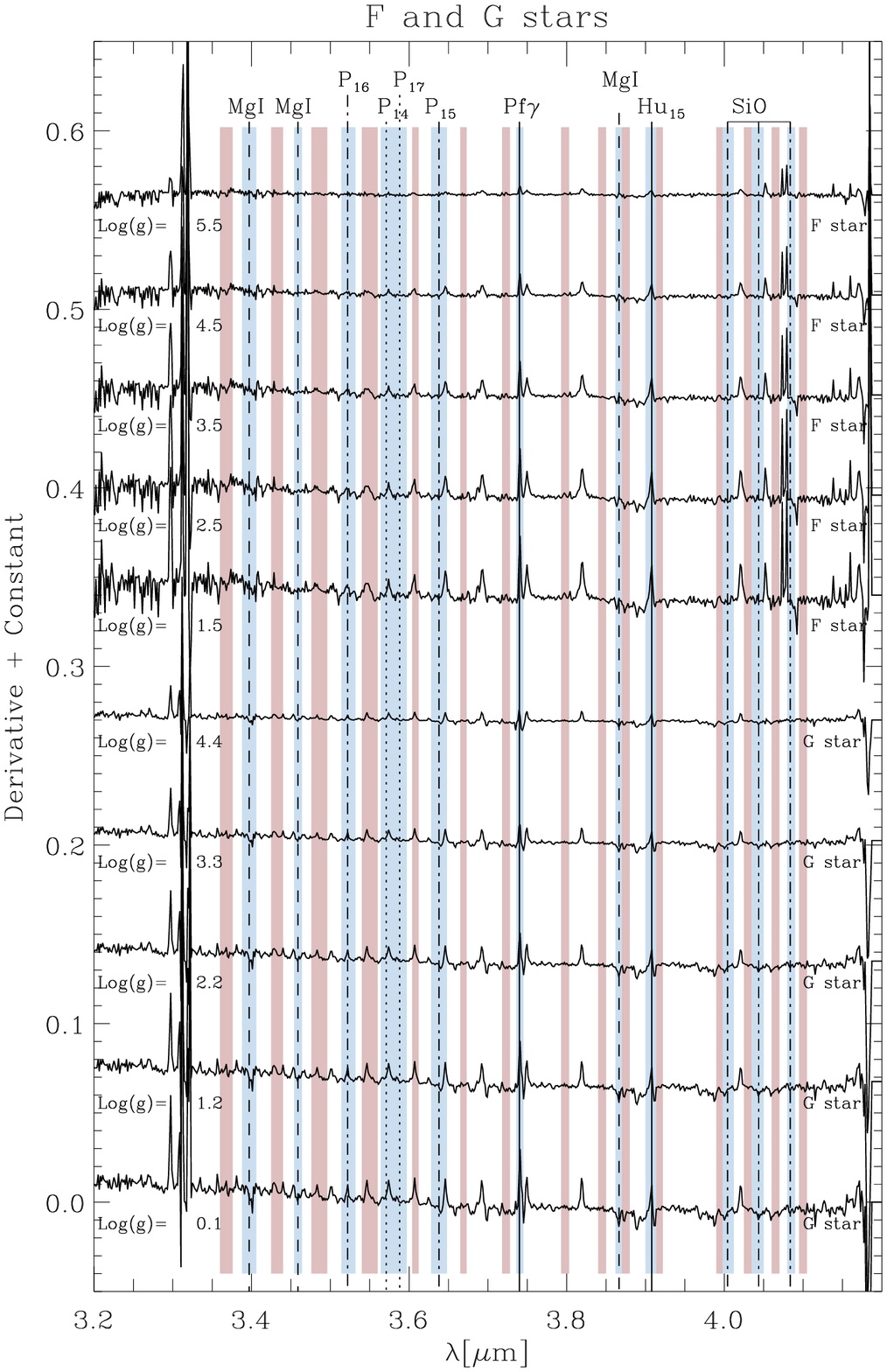}
\caption{L atmospheric window sensitivity map for surface gravity 
of F- (top) and G-type stars (bottom). See Fig.\,\ref{fig:FG_Logg_Y} for 
details.}\label{fig:FG_Logg_L}
\end{figure*}

\begin{figure*}
\includegraphics[width=16truecm,angle=0]{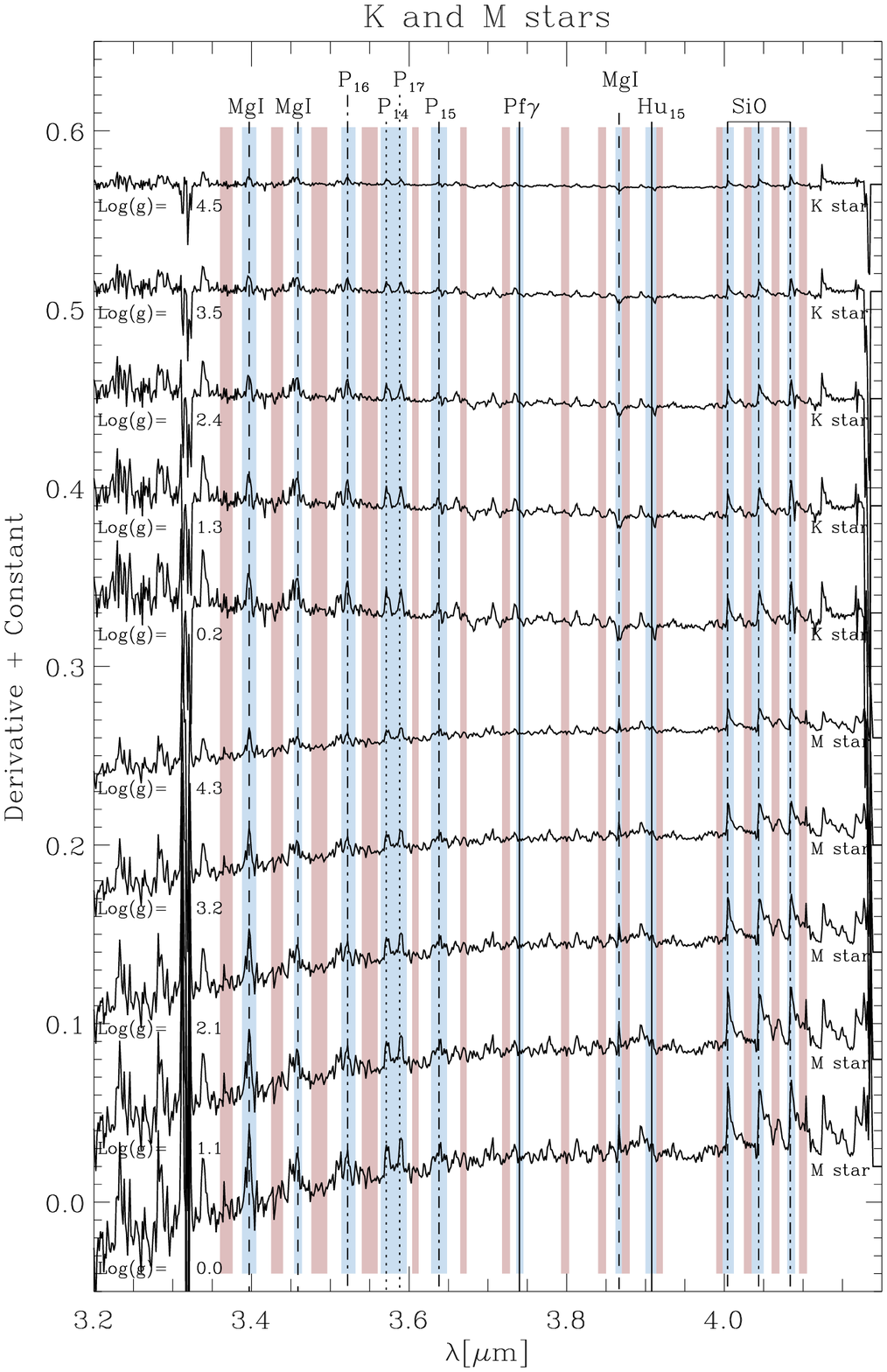}
\caption{L atmospheric window sensitivity map for surface gravity of
  K- (top) and M-type stars (bottom). Fig.\,\ref{fig:KM_Logg_Y} gives
  details.}\label{fig:KM_Logg_L}
\end{figure*}

\clearpage
\section{Broadening effects}\label{app:SigmaV}

\begin{figure*}
\includegraphics[angle=90,width=\textwidth]{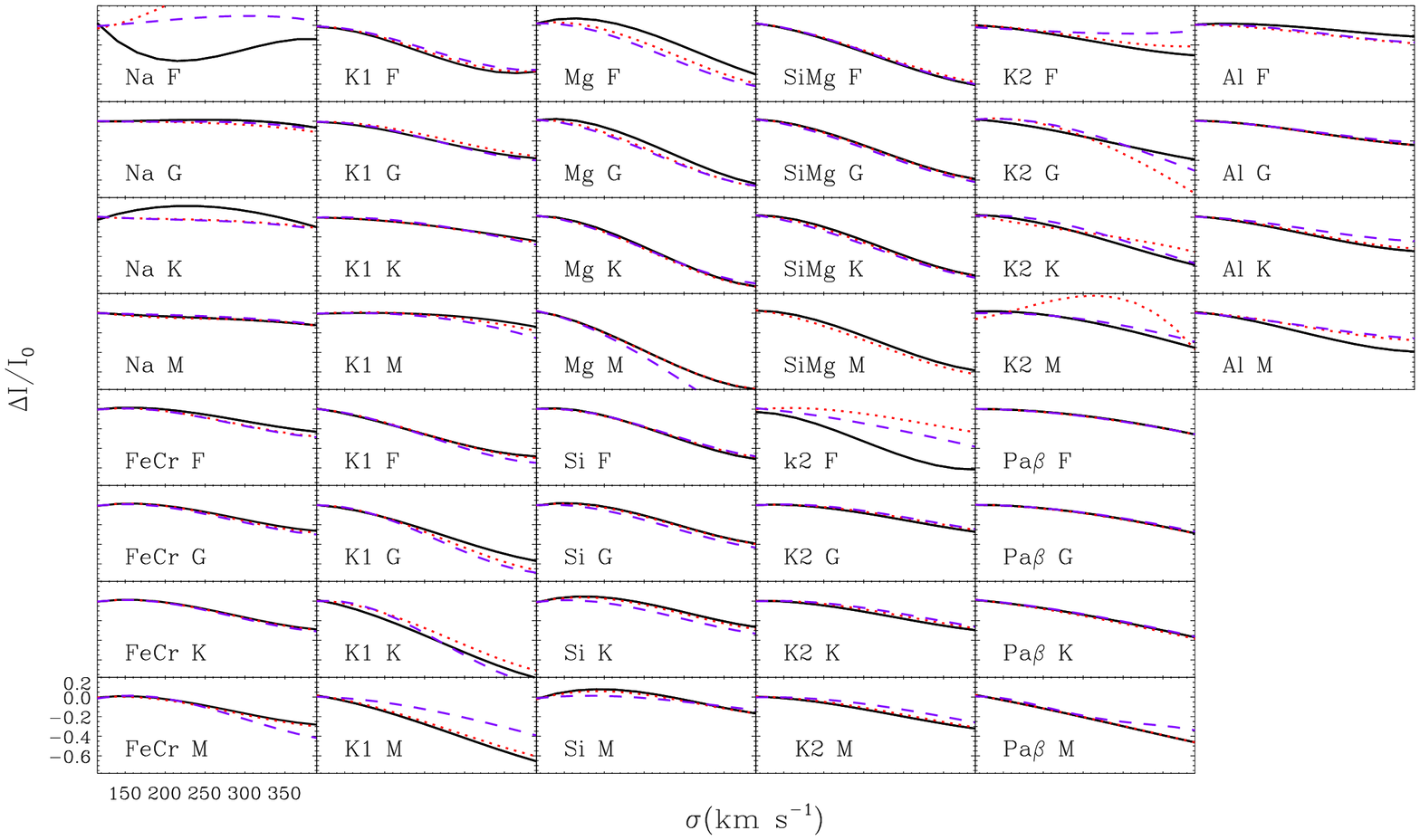}
\caption{Sensitivity of the indices in the J atmospheric window to 
the velocity dispersion broadening. Fig.\,\ref{fig:Y-sigma} gives
details.} \label{fig:J-sigma}
\end{figure*}

\begin{figure*}
\includegraphics[angle=90,width=\textwidth]{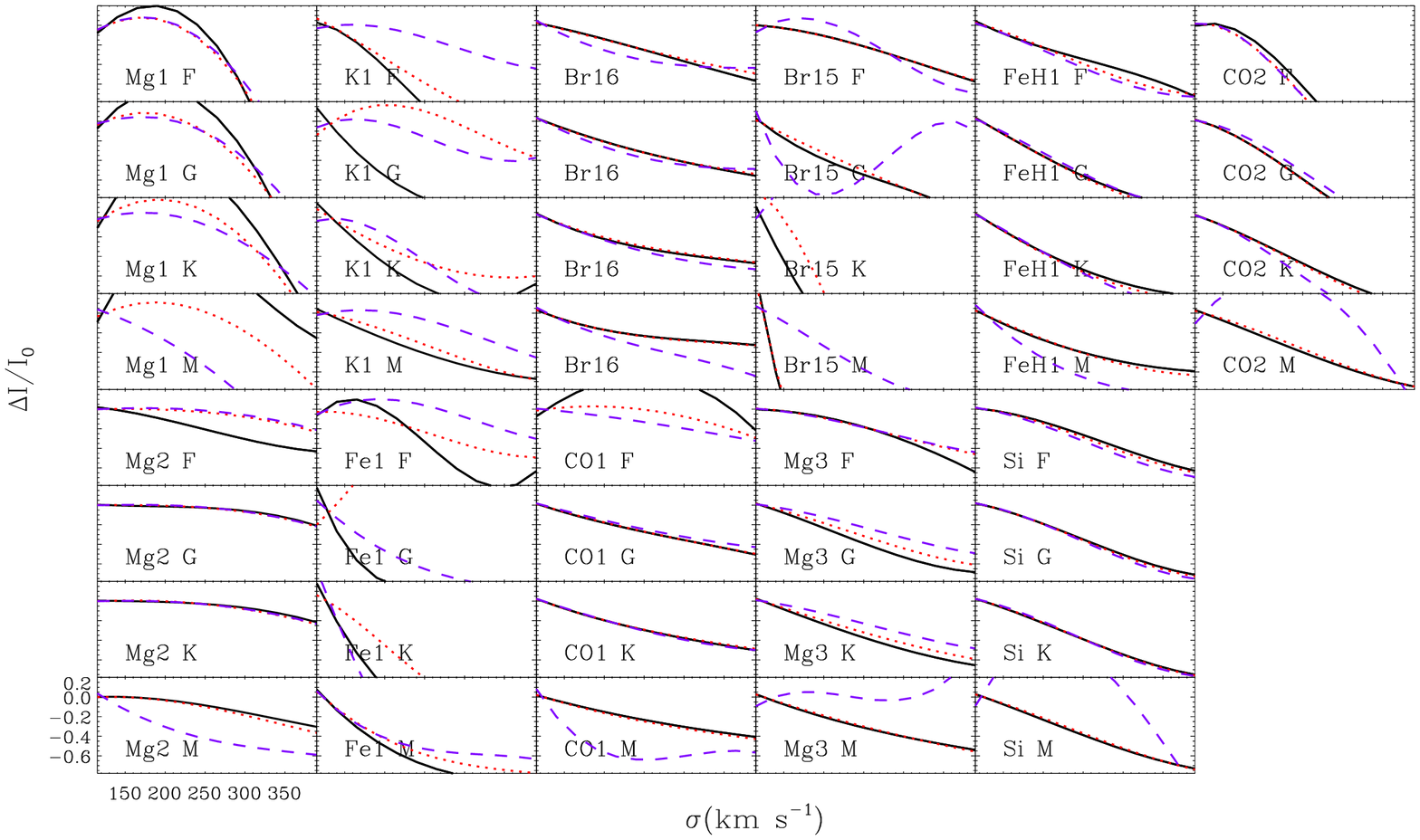}
\includegraphics[angle=90,width=\textwidth]{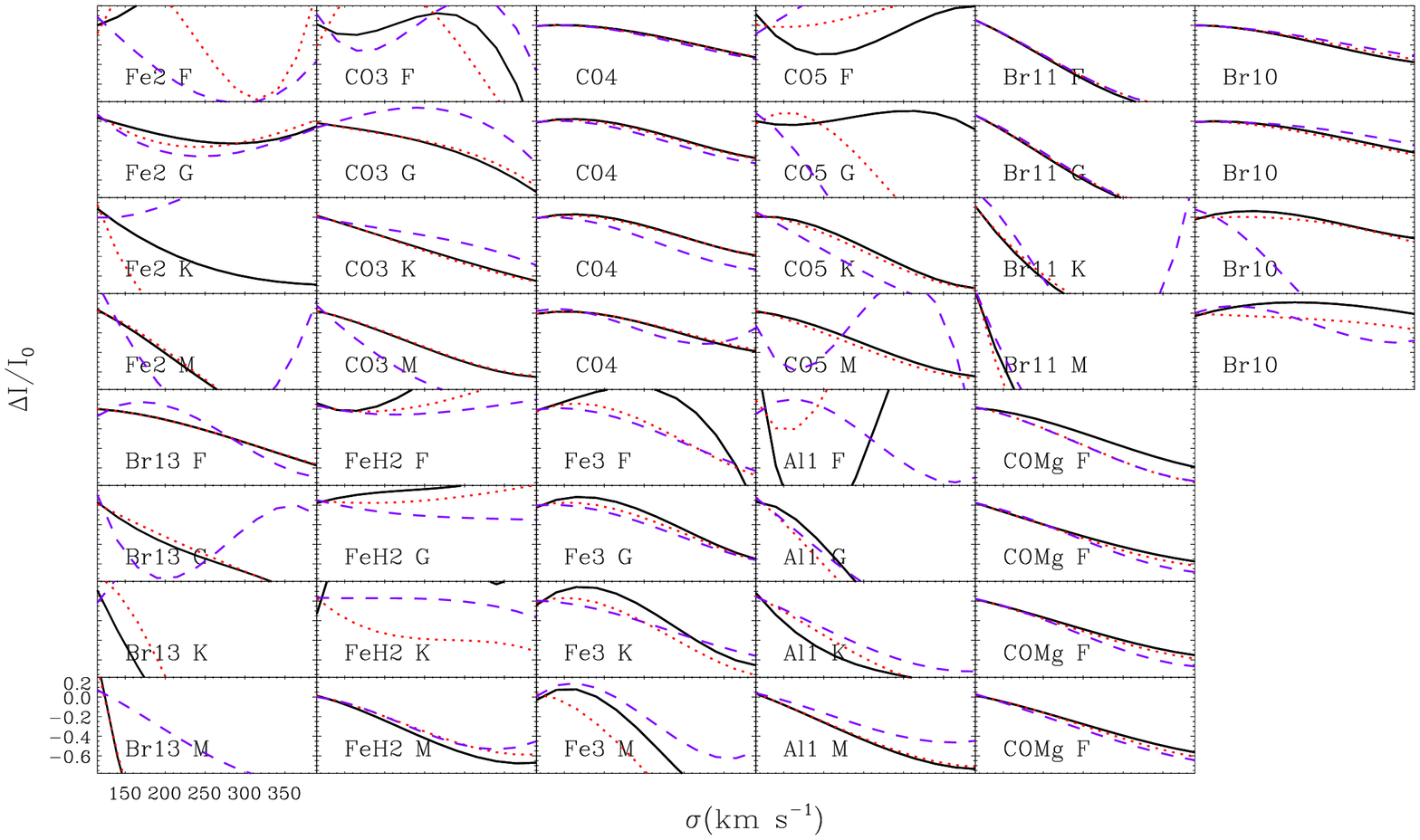}
\caption{Sensitivity of the indices in the H atmospheric window to 
the velocity dispersion broadening. Fig.\,\ref{fig:Y-sigma} gives
details.}\label{fig:H-sigma}
\end{figure*}

\begin{figure*}
\includegraphics[angle=90,width=\textwidth]{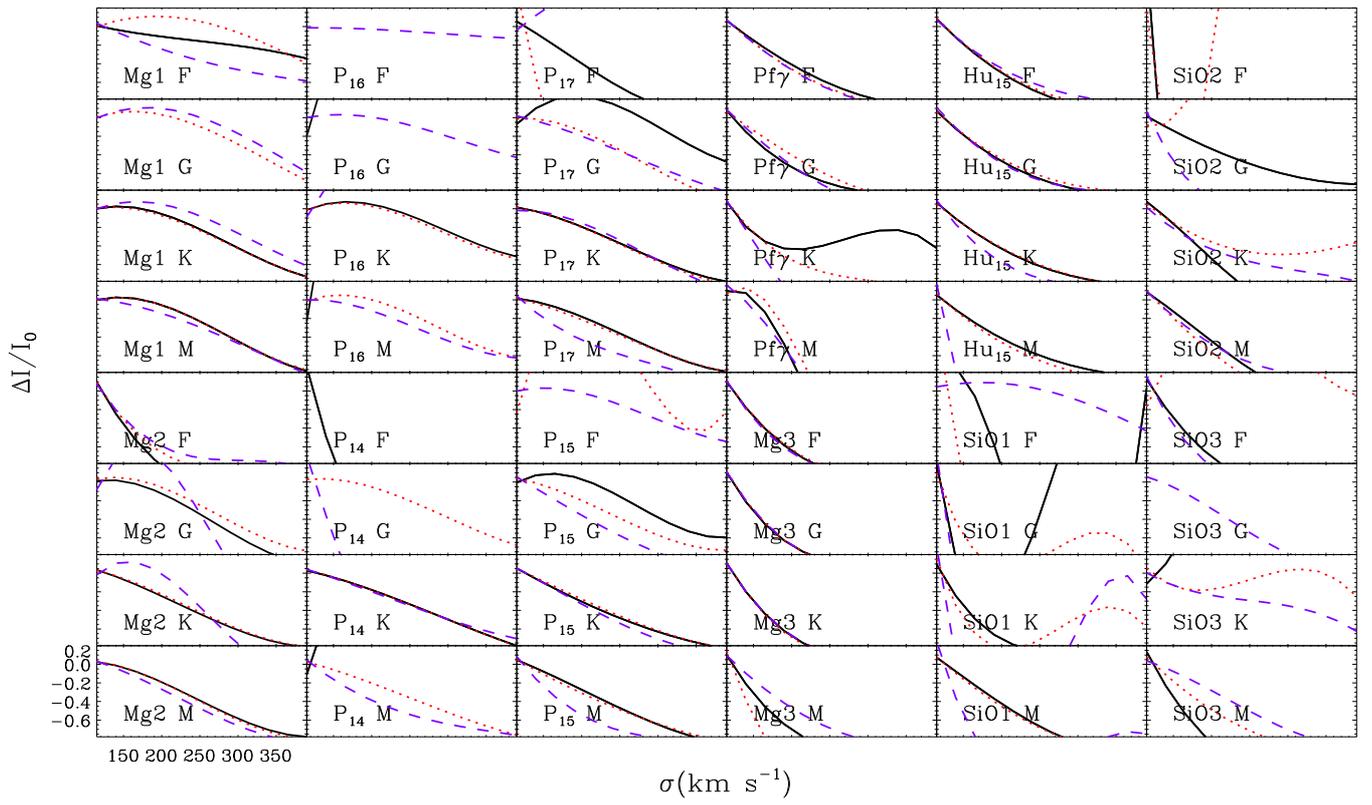}
\caption{Sensitivity of the indices in the L atmospheric window to 
the velocity dispersion broadening. Fig.\,\ref{fig:Y-sigma} gives
details.} \label{fig:L-sigma}
\end{figure*}

\clearpage
\section{Behavior of the spectral indices}\label{app:EWs}

\begin{figure*}
\includegraphics[width=9truecm,angle=0]{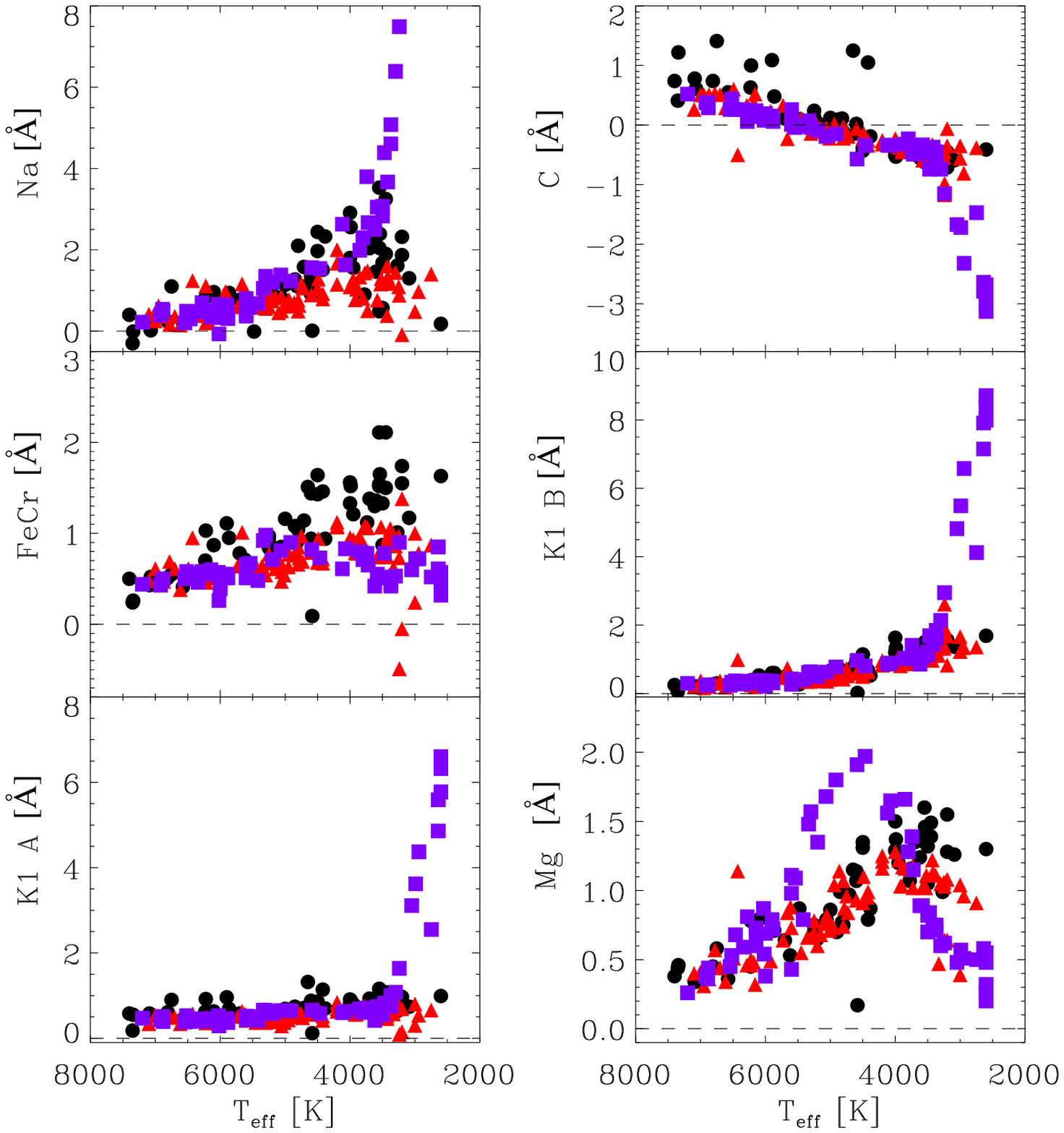}
\includegraphics[width=9truecm,angle=0]{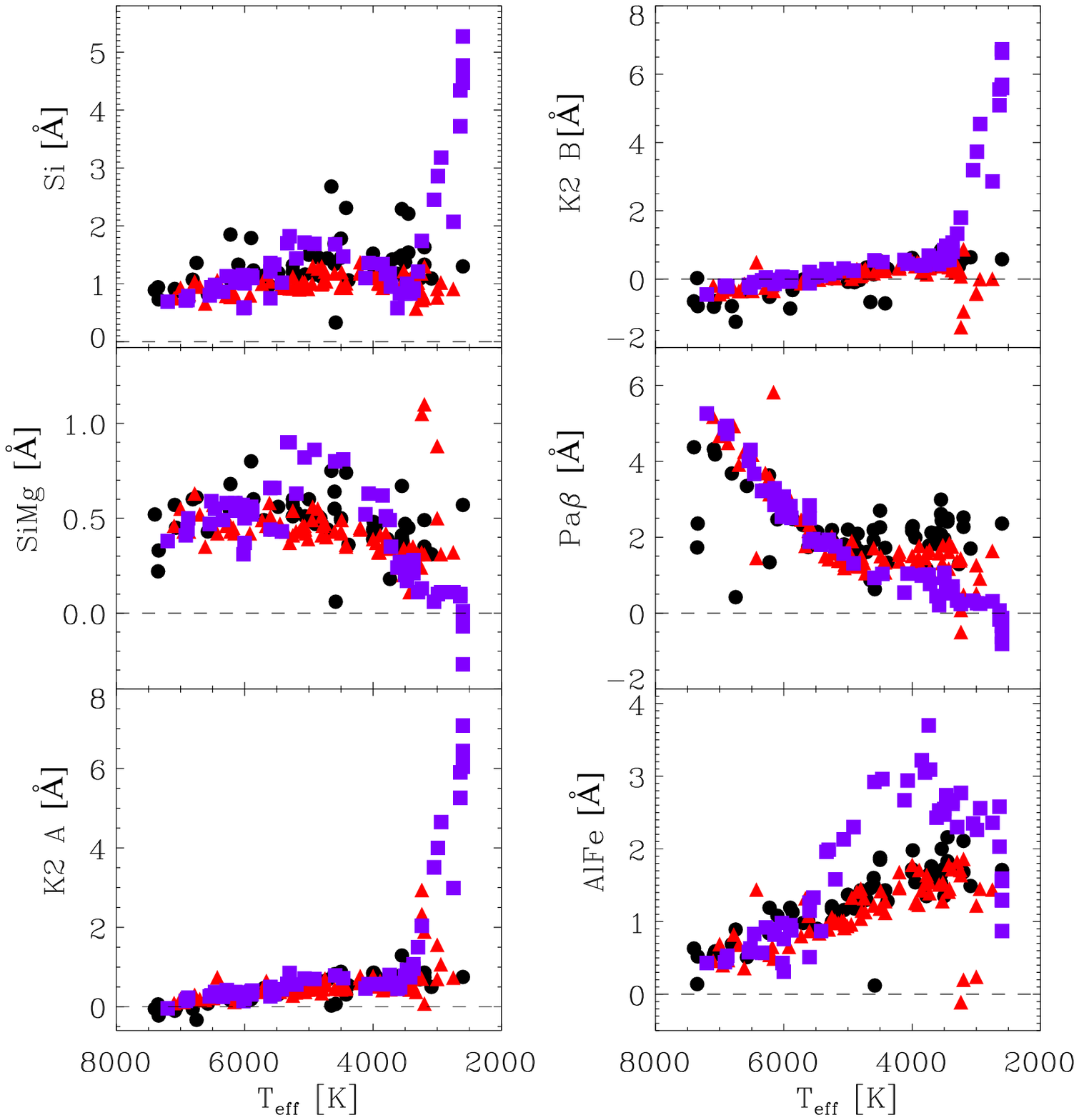}
\caption{Index measurements in the J atmospheric window as 
functions of effective temperature. For details see Fig.\,\ref{fig:Ind_Spt_Y}.}
\label{fig:Ind_Spt_J}
\end{figure*}

\begin{figure*}
\includegraphics[width=9truecm,angle=0]{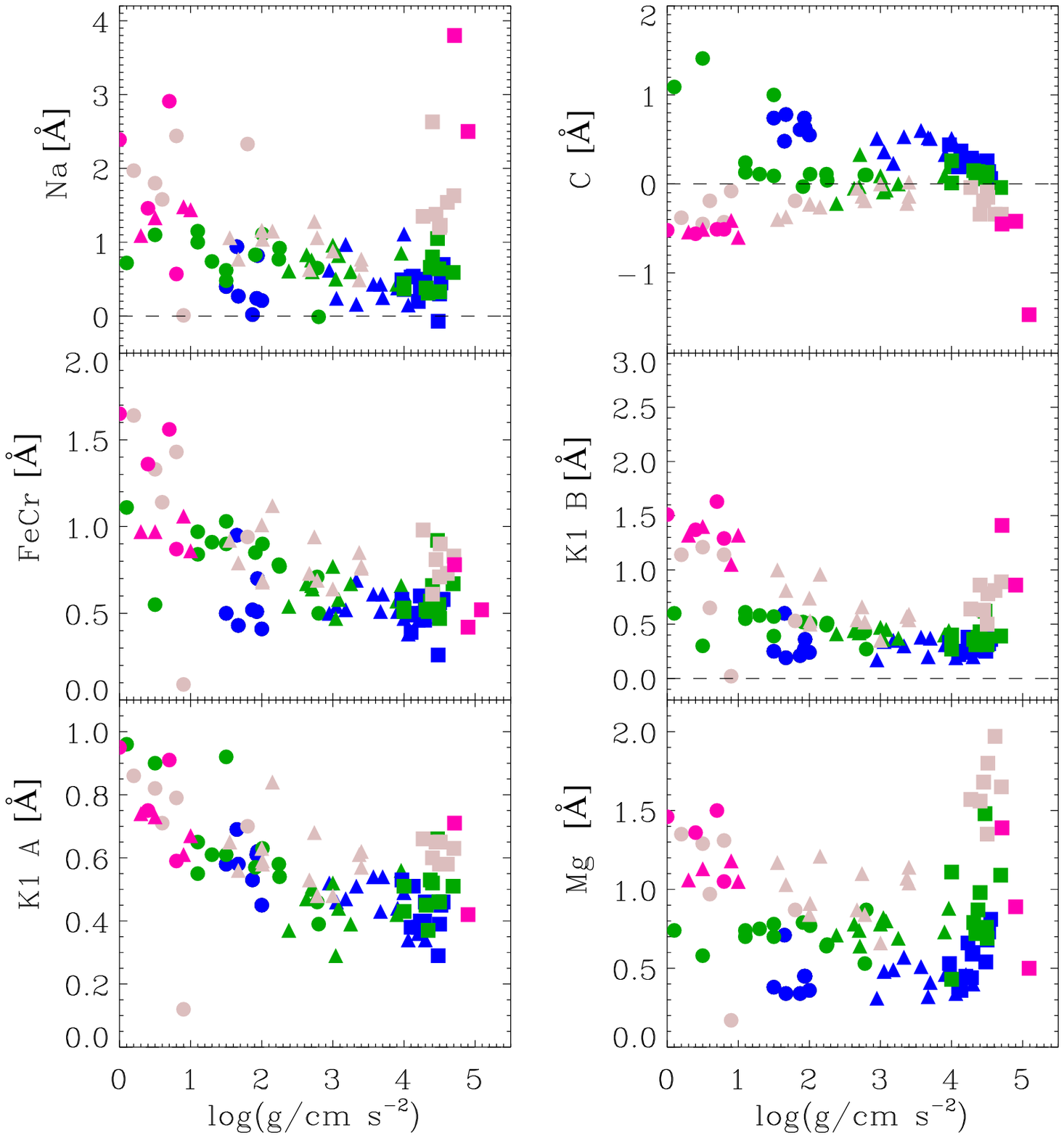}
\includegraphics[width=9truecm,angle=0]{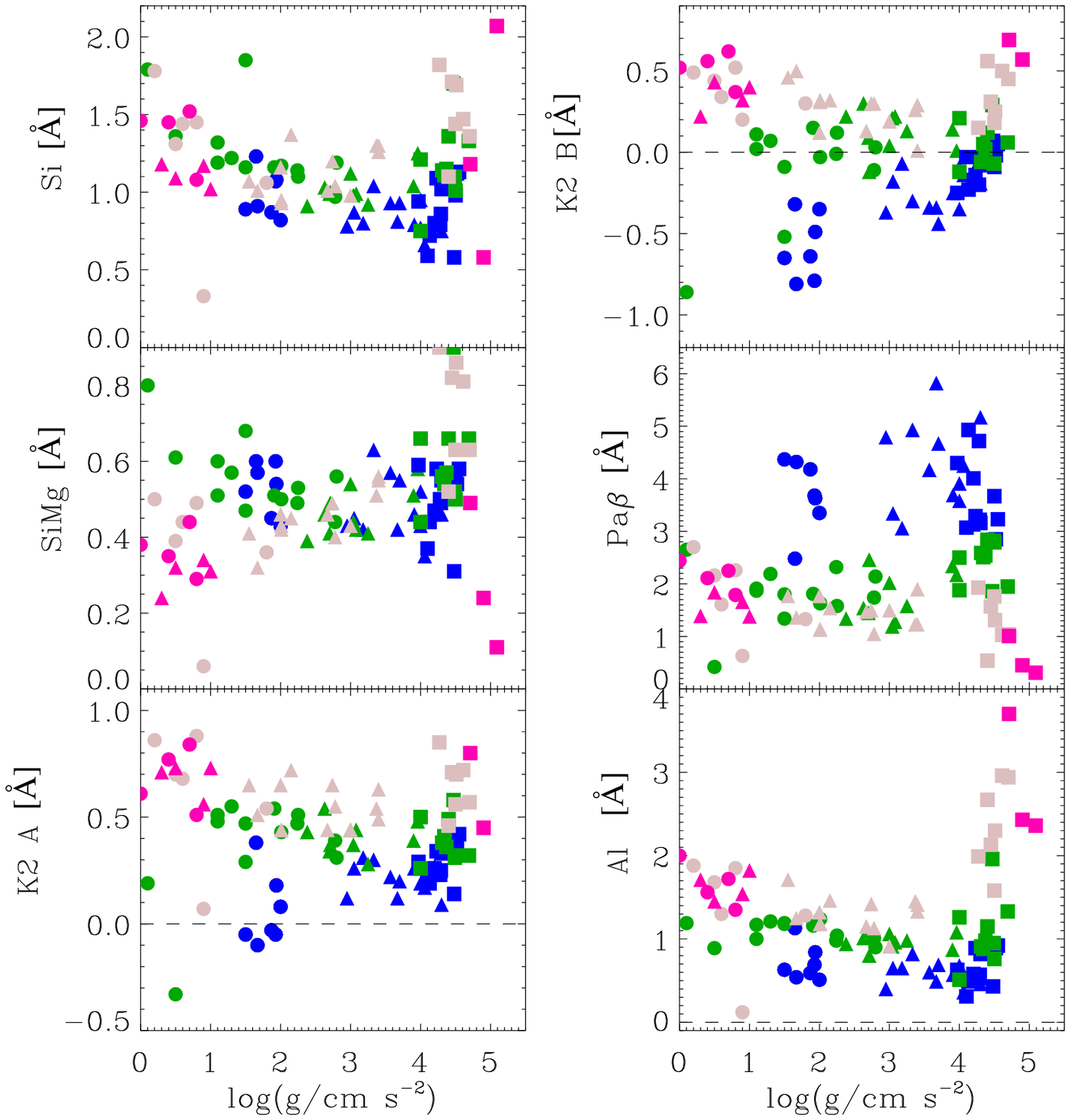}\\
\includegraphics[width=9truecm,angle=0]{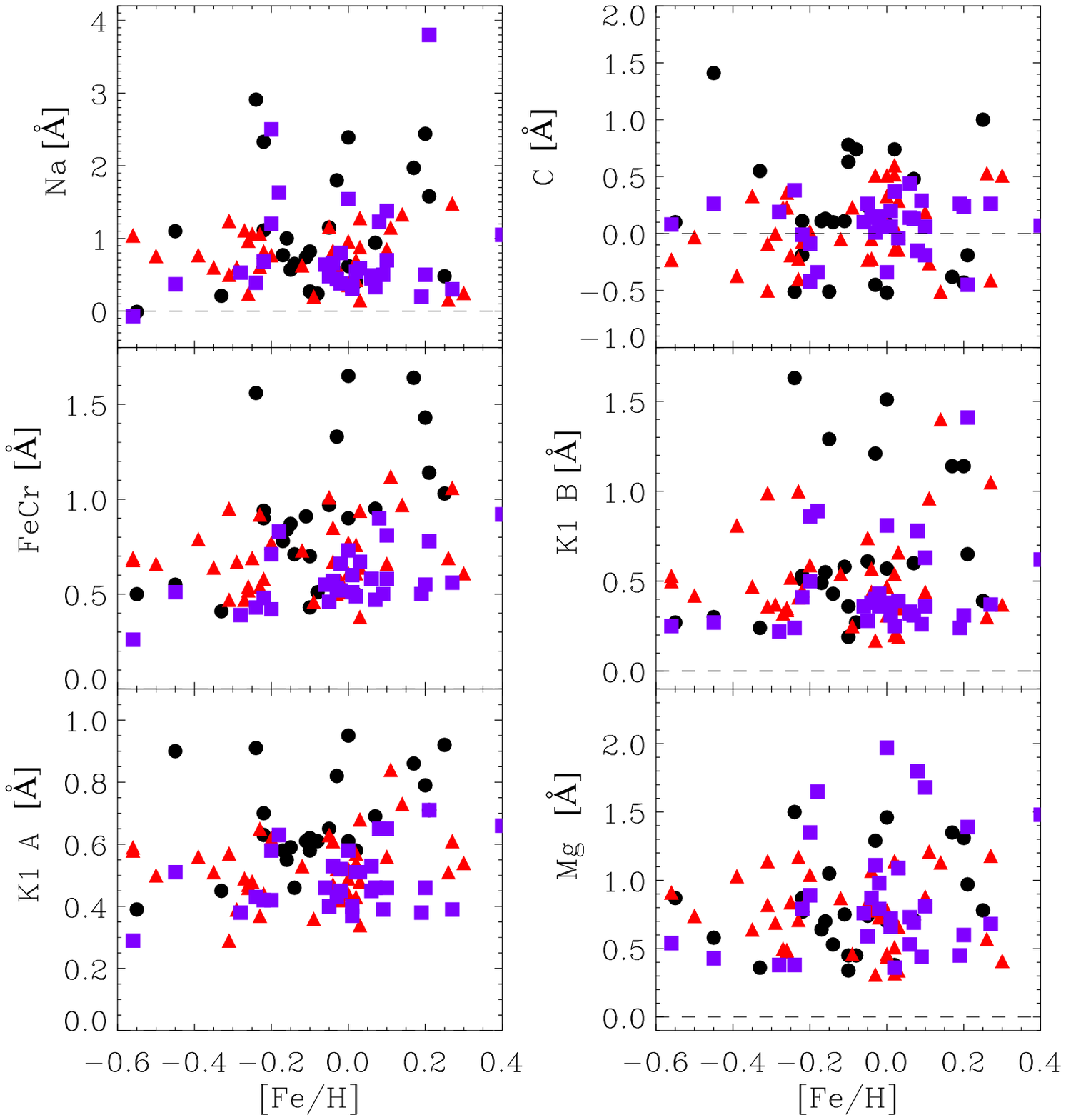}
\includegraphics[width=9truecm,angle=0]{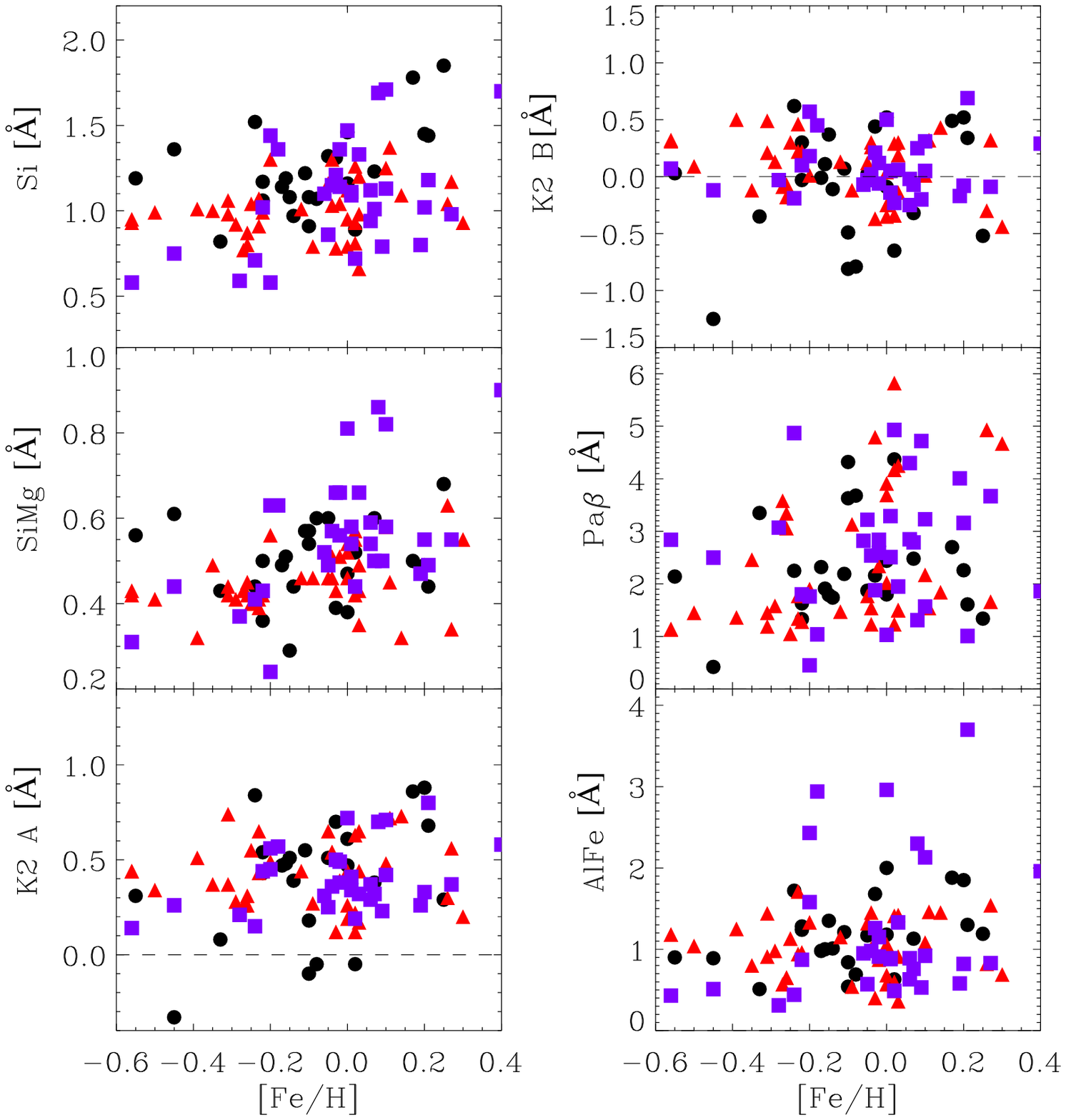}
\caption{Index measurements in the J atmospheric window as functions
  of surface gravity (top panels) and metallicity (bottom
  panels). For details see Fig.\,\ref{fig:Ind_Gr_Y}.}\label{fig:Ind_Gr_J}
\end{figure*}

\begin{figure*}
\includegraphics[width=9truecm,angle=0]{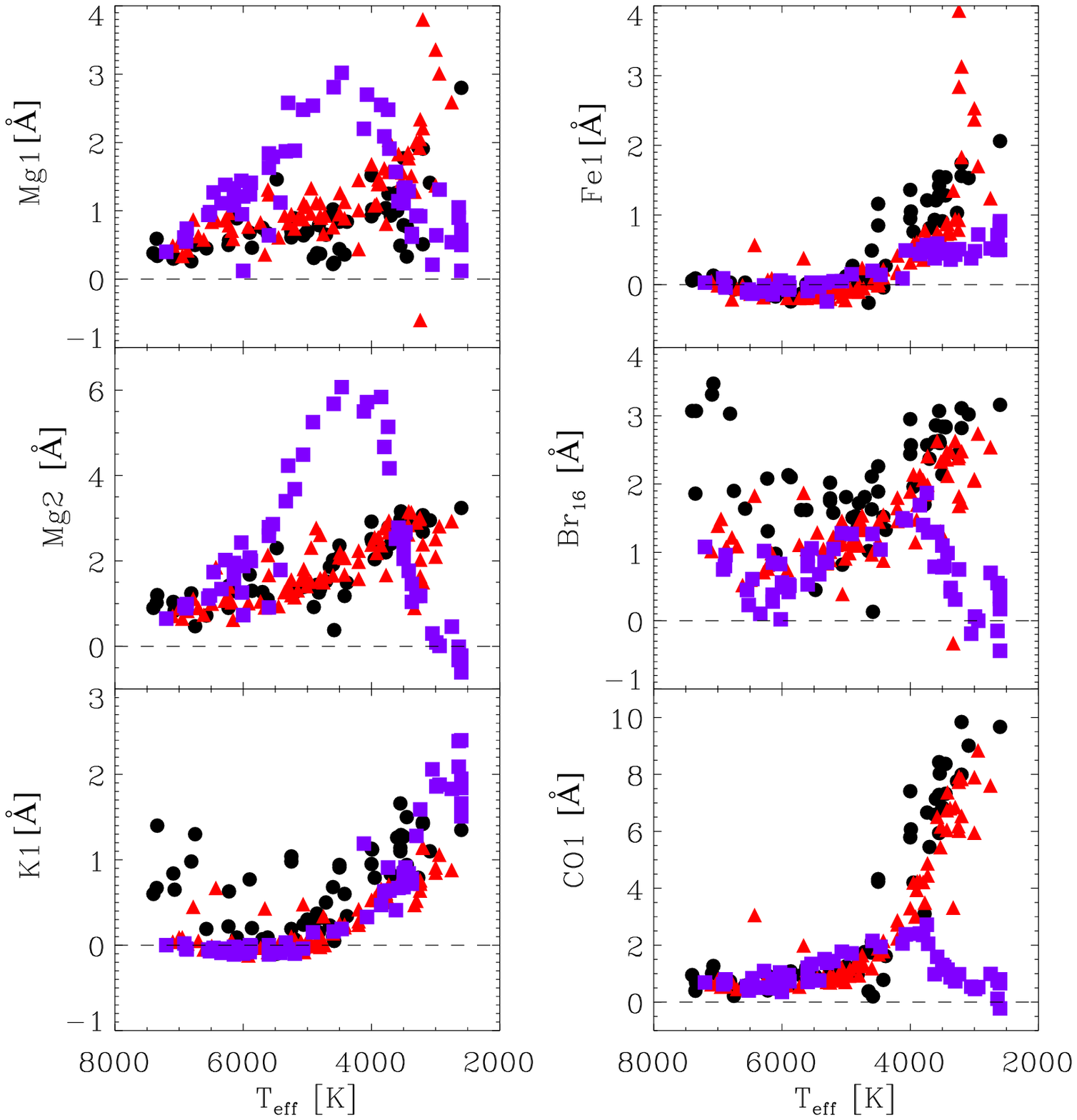}
\includegraphics[width=9truecm,angle=0]{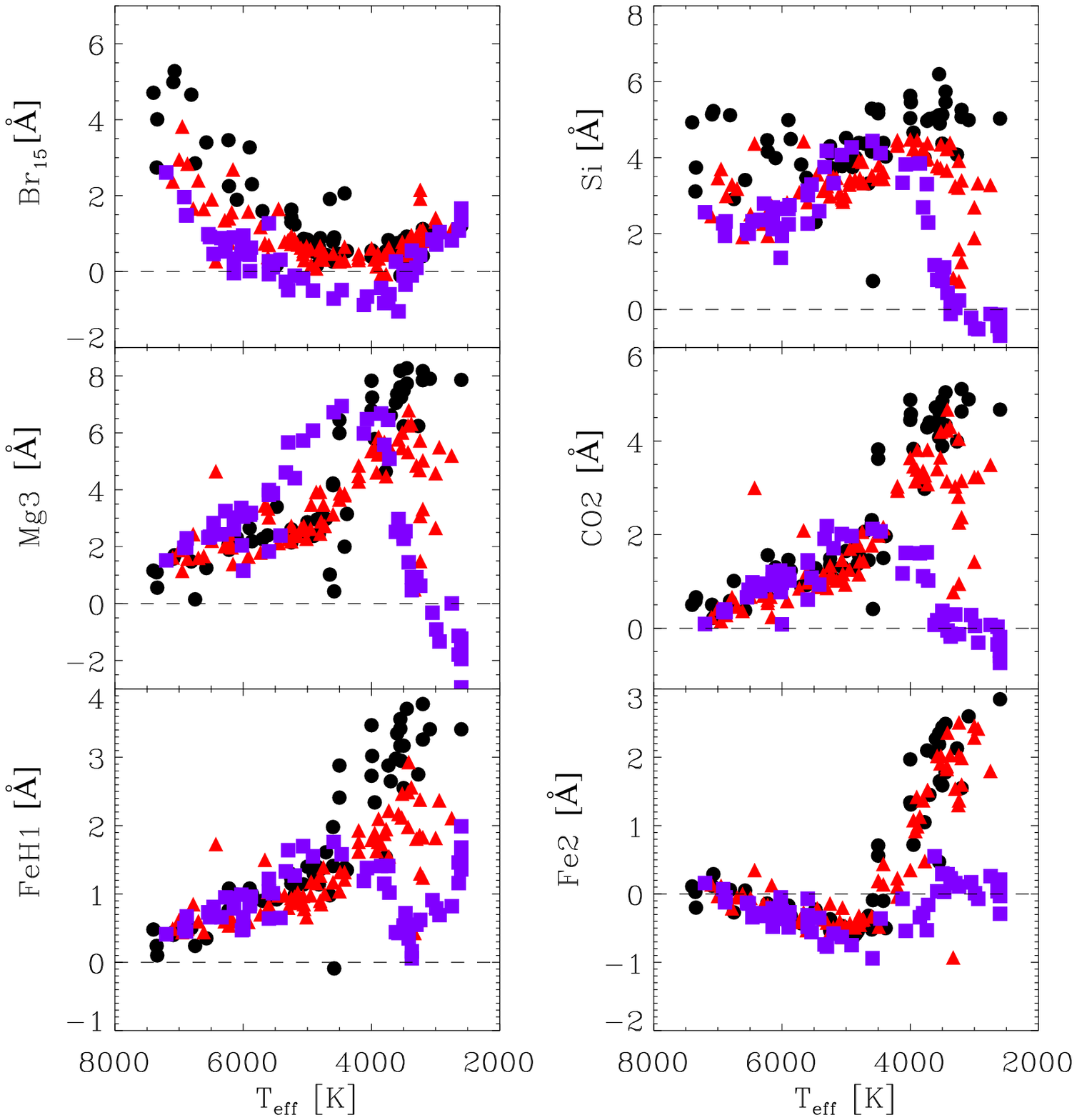}\\
\includegraphics[width=9truecm,angle=0]{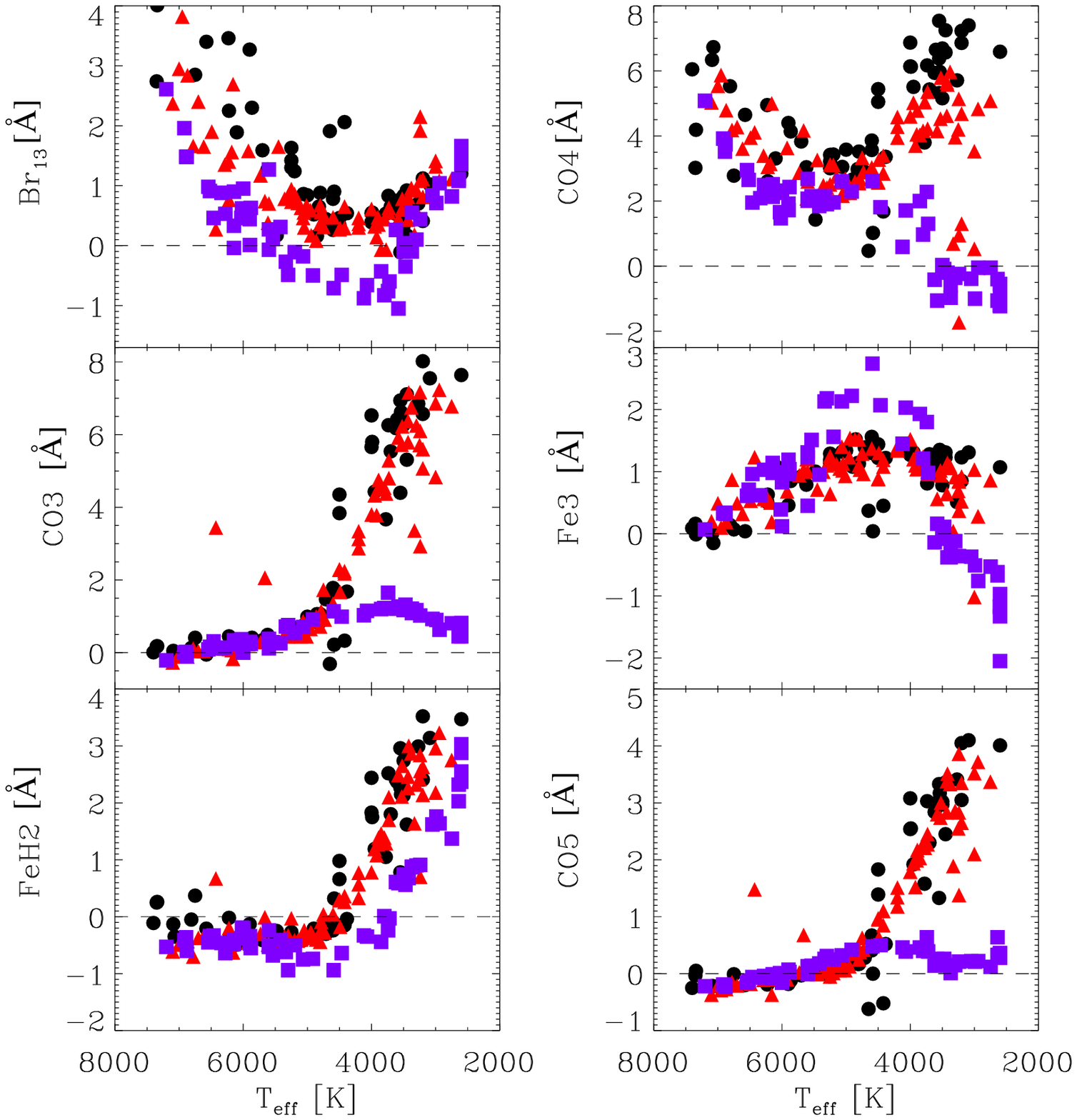}
\includegraphics[width=9truecm,angle=0]{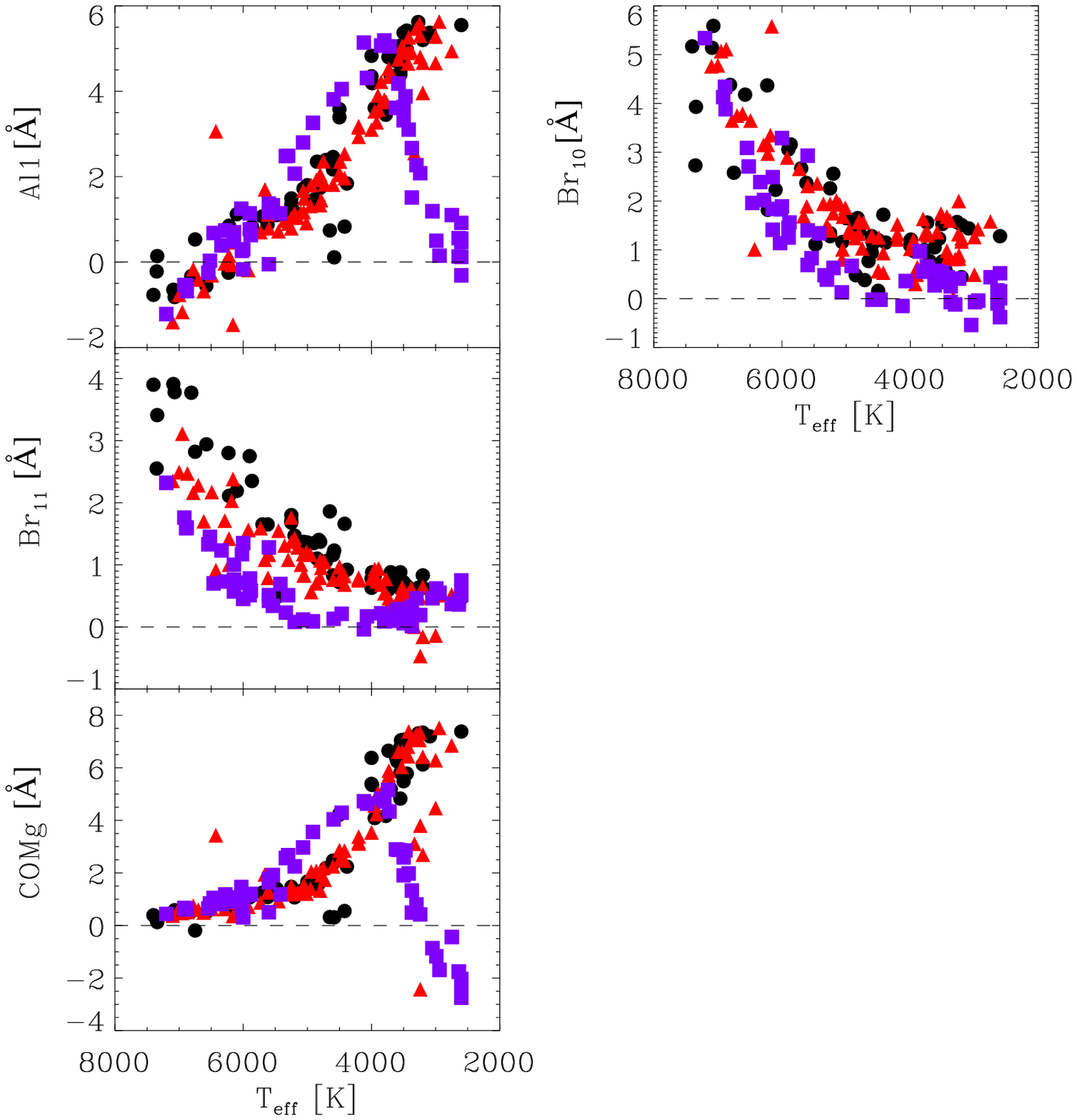}
\caption{Index measurements in the H atmospheric window as functions
  of effective temperature. For details see
  Fig.\,\ref{fig:Ind_Spt_Y}.}\label{fig:Ind_Tef_H}
\label{fig:Ind_Spt_H}
\end{figure*}

\begin{figure*}
\includegraphics[width=9truecm,angle=0]{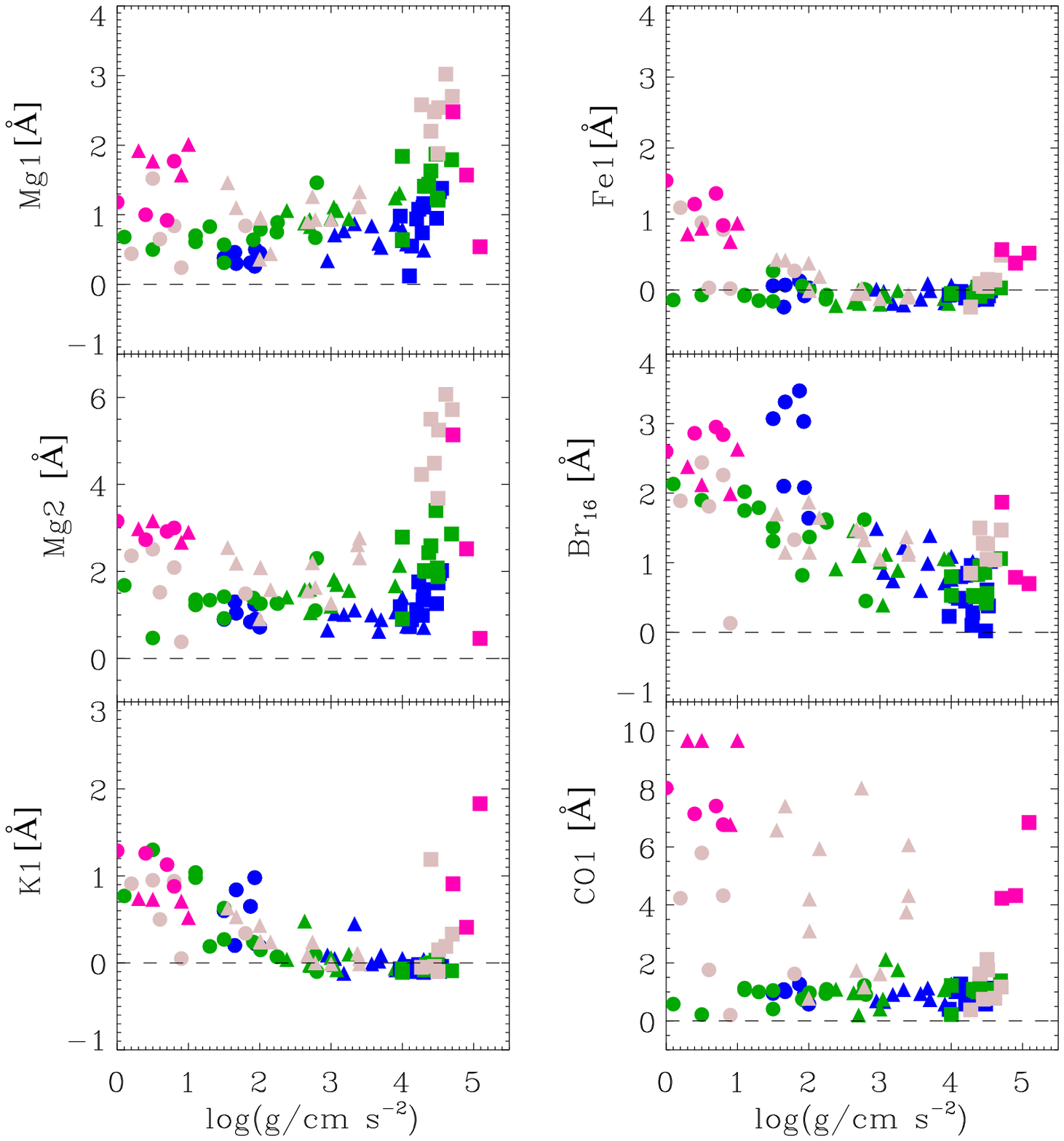}
\includegraphics[width=9truecm,angle=0]{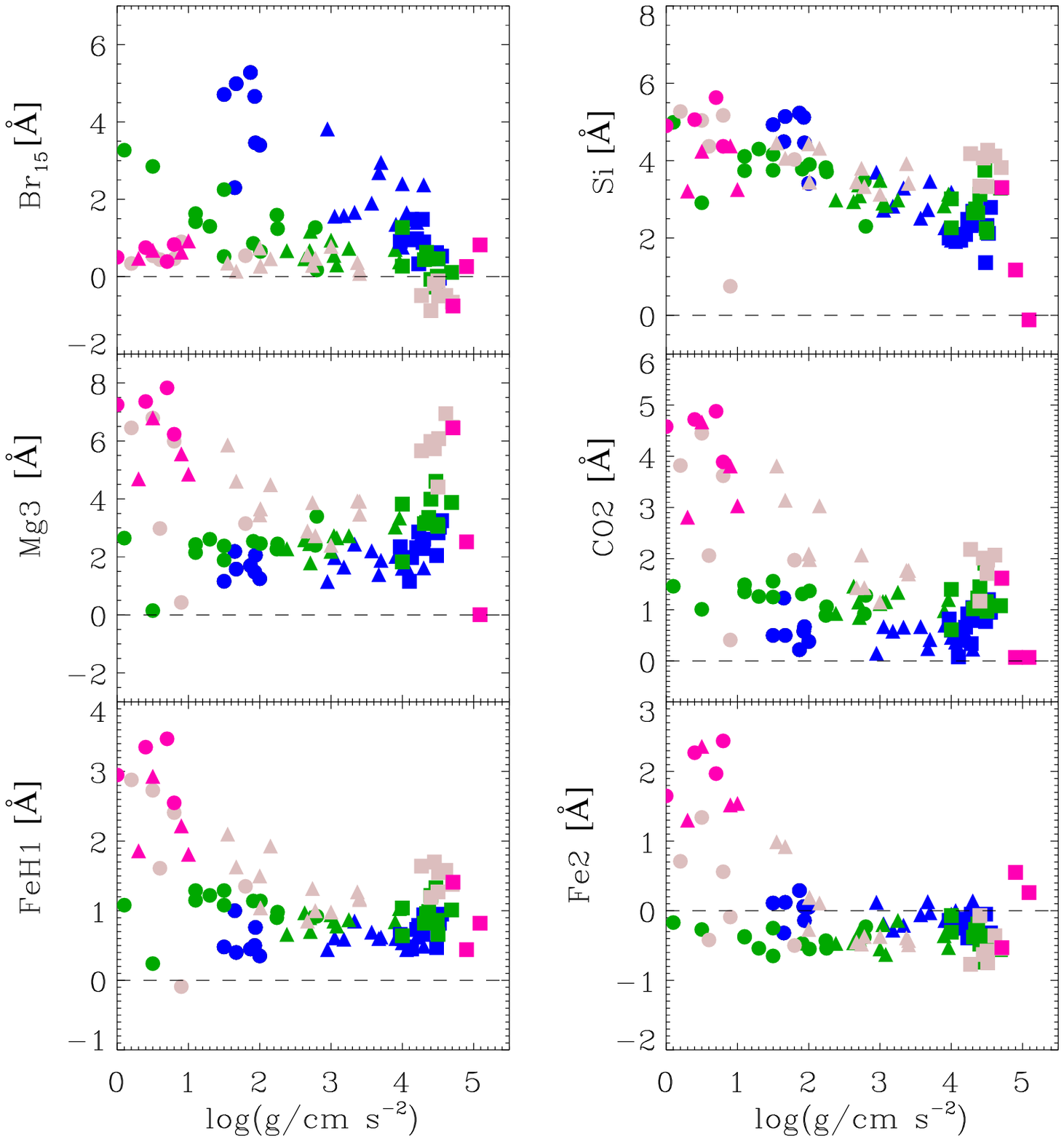}\\
\includegraphics[width=9truecm,angle=0]{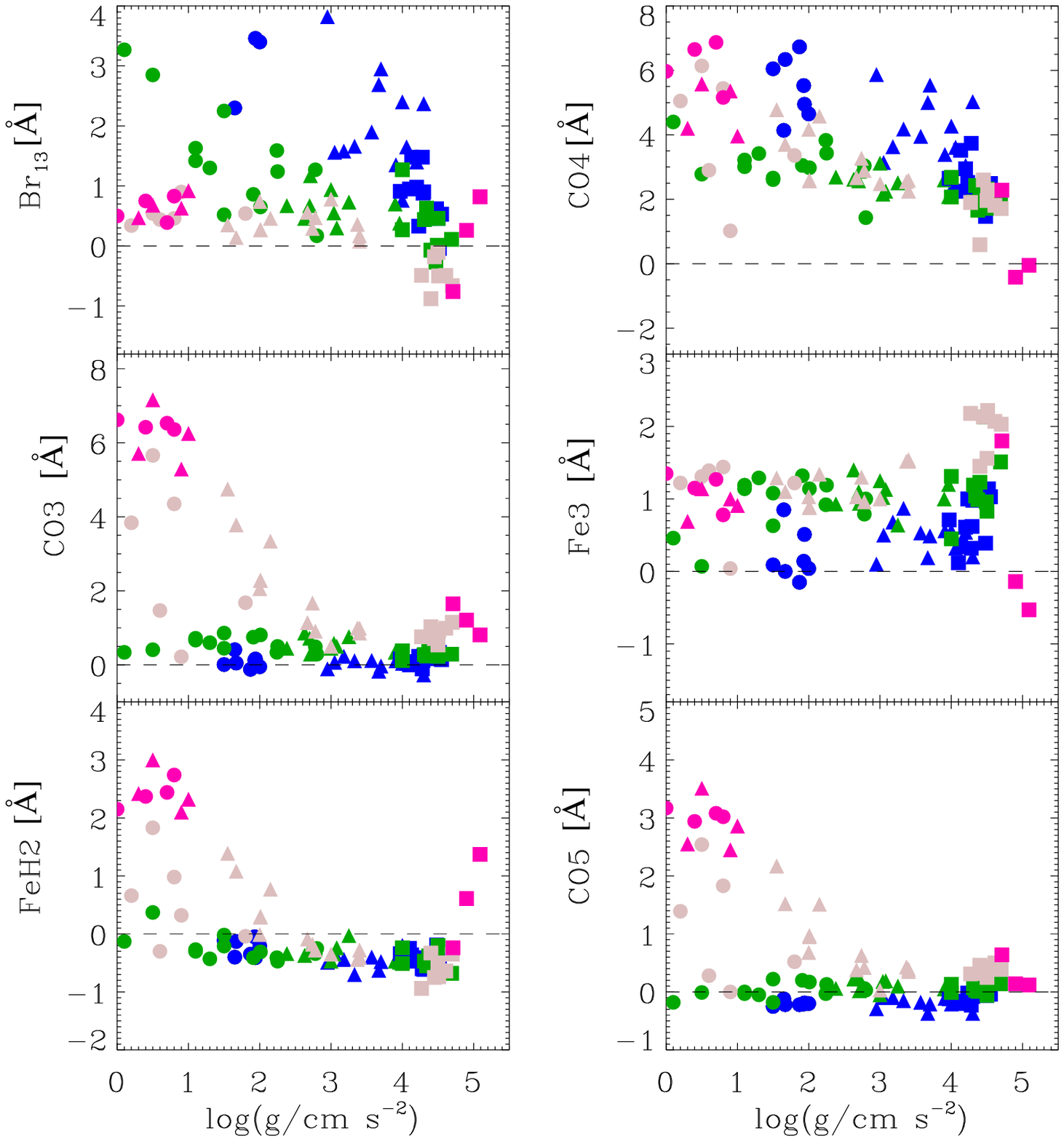}
\includegraphics[width=9truecm,angle=0]{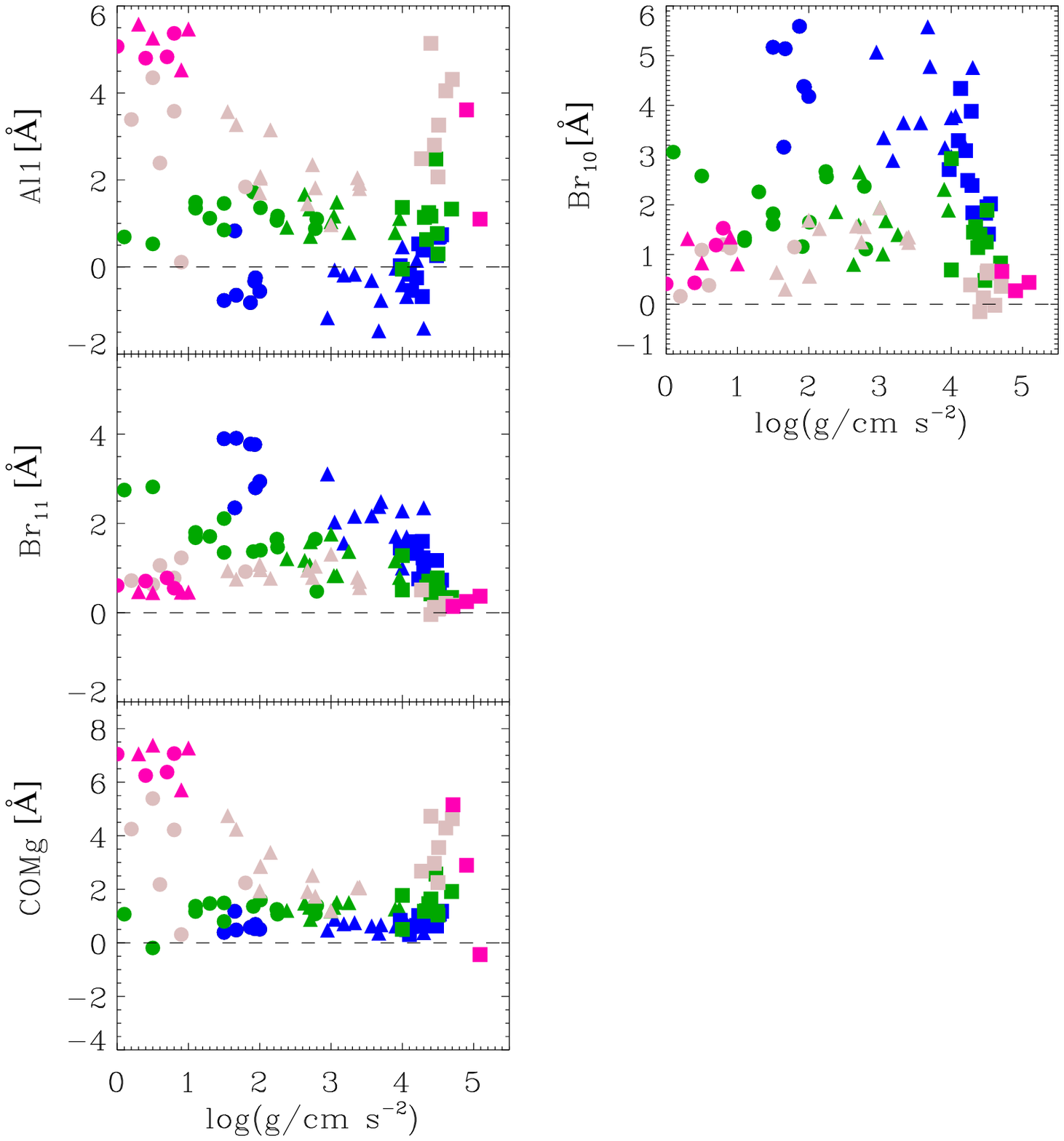}\\
\caption{Index measurements in the H atmospheric window as 
functions of surface gravity. For details see Fig.\,\ref{fig:Ind_Gr_Y}.}
\label{fig:Ind_Gr_H}
\end{figure*}

\begin{figure*}
\includegraphics[width=9truecm,angle=0]{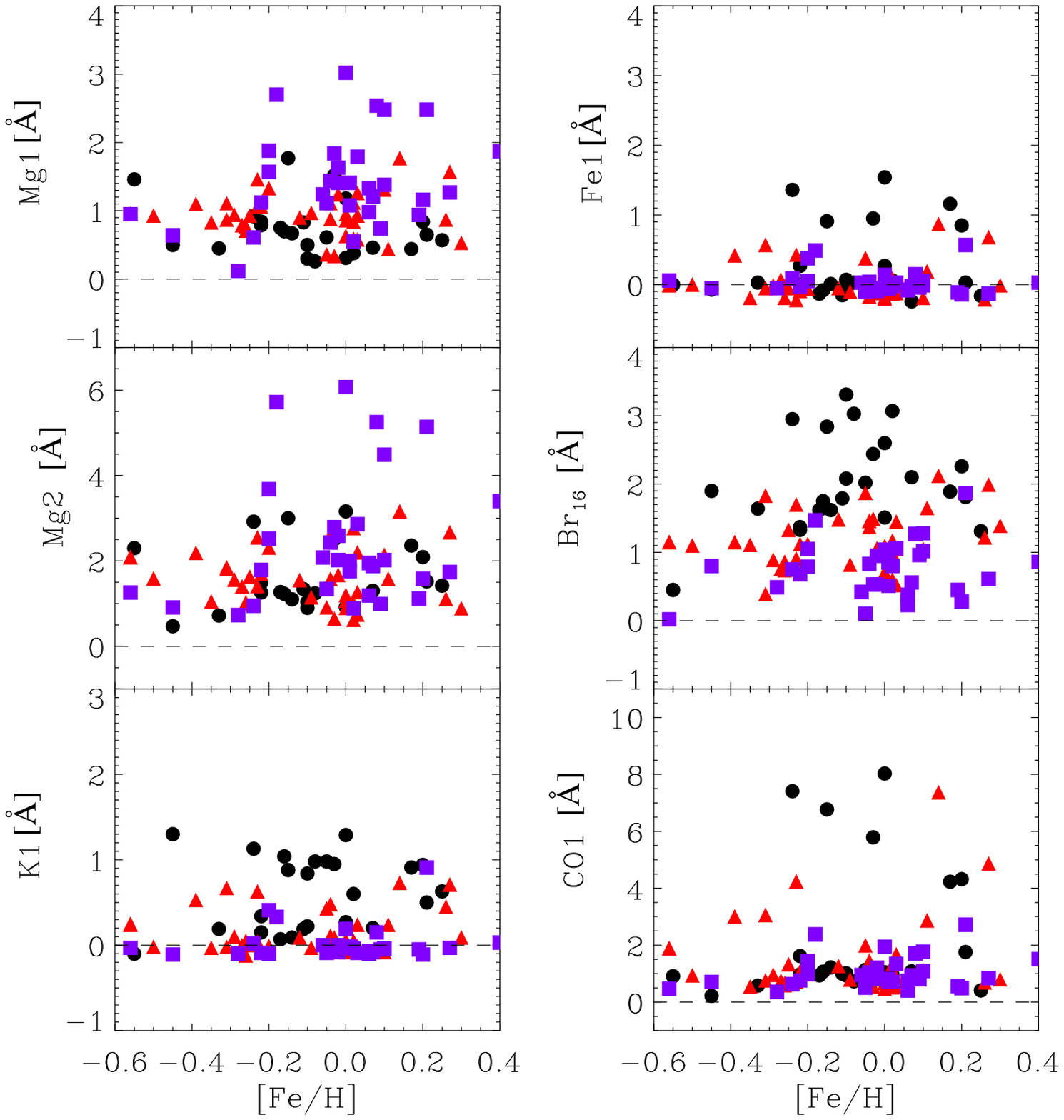}
\includegraphics[width=9truecm,angle=0]{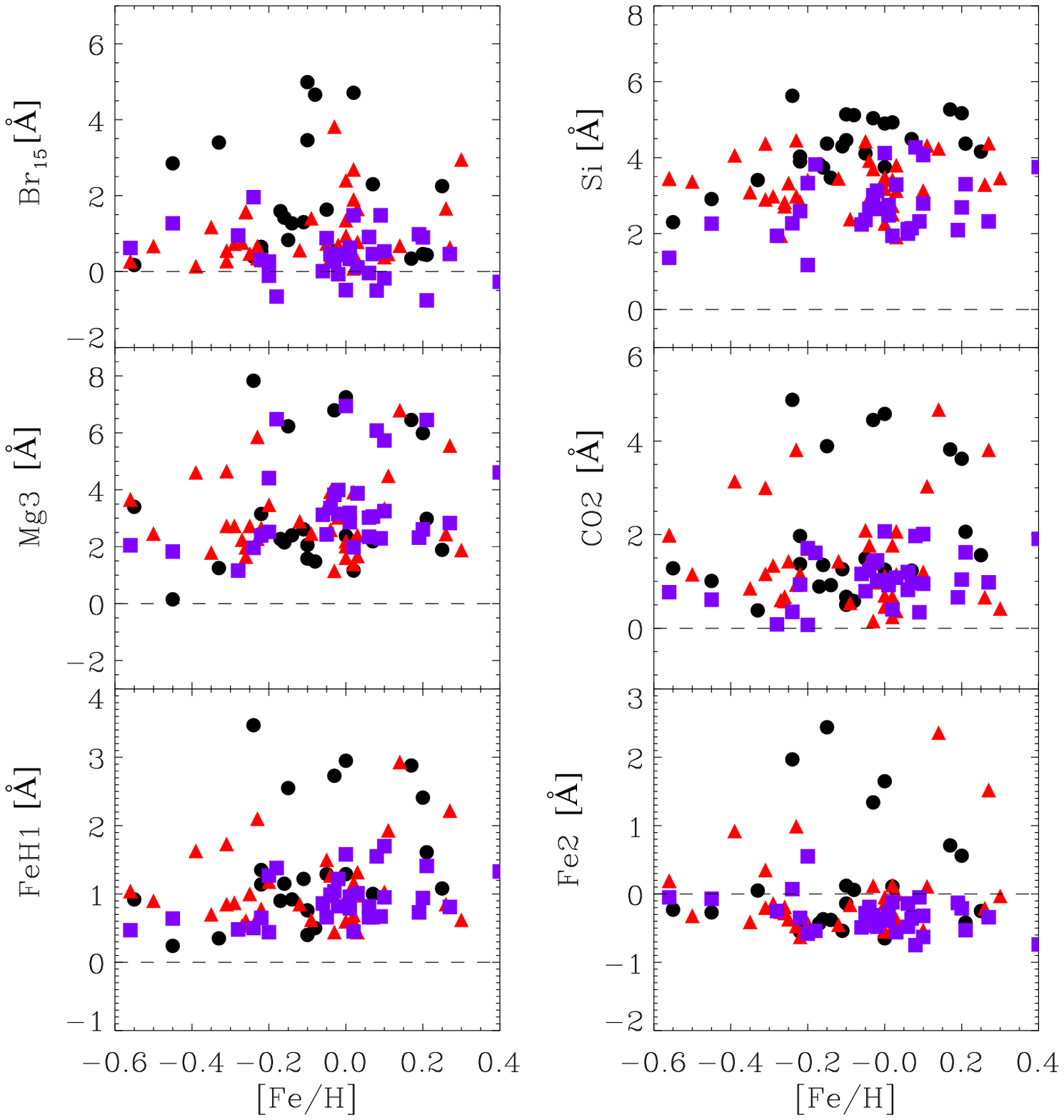}\\
\includegraphics[width=9truecm,angle=0]{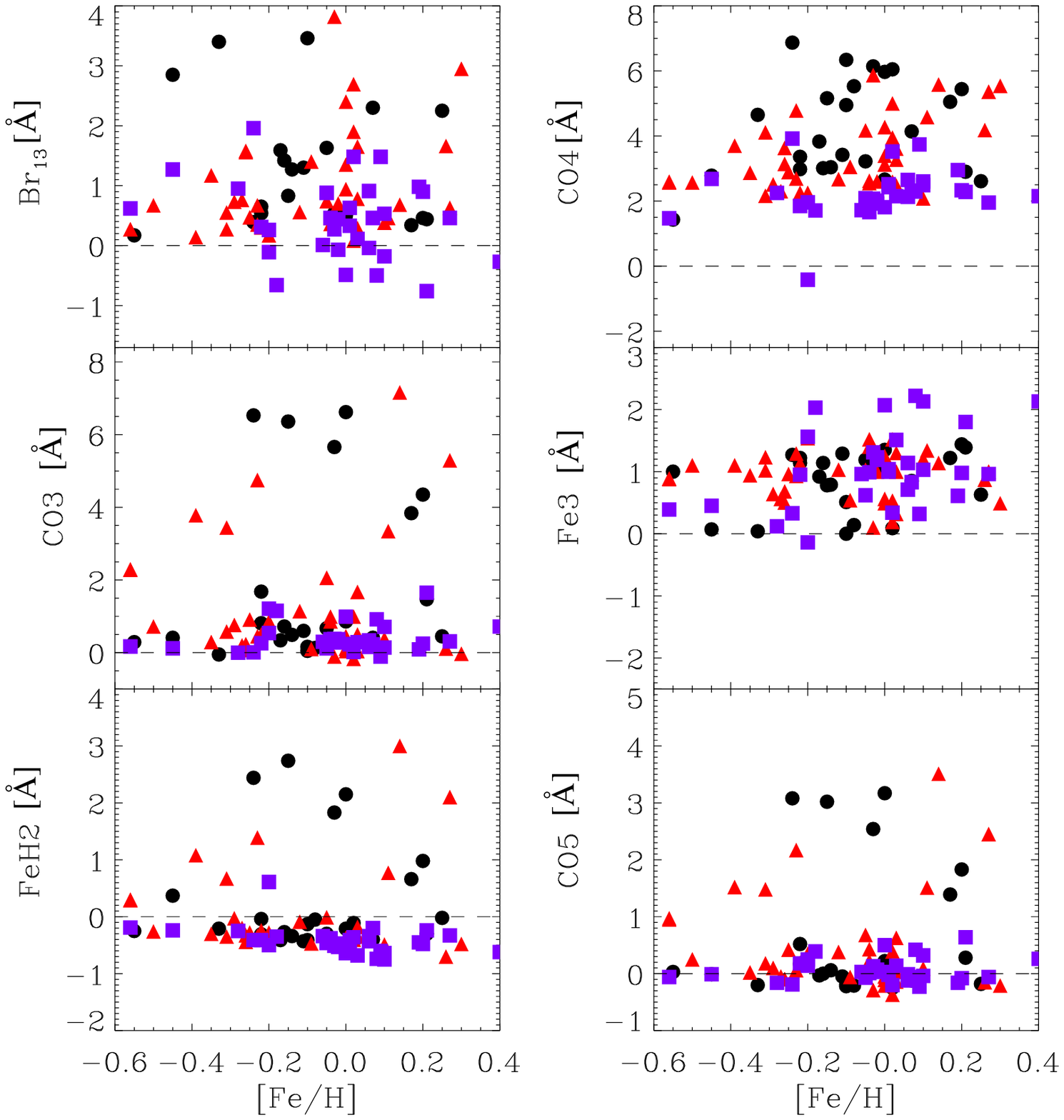}
\includegraphics[width=9truecm,angle=0]{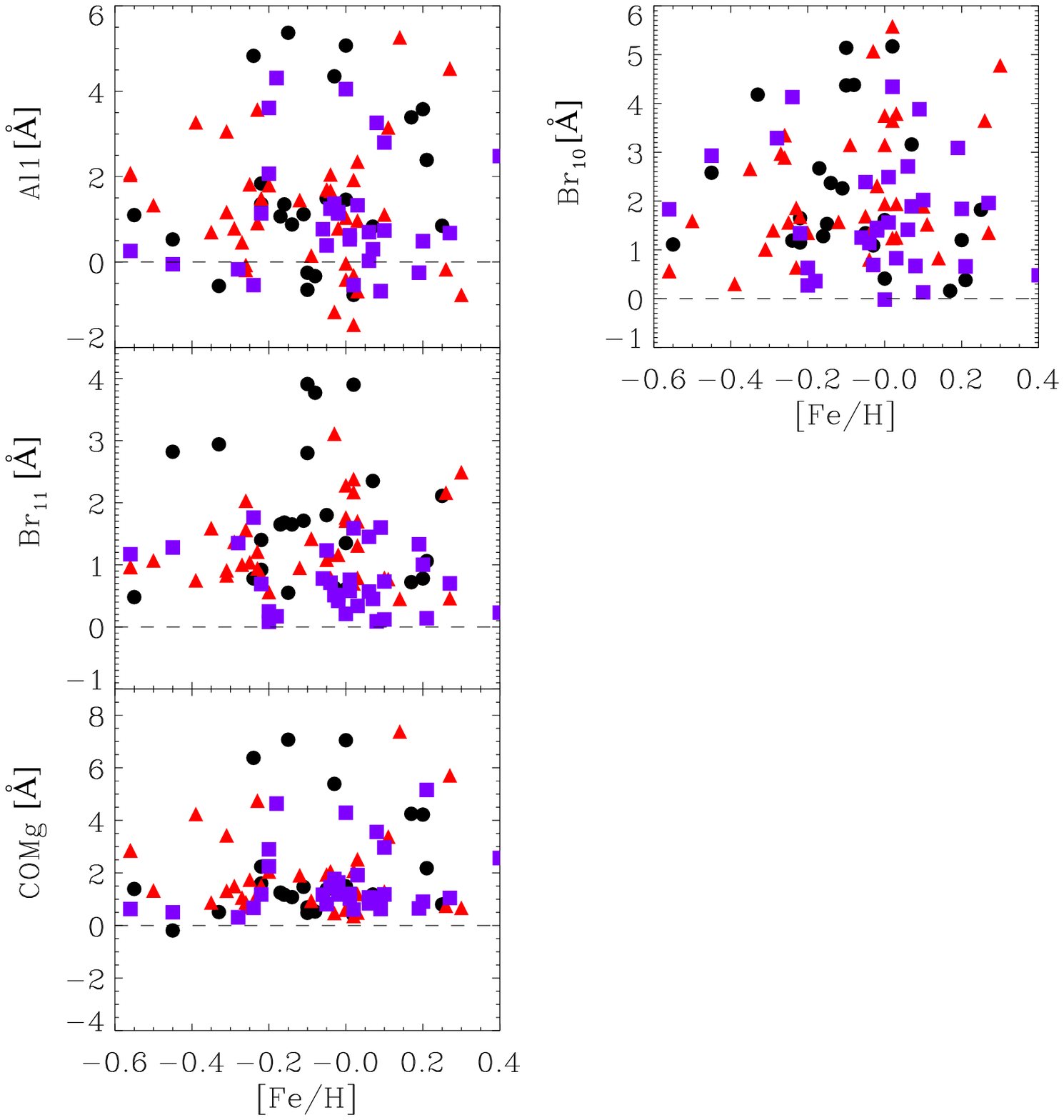}
\caption{Index measurements in the H atmospheric window as 
functions of metallicity. For details see Fig.\,\ref{fig:Ind_Gr_Y}.}
\label{fig:Ind_FeH_H}
\end{figure*}

\begin{figure*}
\includegraphics[width=9truecm,angle=0]{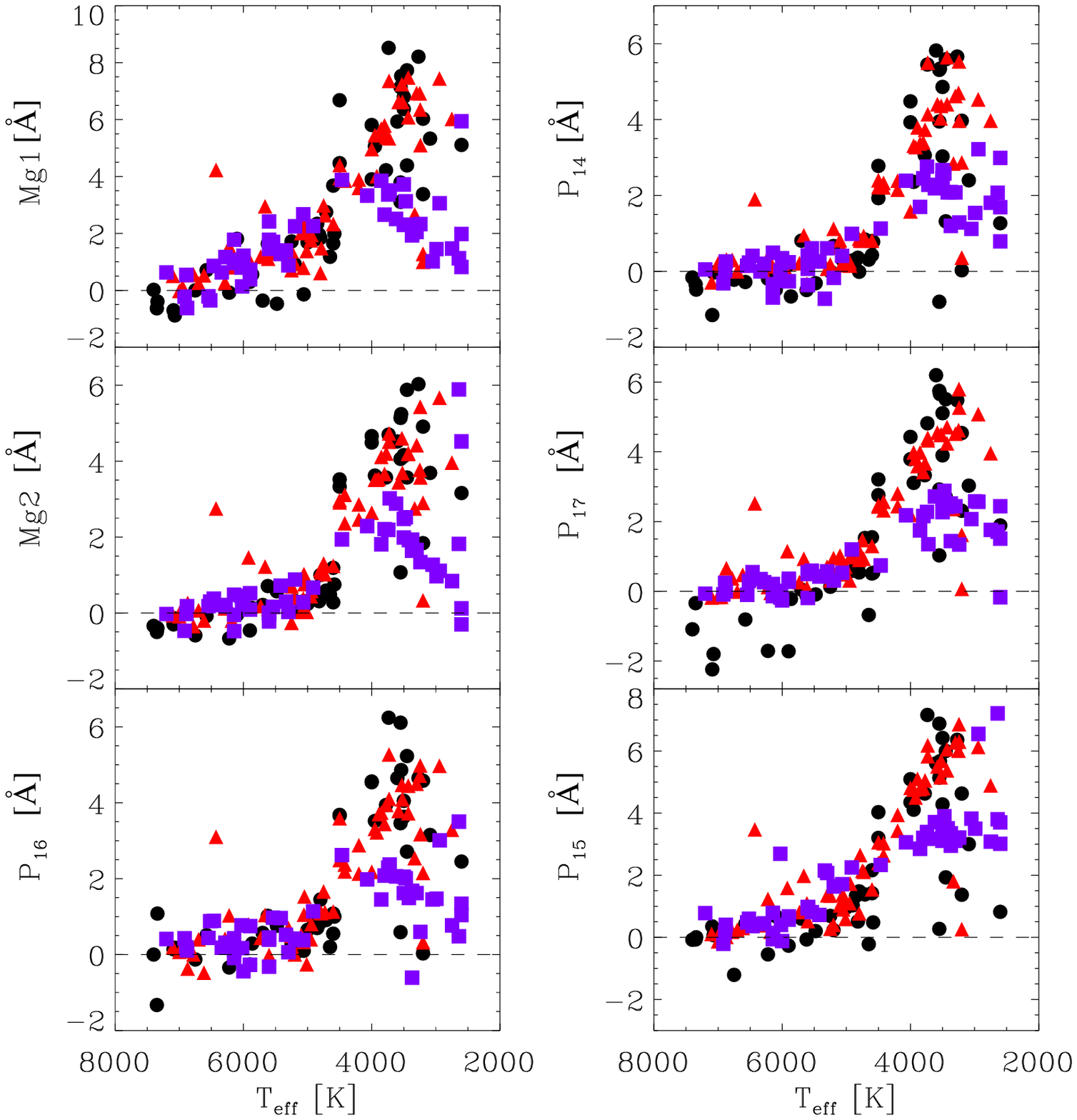}
\includegraphics[width=9truecm,angle=0]{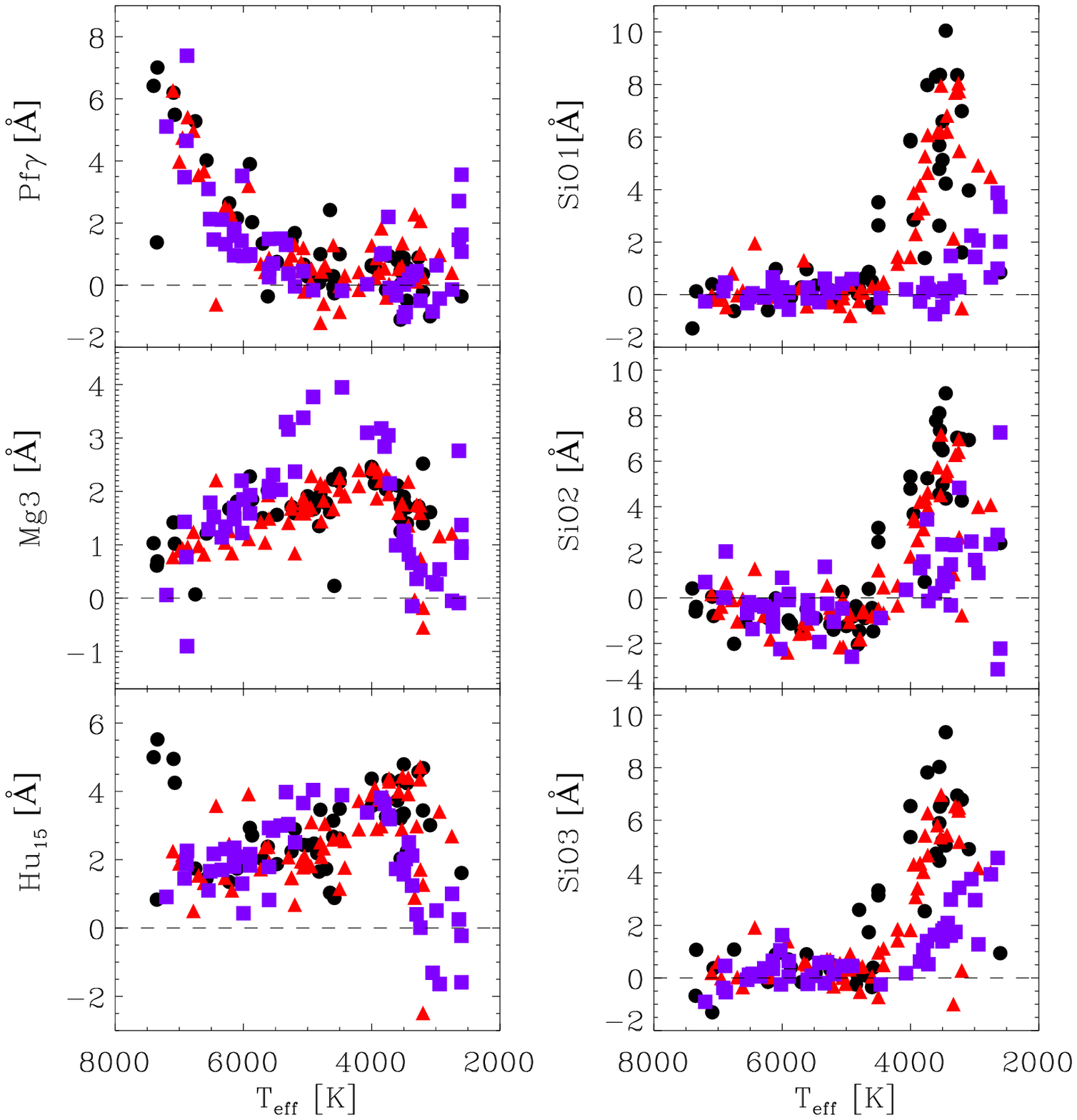}
\caption{Index measurements in the L atmospheric window as 
functions of effective temperature. For details see Fig.\,\ref{fig:Ind_Spt_Y}.}
\label{fig:Ind_Spt_L}
\end{figure*}

\begin{figure*}
\includegraphics[width=9truecm,angle=0]{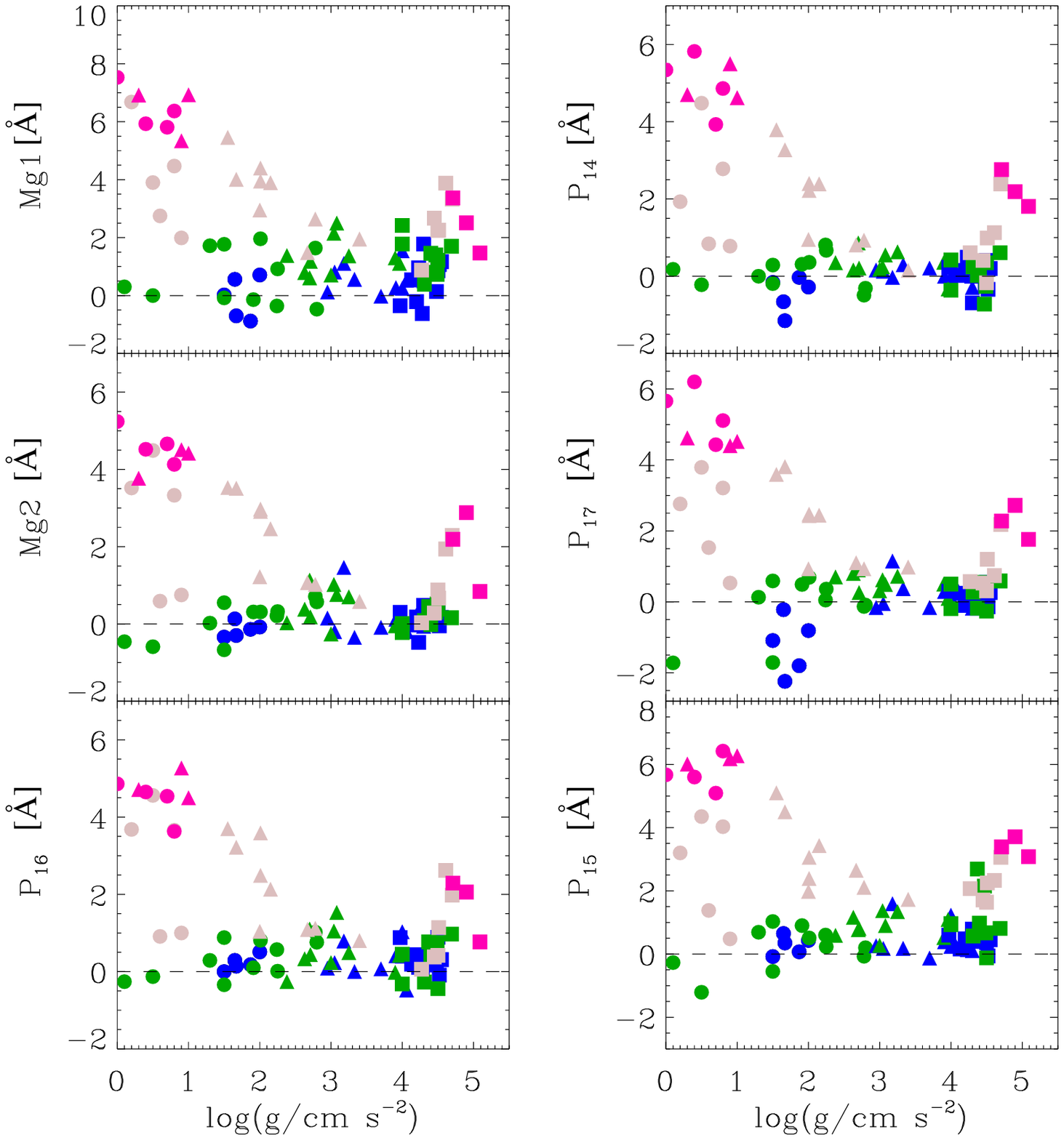}
\includegraphics[width=9truecm,angle=0]{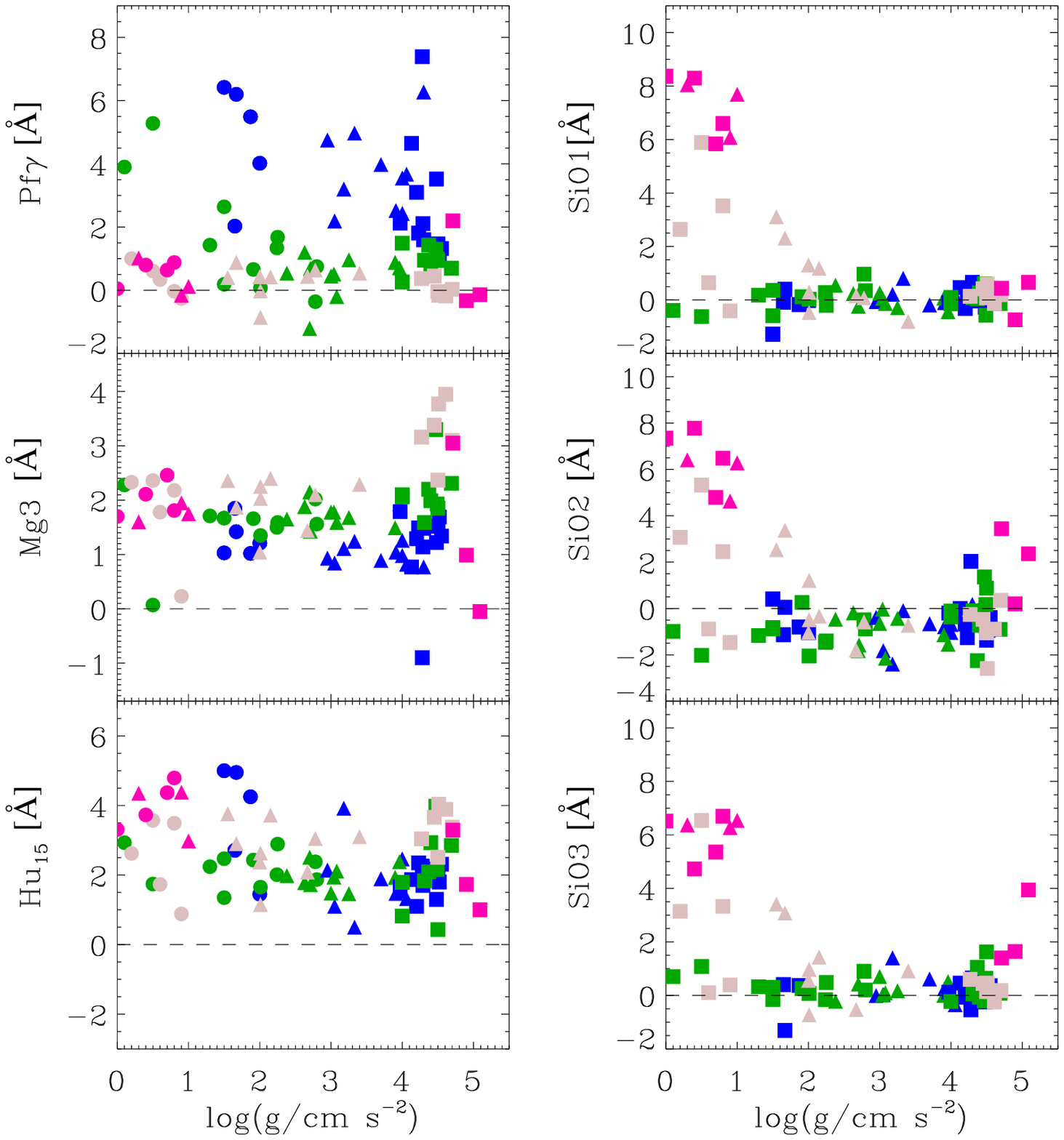}\\
\includegraphics[width=9truecm,angle=0]{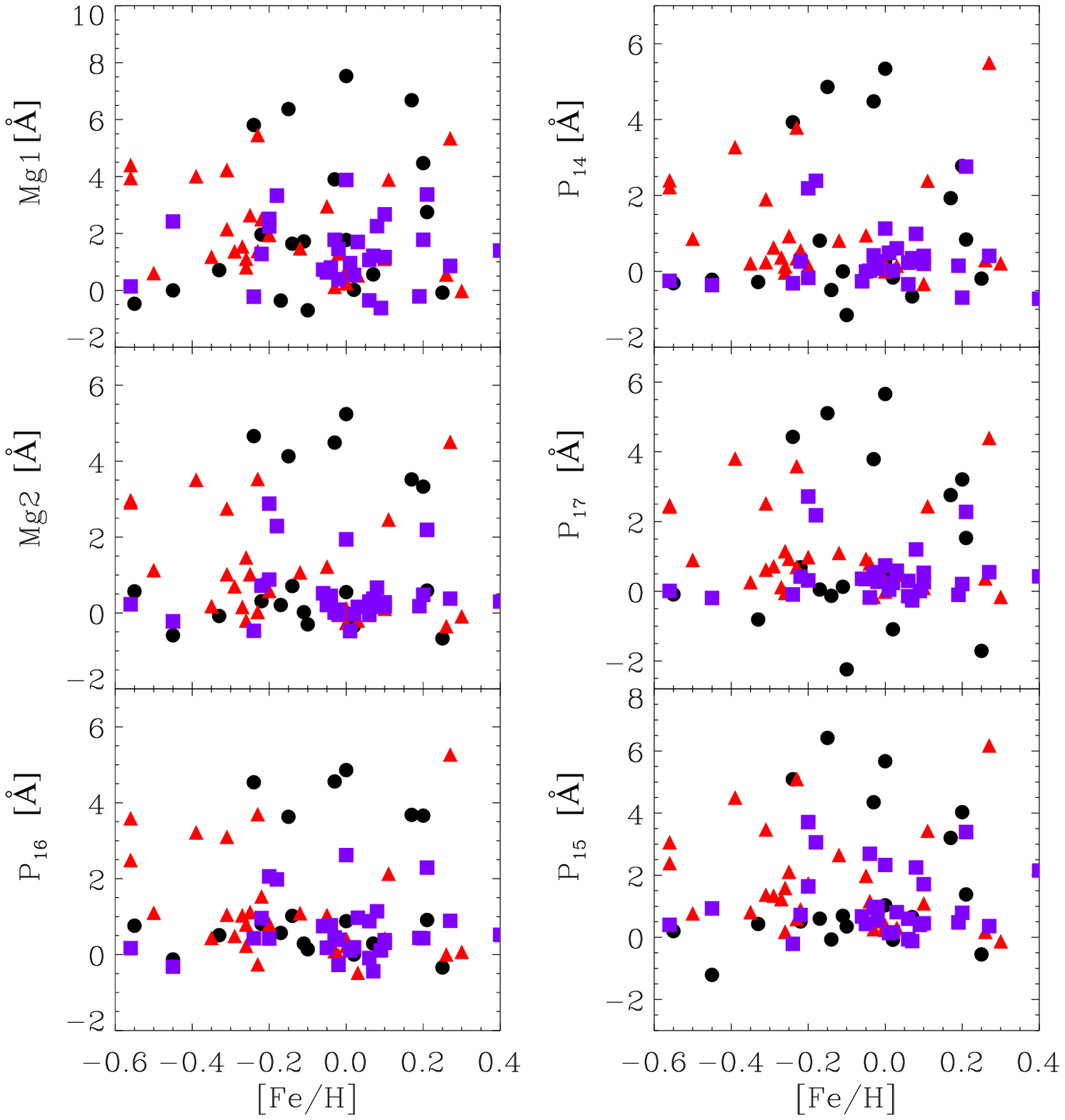}
\includegraphics[width=9truecm,angle=0]{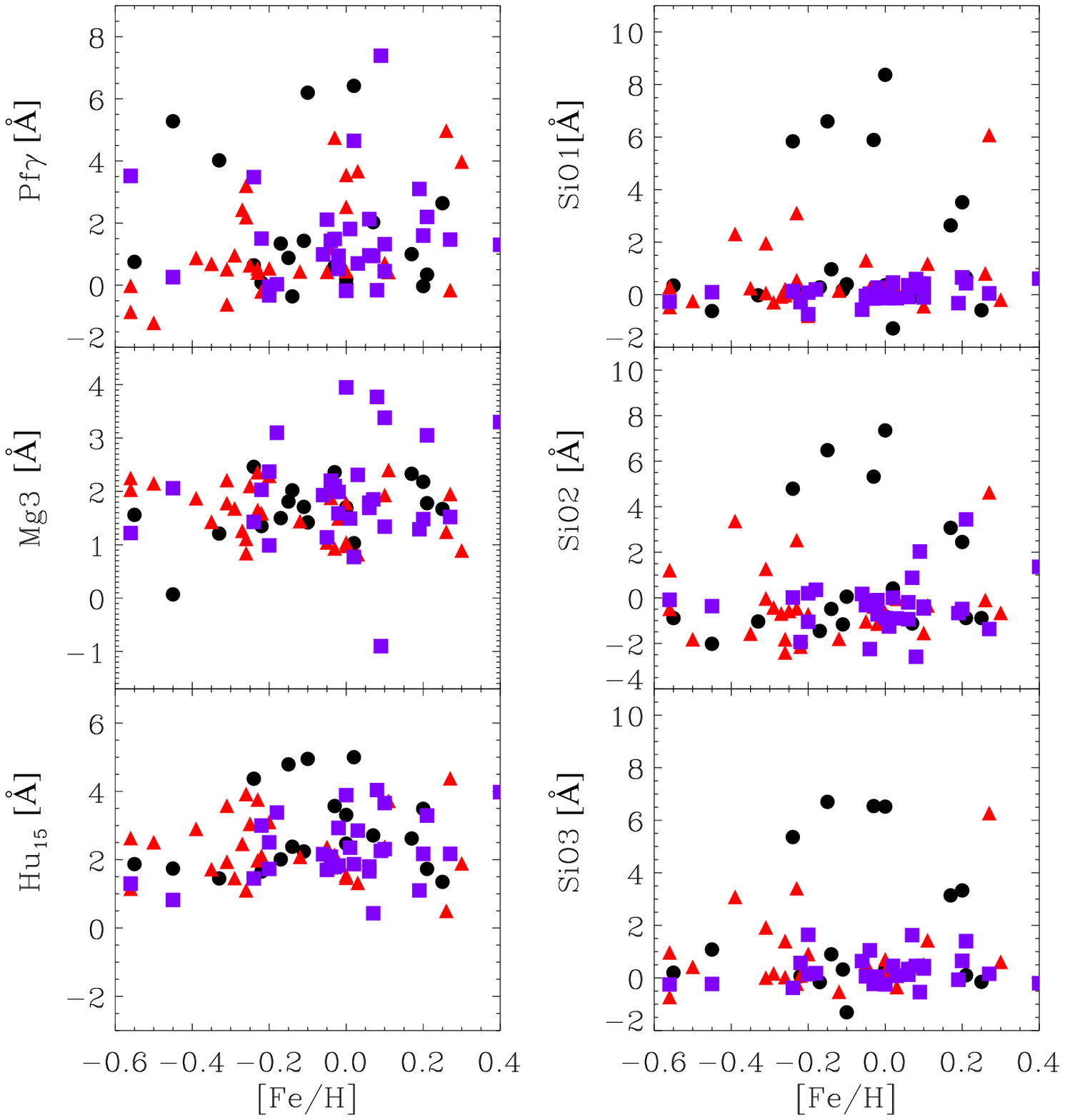}
\caption{Index measurements in the L atmospheric window as 
functions of surface gravity (top panels) and metallicity (bottom
panels). For details see Fig.\,\ref{fig:Ind_Spt_Y}.}
\label{fig:Ind_Gr_L}
\end{figure*}

The behavior of index measurements in the Y, J, H, and L atmospheric 
window as functions of effective temperature, surface gravity 
and metallicity for stars of different luminosity classes are shown in 
Figs.\,\ref{fig:Ind_Spt_Y}, \ref{fig:Ind_Gr_Y}, and 
\ref{fig:Ind_Spt_J}--\ref{fig:Ind_Gr_L}.

\clearpage
\section{Equivalent width tables}\label{app:table}

In this section we report values measured for all the indices in the different bands.
The tables are just an example and their integral version will be published in the online material.

\onecolumn
\begin{center}\tiny
\begin{longtable}{lrrrrrrrrr}
\caption{Equivalent widths of the $Y$-band indices. Part A.}
\label{tab:index_mis_Y_1} \\
\hline
\hline
Star IDs  & CSi & Pa$\epsilon$ & Ti  & TiO\,A   &  TiO\,B  &  FeH   &  Pa$\delta$  &  FeTi  &  Fe  \\
\hline                                                 
\endfirsthead
\multicolumn{10}{c}%
{{\bfseries \tablename\ \thetable{} -- continued from previous page}} \\
\hline
\hline
Star IDs  & CSi & Pa$\epsilon$ & Ti  & TiO\,A   &  TiO\,B  &  FeH   &  Pa$\delta$  &  FeTi  &  Fe   \\
\hline                                                  
\endhead
\hline
\\
\multicolumn{10}{c}{\textbf{Supergiants}}\\

    HD007927 & 0.74 $\pm$ 0.06  &  3.07 $\pm$0.65  &  0.03 $\pm$0.01 &   0.03 $\pm$ 0.01 &   0.56 $\pm$0.01 &   0.49  $\pm$0.02 &  2.92 $\pm$0.03  & -0.05 $\pm$0.03 &  -0.05 $\pm$0.03  \\
    HD006130 & 0.82 $\pm$ 0.05  &  4.11 $\pm$0.46  & -0.05 $\pm$0.03 &  -0.02 $\pm$ 0.02 &   0.41 $\pm$0.01 &   0.22  $\pm$0.03 &  3.44 $\pm$0.08  & -0.03 $\pm$0.07 &  -0.06 $\pm$0.07  \\
    HD135153 & 0.61 $\pm$ 0.13  &  3.92 $\pm$0.46  & -0.05 $\pm$0.03 &  -0.05 $\pm$ 0.03 &   0.32 $\pm$0.02 &   0.07  $\pm$0.05 &  3.49 $\pm$0.07  & -0.02 $\pm$0.05 &  -0.05 $\pm$0.05  \\
    HD173638 & 0.78 $\pm$ 0.07  &  4.34 $\pm$0.42  & -0.04 $\pm$0.01 &  -0.02 $\pm$ 0.02 &   0.41 $\pm$0.01 &   0.16  $\pm$0.04 &  3.58 $\pm$0.10  & -0.07 $\pm$0.05 &  -0.10 $\pm$0.05  \\
    ...      &  ...             &...               &    ...          &     ...           &     ...          &    ...            &    ...           &      ...        &       ...       \\

   \multicolumn{10}{c}{\textbf{Giants}}\\

    HD089025   &    0.89 $\pm$ 0.01  &  3.89 $\pm$ 0.25  &   -0.03 $\pm$ 0.02  &   -0.03 $\pm$ 0.02  &    0.31 $\pm$ 0.01  &    0.19 $\pm$ 0.03  &    2.92 $\pm$ 0.08  &    -0.05 $\pm$ 0.06  &   -0.06 $\pm$ 0.06   \\
    HD027397   &    0.68 $\pm$ 0.02  &  4.89 $\pm$ 0.21  &   -0.05 $\pm$ 0.02  &   -0.06 $\pm$ 0.02  &    0.33 $\pm$ 0.01  &    0.09 $\pm$ 0.03  &    3.51 $\pm$ 0.10  &    -0.01 $\pm$ 0.03  &   -0.01 $\pm$ 0.03   \\
    HD013174   &    0.92 $\pm$ 0.01  &  3.67 $\pm$ 0.23  &   -0.06 $\pm$ 0.01  &   -0.06 $\pm$ 0.02  &    0.29 $\pm$ 0.01  &    0.11 $\pm$ 0.04  &    2.82 $\pm$ 0.09  &    -0.05 $\pm$ 0.04  &   -0.06 $\pm$ 0.04   \\
    HD40535   &    0.89 $\pm$ 0.02  &  3.49 $\pm$ 0.22  &   -0.05 $\pm$ 0.02  &   -0.04 $\pm$ 0.02  &    0.30 $\pm$ 0.01  &    0.05 $\pm$ 0.02  &    2.68 $\pm$ 0.08  &    -0.02 $\pm$ 0.03  &   -0.02 $\pm$ 0.03   \\
     ...      &  ...             &...               &    ...          &     ...           &     ...          &    ...            &    ...           &      ...        &       ...       \\

  \multicolumn{10}{c}{\textbf{Dwarfs}}\\

    HD108519  &   0.73  $\pm$ 0.04  &    4.45 $\pm$ 0.21  &   -0.03 $\pm$ 0.01  &   -0.02  $\pm$ 0.01  &   0.24 $\pm$ 0.02  &    0.11  $\pm$ 0.05  &   3.30  $\pm$ 0.09  &  -0.03 $\pm$ 0.03  &   -0.02  $\pm$  0.03    \\
    HD213135  &   0.74  $\pm$ 0.04  &    3.43 $\pm$ 0.20  &   -0.02 $\pm$ 0.01  &   -0.05  $\pm$ 0.01  &   0.19 $\pm$ 0.00  &    0.04  $\pm$ 0.01  &   2.71  $\pm$ 0.08  &   0.01 $\pm$ 0.03  &   -0.02  $\pm$  0.03    \\
    HD113139  &   0.95  $\pm$ 0.02  &    3.50 $\pm$ 0.22  &   -0.05 $\pm$ 0.02  &   -0.05  $\pm$ 0.01  &   0.25 $\pm$ 0.01  &    0.09  $\pm$ 0.03  &   2.78  $\pm$ 0.09  &  -0.03 $\pm$ 0.02  &   -0.03  $\pm$  0.02    \\
    HD026015  &   0.89  $\pm$ 0.03  &    3.62 $\pm$ 0.18  &   -0.05 $\pm$ 0.02  &   -0.06  $\pm$ 0.02  &   0.29 $\pm$ 0.01  &    0.10  $\pm$ 0.04  &   2.74  $\pm$ 0.09  &   0.02 $\pm$ 0.03  &   -0.03  $\pm$  0.03    \\
    ...      &  ...             &...               &    ...          &     ...           &     ...          &    ...            &    ...           &      ...        &       ...       \\
\hline
\end{longtable}
\end{center}

\onecolumn
\begin{center}\tiny
\begin{longtable}{lrrrrr}
\caption{Equivalent widths of the $Y$-band indices. Part B.}
\label{tab:index_mis_Y_1} \\
\hline
\hline
Star IDs  &  VO  &  Si  &  CN  &  Sr  &  Pa$\gamma$ \\
\hline                                                 
\endfirsthead
\multicolumn{6}{c}%
{{\bfseries \tablename\ \thetable{} -- continued from previous page}} \\
\hline
\hline
Star IDs  &  VO  &  Si  &  CN  &  Sr  &  Pa$\gamma$ \\
\hline                                                  
\endhead
\hline
\\
\multicolumn{6}{c}{\textbf{Supergiants}}\\

    HD007927 &   0.87 $\pm$0.04 &   3.26 $\pm$0.02  &  0.51 $\pm$  0.02  &  1.00 $\pm$0.01  &  2.56 $\pm$0.02 \\
    HD006130 &   0.78 $\pm$0.09 &   2.01 $\pm$0.02  &  0.40 $\pm$  0.01  &  0.70 $\pm$0.03  &  3.11 $\pm$0.06 \\
    HD135153 &   0.68 $\pm$0.06 &   1.73 $\pm$0.02  &  0.43 $\pm$  0.01  &  0.71 $\pm$0.03  &  3.18 $\pm$0.07 \\
    HD173638 &   0.67 $\pm$0.08 &   2.08 $\pm$0.03  &  0.50 $\pm$  0.03  &  0.77 $\pm$0.03  &  3.29 $\pm$0.06 \\
    ...      &     ...         &    ...            &      ...           &        ...       &      ...      \\

    \multicolumn{6}{c}{\textbf{Giants}}\\

    HD089025   &    0.52  $\pm$ 0.11  &   1.50  $\pm$ 0.03  &   0.41 $\pm$ 0.03  &    0.65  $\pm$ 0.03  &   2.98  $\pm$ 0.08 \\
    HD027397   &    0.53  $\pm$ 0.06  &   1.04  $\pm$ 0.03  &   0.35 $\pm$ 0.02  &    0.74  $\pm$ 0.01  &   3.59  $\pm$ 0.03 \\
    HD013174   &    0.53  $\pm$ 0.06  &   1.82  $\pm$ 0.04  &   0.38 $\pm$ 0.03  &    0.51  $\pm$ 0.02  &   2.83  $\pm$ 0.05 \\
    HD40535   &    0.65  $\pm$ 0.05  &   1.49  $\pm$ 0.02  &   0.37 $\pm$ 0.02  &    0.51  $\pm$ 0.01  &   2.70  $\pm$ 0.03 \\
        ...      &     ...         &    ...            &      ...           &        ...       &      ...      \\

  \multicolumn{6}{c}{\textbf{Dwarfs}}\\

    HD108519  &   0.43  $\pm$ 0.04  &   1.61  $\pm$  0.01  &   0.06 $\pm$  0.01  &    0.51  $\pm$  0.01  &   3.40 $\pm$ 0.03   \\
    HD213135  &   0.40  $\pm$ 0.05  &   1.19  $\pm$  0.02  &   0.24 $\pm$  0.02  &    0.37  $\pm$  0.01  &   2.86 $\pm$ 0.02   \\
    HD113139  &   0.34  $\pm$ 0.03  &   1.34  $\pm$  0.02  &   0.29 $\pm$  0.02  &    0.40  $\pm$  0.02  &   2.88 $\pm$ 0.05   \\
    HD026015  &   0.44  $\pm$ 0.03  &   1.21  $\pm$  0.02  &   0.34 $\pm$  0.02  &    0.53  $\pm$  0.03  &   2.80 $\pm$ 0.08   \\
        ...      &     ...         &    ...            &      ...           &        ...       &      ...      \\
\hline
\end{longtable}
\end{center}

\onecolumn
\begin{center}\tiny
\begin{landscape}
\begin{longtable}{lrrrrrrrrrrrr}
\caption{Equivalent widths of the $J$-band indices.}
\label{tab:index_mis_J_1} \\
\hline
\hline
Star IDs  & Na   &  FeCr  &    FeK   &     C   &     K  &     Mg   &    Si  &   SiMg  &    KMg   &   CrK  &    Pab  &   Al \\
\hline                                                 
\endfirsthead
\multicolumn{13}{c}%
{{\bfseries \tablename\ \thetable{} -- continued from previous page}} \\
\hline
\hline
Star IDs  & Na   &  FeCr  &    FeK   &     C   &     K  &     Mg   &    Si  &   SiMg  &    KMg   &   CrK  &    Pab  &   Al\\
\hline                                                  
\endhead
\hline
\\
\multicolumn{13}{c}{\textbf{Supergiants}}\\

    HD007927  &  -0.01 $\pm$  0.30  &    0.26 $\pm$  0.03  &    0.55 $\pm$ 0.07  &    1.22  $\pm$ 1.44  &   0.17  $\pm$ 0.04  &   0.46  $\pm$ 0.06   &   0.73  $\pm$  0.01   &   0.33 $\pm$ 0.02  &   -0.22 $\pm$  0.09   &   -0.79  $\pm$  0.19  &   2.36 $\pm$ 0.03   &    0.52  $\pm$ 0.02   \\
    HD006130  &   0.40 $\pm$  0.24  &    0.50 $\pm$  0.02  &    0.58 $\pm$ 0.05  &    0.74  $\pm$ 1.46  &   0.25  $\pm$ 0.04  &   0.38  $\pm$ 0.06   &   0.89  $\pm$  0.01   &   0.52 $\pm$ 0.04  &   -0.05 $\pm$  0.06   &   -0.65  $\pm$  0.17  &   4.37 $\pm$ 0.07   &    0.63  $\pm$ 0.02   \\
    HD135153  &   0.02 $\pm$  0.36  &    0.52 $\pm$  0.04  &    0.53 $\pm$ 0.09  &    0.61  $\pm$ 1.60  &   0.21  $\pm$ 0.05  &   0.34  $\pm$ 0.06   &   0.87  $\pm$  0.02   &   0.45 $\pm$ 0.04  &   -0.03 $\pm$  0.07   &   -0.64  $\pm$  0.17  &   4.18 $\pm$ 0.07   &    0.59  $\pm$ 0.07   \\
    HD173638  &   0.27 $\pm$  0.30  &    0.43 $\pm$  0.03  &    0.58 $\pm$ 0.05  &    0.78  $\pm$ 1.50  &   0.19  $\pm$ 0.03  &   0.34  $\pm$ 0.05   &   0.91  $\pm$  0.02   &   0.57 $\pm$ 0.04  &   -0.10 $\pm$  0.09   &   -0.81  $\pm$  0.22  &   4.32 $\pm$ 0.06   &    0.54  $\pm$ 0.03   \\
        ...      &     ...         &    ...            &      ...           &        ...       &      ...   &  ...      &     ...         &    ...            &      ...           &        ...       &      ...  &      ...    \\

\multicolumn{13}{c}{\textbf{Giants}}\\

    HD089025  &   0.62  $\pm$ 0.18  &    0.50   $\pm$ 0.01  &    0.52   $\pm$ 0.02  &    0.51  $\pm$  1.52  &     0.17  $\pm$ 0.03   &     0.31  $\pm$ 0.05   &     0.78  $\pm$ 0.02   &     0.43  $\pm$  0.04  &     0.12   $\pm$ 0.03   &   -0.37   $\pm$0.11   &    4.79  $\pm$  0.07  &     0.40   $\pm$ 0.02   \\
    HD027397  &   0.41  $\pm$ 0.19  &    0.48   $\pm$ 0.02  &    0.34   $\pm$ 0.05  &    0.26  $\pm$  1.68  &     0.20  $\pm$ 0.02   &     0.40  $\pm$ 0.02   &     0.75  $\pm$ 0.01   &     0.46  $\pm$  0.02  &     0.09   $\pm$ 0.03   &   -0.19   $\pm$0.06   &    5.17  $\pm$  0.11  &     0.46   $\pm$ 0.03   \\
    HD013174  &   0.25  $\pm$ 0.22  &    0.61   $\pm$ 0.02  &    0.54   $\pm$ 0.05  &    0.51  $\pm$  1.55  &     0.37  $\pm$ 0.03   &     0.41  $\pm$ 0.02   &     0.93  $\pm$ 0.02   &     0.55  $\pm$  0.04  &     0.20   $\pm$ 0.03   &   -0.44   $\pm$0.09   &    4.67  $\pm$  0.05  &     0.69   $\pm$ 0.02   \\
     HD40535  &   0.36  $\pm$ 0.20  &    0.47   $\pm$ 0.01  &    0.51   $\pm$ 0.03  &    0.51  $\pm$  1.51  &     0.21  $\pm$ 0.02   &     0.36  $\pm$ 0.03   &     0.82  $\pm$ 0.01   &     0.43  $\pm$  0.02  &     0.20   $\pm$ 0.02   &   -0.30   $\pm$0.07   &    4.48  $\pm$  0.08  &     0.47   $\pm$ 0.01   \\
        ...      &     ...         &    ...            &      ...           &        ...       &      ...   &  ...      &     ...         &    ...            &      ...           &        ...       &      ...  &      ...    \\

\multicolumn{13}{c}{\textbf{Dwarf}}\\

    HD108519  &   0.22 $\pm$ 0.20   &    0.44 $\pm$ 0.02   &    0.47 $\pm$  0.04   &    0.52 $\pm$ 1.65   &    0.30 $\pm$0.02   &    0.26 $\pm$ 0.02   &    0.69 $\pm$ 0.01   &    0.38 $\pm$ 0.04   &   -0.04 $\pm$ 0.04   &   -0.45 $\pm$  0.09   &    5.26 $\pm$ 0.12   &    0.43 $\pm$  0.02   \\
    HD213135  &   0.39 $\pm$ 0.19   &    0.43 $\pm$ 0.01   &    0.43 $\pm$  0.03   &    0.38 $\pm$ 1.55   &    0.24 $\pm$0.03   &    0.38 $\pm$ 0.03   &    0.71 $\pm$ 0.01   &    0.41 $\pm$ 0.03   &    0.15 $\pm$ 0.03   &   -0.19 $\pm$  0.07   &    4.87 $\pm$ 0.08   &    0.44 $\pm$  0.02   \\
    HD113139  &   0.54 $\pm$ 0.14   &    0.49 $\pm$ 0.02   &    0.51 $\pm$  0.04   &    0.37 $\pm$ 1.49   &    0.25 $\pm$0.02   &    0.36 $\pm$ 0.02   &    0.72 $\pm$ 0.01   &    0.44 $\pm$ 0.03   &    0.19 $\pm$ 0.03   &   -0.23 $\pm$  0.07   &    4.93 $\pm$ 0.09   &    0.49 $\pm$  0.03   \\
    HD026015  &   0.50 $\pm$ 0.17   &    0.50 $\pm$ 0.01   &    0.39 $\pm$  0.03   &    0.29 $\pm$ 1.59   &    0.26 $\pm$0.01   &    0.44 $\pm$ 0.03   &    0.79 $\pm$ 0.01   &    0.50 $\pm$ 0.04   &    0.23 $\pm$ 0.03   &   -0.20 $\pm$  0.06   &    4.72 $\pm$ 0.10   &    0.53 $\pm$  0.01   \\
        ...      &     ...         &    ...            &      ...           &        ...       &      ...   &  ...      &     ...         &    ...            &      ...           &        ...       &      ...  &      ...    \\
\hline
\end{longtable}
\end{landscape}
\end{center}

\onecolumn
\begin{center}\tiny
\begin{landscape}
\begin{longtable}{lrrrrrrrrrrr}
\caption{Equivalent widths of the $H$-band indices. Part A.}
\label{tab:index_mis_Ha_1} \\

\hline
\hline
Star IDs  & Mg1  &    Mg2   &    K1  &    Fe1   &  Br$_{16}$  &    CO1  &   Br$_{15}$   &   Mg3  &   FeH1  &    HMg   &   CO2S ts \\
\hline                                                 
\endfirsthead
\multicolumn{12}{c}%
{{\bfseries \tablename\ \thetable{} -- continued from previous page}} \\
\hline
\hline
Star IDs  & Mg1  &    Mg2   &    K1  &    Fe1   &  Br$_{16}$  &    CO1  &   Br$_{15}$   &   Mg3  &   FeH1  &    HMg   &   CO2S ts \\
\hline                                                  
\endhead
\hline
\\
\multicolumn{12}{c}{\textbf{Supergiants}}\\

    HD007927 &   0.34 $\pm$  0.55  &    1.20 $\pm$0.60   &    1.40 $\pm$0.07   &    0.08 $\pm$ 0.09   &    3.07 $\pm$ 0.11   &    0.69 $\pm$ 0.12   &    4.01 $\pm$ 0.02   &    0.56 $\pm$ 0.22   &    0.10 $\pm$ 0.12   &    3.74 $\pm$ 0.17   &    0.66  $\pm$ 0.09    \\
    HD006130 &   0.38 $\pm$  0.13  &    0.90 $\pm$0.32   &    0.60 $\pm$0.06   &    0.06 $\pm$ 0.15   &    3.07 $\pm$ 0.25   &    0.95 $\pm$ 0.26   &    4.71 $\pm$ 0.03   &    1.16 $\pm$ 0.49   &    0.48 $\pm$ 0.26   &    4.93 $\pm$ 0.37   &    0.50  $\pm$ 0.11    \\
    HD135153 &   0.31 $\pm$  0.17  &    0.84 $\pm$0.32   &    0.65 $\pm$0.06   &    0.13 $\pm$ 0.15   &    3.47 $\pm$ 0.23   &    1.27 $\pm$ 0.24   &    5.28 $\pm$ 0.04   &    1.70 $\pm$ 0.46   &    0.45 $\pm$ 0.25   &    5.23 $\pm$ 0.35   &    0.22  $\pm$ 0.14    \\
    HD173638 &   0.30 $\pm$  0.18  &    1.04 $\pm$0.37   &    0.84 $\pm$0.07   &    0.07 $\pm$ 0.18   &    3.31 $\pm$ 0.31   &    1.01 $\pm$ 0.32   &    4.99 $\pm$ 0.03   &    1.58 $\pm$ 0.61   &    0.40 $\pm$ 0.32   &    5.14 $\pm$ 0.46   &    0.50  $\pm$ 0.11    \\
        ...      &     ...         &    ...            &      ...           &        ...       &      ...       &     ...         &    ...            &      ...           &        ...       &      ...  &      ...    \\

\multicolumn{12}{c}{\textbf{Giants}}\\
    
    HD089025 &   0.34 $\pm$  0.08  &    0.65 $\pm$0.23    &    0.09 $\pm$ 0.03   &    0.01 $\pm$0.12   &    1.49 $\pm$ 0.22   &    0.54 $\pm$ 0.22   &    3.82 $\pm$ 0.04   &    1.15 $\pm$ 0.42   &    0.44 $\pm$ 0.23   &    3.70 $\pm$0.33   &    0.15 $\pm$ 0.17   \\
    HD027397 &   0.49 $\pm$  0.08  &    0.71 $\pm$0.26    &    0.04 $\pm$ 0.04   &    0.05 $\pm$0.09   &    1.02 $\pm$ 0.16   &    0.63 $\pm$ 0.16   &    2.37 $\pm$ 0.04   &    1.62 $\pm$ 0.30   &    0.49 $\pm$ 0.16   &    2.46 $\pm$0.24   &    0.23 $\pm$ 0.15   \\
    HD013174 &   0.53 $\pm$  0.10  &    0.89 $\pm$0.31    &    0.09 $\pm$ 0.04   &   -0.01 $\pm$0.09   &    1.39 $\pm$ 0.15   &    0.80 $\pm$ 0.16   &    2.95 $\pm$ 0.03   &    1.88 $\pm$ 0.30   &    0.62 $\pm$ 0.16   &    3.46 $\pm$0.23   &    0.42 $\pm$ 0.12   \\
     HD40535 &   0.42 $\pm$  0.08  &    0.83 $\pm$0.29    &    0.02 $\pm$ 0.04   &    0.01 $\pm$0.14   &    1.08 $\pm$ 0.24   &    0.55 $\pm$ 0.25   &    2.84 $\pm$ 0.05   &    1.58 $\pm$ 0.47   &    0.49 $\pm$ 0.25   &    2.99 $\pm$0.37   &    0.28 $\pm$ 0.21   \\
        ...      &     ...         &    ...            &      ...           &        ...       &      ...       &     ...         &    ...            &      ...           &        ...       &      ...  &      ...    \\

\multicolumn{12}{c}{\textbf{Dwarfs}}\\

    HD108519 &   0.40 $\pm$0.06   &    0.65 $\pm$ 0.22   &   -0.00 $\pm$0.02   &    0.03 $\pm$ 0.06    &    1.08 $\pm$ 0.09   &    0.69 $\pm$ 0.09    &    2.61 $\pm$ 0.02   &    1.52 $\pm$ 0.18   &    0.41 $\pm$ 0.10   &    2.56 $\pm$ 0.14   &    0.09 $\pm$ 0.09   \\
    HD213135 &   0.61 $\pm$0.08   &    0.95 $\pm$ 0.34   &    0.02 $\pm$0.03   &    0.09 $\pm$ 0.06    &    0.75 $\pm$ 0.11   &    0.63 $\pm$ 0.11    &    1.96 $\pm$ 0.05   &    1.96 $\pm$ 0.20   &    0.50 $\pm$ 0.11   &    2.27 $\pm$ 0.17   &    0.35 $\pm$ 0.21   \\ 
    HD113139 &   0.55 $\pm$0.11   &    0.89 $\pm$ 0.30   &   -0.04 $\pm$0.04   &   -0.02 $\pm$ 0.07    &    0.80 $\pm$ 0.11   &    0.71 $\pm$ 0.11    &    1.48 $\pm$ 0.04   &    1.98 $\pm$ 0.21   &    0.45 $\pm$ 0.11   &    1.94 $\pm$ 0.17   &    0.40 $\pm$ 0.14   \\ 
    HD026015 &   0.74 $\pm$0.10   &    0.99 $\pm$ 0.36   &   -0.05 $\pm$0.05   &   -0.04 $\pm$ 0.09    &    0.96 $\pm$ 0.15   &    0.80 $\pm$ 0.15    &    1.48 $\pm$ 0.05   &    2.29 $\pm$ 0.28   &    0.67 $\pm$ 0.15   &    2.32 $\pm$ 0.23   &    0.34 $\pm$ 0.21   \\ 
        ...      &     ...         &    ...            &      ...           &        ...       &      ...       &     ...         &    ...            &      ...           &        ...       &      ...  &      ...    \\

\hline
\end{longtable}
\end{landscape}
\end{center}

\onecolumn
\begin{center}\tiny
\begin{landscape}
\begin{longtable}{lrrrrrrrrrrr}
\caption{Equivalent widths of the $H$-band indices. Part B.}
\label{tab:index_mis_Hb_1} \\

\hline
\hline
Star IDs  & Fe2   &  Br$_{13}$  &   CO3  &   FeH2   &   CO4  &    Fe3   &   CO5   &   Al1   & Br$_{11}$  &   COMg   &  Br$_{10}$ \\
\hline                                                 
\endfirsthead
\multicolumn{12}{c}%
{{\bfseries \tablename\ \thetable{} -- continued from previous page}} \\
\hline
\hline
Star IDs  & Fe2   &  Br$_{13}$  &   CO3  &   FeH2   &   CO4  &    Fe3   &   CO5   &   Al1   & Br$_{11}$  &   COMg   &  Br$_{10}$ \\
\hline                                                  
\endhead
\hline
\\
\multicolumn{12}{c}{\textbf{Supergiants}}\\

    HD007927 &  -0.20 $\pm$0.04   &    4.01 $\pm$0.02   &     0.18 $\pm$0.02   &     0.26 $\pm$ 0.03    &     4.19 $\pm$ 0.04   &    -0.01 $\pm$ 0.04   &     0.05 $\pm$ 0.02   &     0.14 $\pm$ 0.05   &     3.41 $\pm$ 0.06   &     0.13 $\pm$ 0.02   &     3.93 $\pm$ 0.14    \\  
    HD006130 &   0.11 $\pm$0.05   &    4.71 $\pm$0.03   &     0.01 $\pm$0.03   &    -0.11 $\pm$ 0.05    &     6.05 $\pm$ 0.06   &     0.09 $\pm$ 0.06   &    -0.25 $\pm$ 0.05   &    -0.77 $\pm$ 0.15   &     3.90 $\pm$ 0.11   &     0.39 $\pm$ 0.04   &     5.17 $\pm$ 0.14    \\  
    HD135153 &   0.29 $\pm$0.06   &    5.28 $\pm$0.04   &    -0.12 $\pm$0.05   &    -0.35 $\pm$ 0.07    &     6.73 $\pm$ 0.09   &    -0.15 $\pm$ 0.10   &    -0.22 $\pm$ 0.04   &    -0.82 $\pm$ 0.10   &     3.78 $\pm$ 0.10   &     0.58 $\pm$ 0.04   &     5.59 $\pm$ 0.12    \\  
    HD173638 &   0.12 $\pm$0.05   &    4.99 $\pm$0.03   &     0.05 $\pm$0.04   &    -0.13 $\pm$ 0.06    &     6.34 $\pm$ 0.07   &    -0.00 $\pm$ 0.07   &    -0.22 $\pm$ 0.03   &    -0.65 $\pm$ 0.09   &     3.91 $\pm$ 0.10   &     0.48 $\pm$ 0.03   &     5.14 $\pm$ 0.17    \\  
        ...      &     ...         &    ...            &      ...           &        ...       &      ...       &     ...         &    ...            &      ...           &        ...       &      ...  &      ...    \\
\multicolumn{12}{c}{\textbf{Giants}}\\

    HD089025 &   0.12 $\pm$ 0.08   &    3.82 $\pm$ 0.04   &   -0.11 $\pm$0.06   &   -0.49 $\pm$0.10   &    5.87 $\pm$ 0.13   &    0.10 $\pm$ 0.14   &   -0.29 $\pm$ 0.06   &   -1.17 $\pm$ 0.17   &    3.11 $\pm$ 0.11   &    0.47 $\pm$ 0.03   &    5.07 $\pm$ 0.12   \\
    HD027397 &   0.14 $\pm$ 0.07   &    2.37 $\pm$ 0.04   &   -0.27 $\pm$0.07   &   -0.61 $\pm$0.11   &    5.03 $\pm$ 0.14   &    0.20 $\pm$ 0.15   &   -0.37 $\pm$ 0.05   &   -1.41 $\pm$ 0.13   &    2.35 $\pm$ 0.10   &    0.38 $\pm$ 0.01   &    4.76 $\pm$ 0.06   \\
    HD013174 &  -0.03 $\pm$ 0.06   &    2.95 $\pm$ 0.03   &   -0.03 $\pm$0.06   &   -0.48 $\pm$0.09   &    5.54 $\pm$ 0.11   &    0.49 $\pm$ 0.12   &   -0.21 $\pm$ 0.04   &   -0.77 $\pm$ 0.11   &    2.49 $\pm$ 0.10   &    0.67 $\pm$ 0.03   &    4.78 $\pm$ 0.08   \\
     HD40535 &   0.03 $\pm$ 0.10   &    2.84 $\pm$ 0.05   &   -0.11 $\pm$0.06   &   -0.43 $\pm$0.10   &    4.79 $\pm$ 0.11   &    0.19 $\pm$ 0.11   &   -0.27 $\pm$ 0.05   &   -0.67 $\pm$ 0.14   &    2.47 $\pm$ 0.09   &    0.51 $\pm$ 0.02   &    5.11 $\pm$ 0.10   \\
        ...      &     ...         &    ...            &      ...           &        ...       &      ...       &     ...         &    ...            &      ...           &        ...       &      ...  &      ...    \\
\multicolumn{12}{c}{\textbf{Dwarfs}}\\

    HD108519  &   0.16 $\pm$0.04   &    2.61 $\pm$0.02    &   -0.21 $\pm$0.04   &   -0.53 $\pm$0.06   &    5.08 $\pm$0.08   &    0.07 $\pm$ 0.09   &   -0.22 $\pm$0.05   &   -1.22 $\pm$ 0.14   &    2.32 $\pm$ 0.11   &    0.44 $\pm$ 0.02   &    5.34 $\pm$ 0.08   \\
    HD213135  &   0.07 $\pm$0.09   &    1.96 $\pm$0.05    &    0.01 $\pm$0.07   &   -0.41 $\pm$0.11   &    3.92 $\pm$0.13   &    0.33 $\pm$ 0.14   &   -0.19 $\pm$0.04   &   -0.54 $\pm$ 0.10   &    1.76 $\pm$ 0.08   &    0.67 $\pm$ 0.03   &    4.13 $\pm$ 0.08   \\
    HD113139  &  -0.12 $\pm$0.06   &    1.48 $\pm$0.04    &    0.02 $\pm$0.04   &   -0.36 $\pm$0.07   &    3.52 $\pm$0.08   &    0.34 $\pm$ 0.08   &   -0.20 $\pm$0.04   &   -0.54 $\pm$ 0.10   &    1.59 $\pm$ 0.08   &    0.61 $\pm$ 0.03   &    4.34 $\pm$ 0.09   \\
    HD026015  &  -0.05 $\pm$0.09   &    1.48 $\pm$0.05    &   -0.11 $\pm$0.08   &   -0.60 $\pm$0.13   &    3.74 $\pm$0.17   &    0.32 $\pm$ 0.17   &   -0.23 $\pm$0.04   &   -0.68 $\pm$ 0.11   &    1.60 $\pm$ 0.08   &    0.63 $\pm$ 0.04   &    3.88 $\pm$ 0.09   \\
        ...      &     ...         &    ...            &      ...           &        ...       &      ...       &     ...         &    ...            &      ...           &        ...       &      ...  &      ...    \\
\hline
\end{longtable}
\end{landscape}
\end{center}

\onecolumn
\begin{center}\tiny
\begin{landscape}
\begin{longtable}{lrrrrrrrrrrrr}
\caption{Equivalent widths of the $L$-band indices.}
\label{tab:index_mis_L_1} \\

\hline
\hline
Star IDs  & Mg1 &  Mg2 &P$_{16}$ &P$_{14}$ & P$_{17}$ & P$_{15}$  & Pf$\gamma$ & Mg3 & Hu$_{15}$ & SiO1 & SiO2 & SiO3 \\
\hline                                                 
\endfirsthead
\multicolumn{13}{c}%
{{\bfseries \tablename\ \thetable{} -- continued from previous page}} \\
\hline
\hline
Star IDs  & Mg1 &  Mg2 &P$_{16}$ &P$_{14}$ & P$_{17}$ & P$_{15}$  & Pf$\gamma$ & Mg3 & Hu$_{15}$ & SiO1 & SiO2 & SiO3 \\
\hline                                                  
\endhead
\hline
\\
\multicolumn{13}{c}{\textbf{Supergiants}}\\

    HD007927 &  -0.39 $\pm$0.77   &   -0.40 $\pm$0.28   &    1.08 $\pm$ 0.96   &   -0.48 $\pm$ 1.02   &   -3.30 $\pm$ 0.82   &   -0.00 $\pm$ 0.62    &    7.01 $\pm$ 0.29    &    0.69 $\pm$ 0.24   &    5.52 $\pm$ 0.32   &    0.13 $\pm$ 0.40   &   -0.40 $\pm$ 0.42   &    1.07  $\pm$ 0.45   \\
    HD006130 &   0.02 $\pm$0.94   &   -0.34 $\pm$0.19   &    0.00 $\pm$ 0.59   &   -0.16 $\pm$ 0.57   &   -1.09 $\pm$ 0.45   &   -0.08 $\pm$ 0.27    &    6.42 $\pm$ 0.21    &    1.03 $\pm$ 0.17   &    5.00 $\pm$ 0.26   &   -1.28 $\pm$ 0.41   &    0.41 $\pm$ 0.68   &    -                  \\
    HD135153 &  -0.88 $\pm$1.03   &   -0.14 $\pm$0.25   &    0.18 $\pm$ 0.31   &   -0.03 $\pm$ 0.39   &   -1.80 $\pm$ 0.32   &    0.07 $\pm$ 0.33    &    5.49 $\pm$ 0.19    &    1.02 $\pm$ 0.09   &    4.25 $\pm$ 0.12   &   -0.17 $\pm$ 0.22   &   -0.80 $\pm$ 0.18   &    0.37  $\pm$ 0.12   \\
    HD173638 &  -0.70 $\pm$0.97   &   -0.30 $\pm$0.23   &    0.14 $\pm$ 0.61   &   -1.15 $\pm$ 0.64   &   -2.24 $\pm$ 0.51   &    0.35 $\pm$ 0.40    &    6.20 $\pm$ 0.27    &    1.42 $\pm$ 0.26   &    4.95 $\pm$ 0.44   &    0.40 $\pm$ 0.32   &    0.05 $\pm$ 0.44   &   -1.31  $\pm$ 0.69   \\
        ...      &     ...         &    ...            &      ...           &        ...       &      ...       &     ...         &    ...            &      ...           &        ...       &      ...  &      ...  &      ...  \\

      \multicolumn{13}{c}{\textbf{Giants}}\\
      
    HD089025 &   0.12 $\pm$ 0.58   &    0.15 $\pm$ 0.12   &    0.09 $\pm$ 0.16   &    0.16 $\pm$ 0.13   &   -0.16 $\pm$ 0.11   &    0.26 $\pm$ 0.09   &   4.75 $\pm$ 0.18   &    0.93 $\pm$ 0.09   &    2.14 $\pm$ 0.19   &   -0.06 $\pm$ 0.15   &   -0.39 $\pm$ 0.24   &   -0.01 $\pm$   0.29      \\
    HD027397 &   0.51 $\pm$ 0.31   &   -0.06 $\pm$ 0.06   &    0.19 $\pm$ 0.13   &   -0.29 $\pm$ 0.09   &   -0.18 $\pm$ 0.07   &    0.11 $\pm$ 0.07   &   6.27 $\pm$ 0.11   &    0.77 $\pm$ 0.08   &    2.25 $\pm$ 0.12   &   -0.01 $\pm$ 0.18   &    0.17 $\pm$ 0.17   &    0.19 $\pm$   0.15      \\
    HD013174 &  -0.02 $\pm$ 0.63   &   -0.09 $\pm$ 0.11   &    0.07 $\pm$ 0.20   &    0.21 $\pm$ 0.16   &   -0.16 $\pm$ 0.13   &   -0.13 $\pm$ 0.09   &   3.98 $\pm$ 0.17   &    0.89 $\pm$ 0.15   &    1.89 $\pm$ 0.17   &   -0.19 $\pm$ 0.15   &   -0.66 $\pm$ 0.22   &    0.61 $\pm$   0.23      \\
     HD40535 &   0.54 $\pm$ 0.51   &    0.26 $\pm$ 0.16   &   -0.37 $\pm$ 0.42   &    0.31 $\pm$ 0.28   &    0.66 $\pm$ 0.23   &    0.01 $\pm$ 0.21   &   5.41 $\pm$ 0.18   &    0.96 $\pm$ 0.23   &    1.97 $\pm$ 0.33   &   -0.47 $\pm$ 0.35   &    0.66 $\pm$ 0.27   &   -2.05 $\pm$   0.37      \\
        ...      &     ...         &    ...            &      ...           &        ...       &      ...       &     ...         &    ...            &      ...           &        ...       &      ...  &      ...  &      ...  \\

      \multicolumn{13}{c}{\textbf{Dwarfs}}\\

    HD108519 &    0.63 $\pm$0.77    &   -0.03 $\pm$ 0.16   &    0.41 $\pm$0.37   &    0.05 $\pm$ 0.30   &   -0.07 $\pm$ 0.24   &    0.78 $\pm$ 0.30   &     5.11 $\pm$ 0.37   &      0.06 $\pm$0.18   &    0.91 $\pm$ 0.37   &   -0.25 $\pm$ 0.31   &    0.69 $\pm$ 0.51   &   -0.91  $\pm$   0.80       \\
    HD213135 &   -0.22 $\pm$0.62    &   -0.47 $\pm$ 0.23   &    0.43 $\pm$0.41   &   -0.31 $\pm$ 0.30   &   -0.09 $\pm$ 0.24   &   -0.21 $\pm$ 0.18   &     3.48 $\pm$ 0.23   &      1.43 $\pm$0.16   &    1.45 $\pm$ 0.31   &    0.14 $\pm$ 0.35   &    0.01 $\pm$ 0.49   &   -0.38  $\pm$   0.66       \\
    HD113139 &    0.54 $\pm$0.61    &   -0.03 $\pm$ 0.16   &    0.19 $\pm$0.22   &    0.02 $\pm$ 0.18   &    0.24 $\pm$ 0.14   &    0.16 $\pm$ 0.15   &     4.65 $\pm$ 0.18   &      0.77 $\pm$0.07   &    1.87 $\pm$ 0.17   &    0.47 $\pm$ 0.23   &    0.00 $\pm$ 0.22   &    0.46  $\pm$   0.24       \\
    HD026015 &   -0.62 $\pm$0.67    &    0.18 $\pm$ 0.13   &    0.11 $\pm$0.16   &    0.26 $\pm$ 0.10   &    0.01 $\pm$ 0.08   &    0.40 $\pm$ 0.06   &     7.39 $\pm$ 0.20   &     -0.90 $\pm$0.07   &    2.26 $\pm$ 0.11   &    0.42 $\pm$ 0.40   &    2.03 $\pm$ 0.32   &   -0.54  $\pm$   0.15       \\
        ...      &     ...         &    ...            &      ...           &        ...       &      ...       &     ...         &    ...            &      ...           &        ...       &      ...  &      ...  &      ...  \\

\hline
\end{longtable}
\end{landscape}
\end{center}

\end{appendix}

\end{document}